\documentclass[10pt,journal,compsoc]{IEEEtran}

\makeatletter
\def\endthebibliography{%
	\def\@noitemerr{\@latex@warning{Empty `thebibliography' environment}}%
	\endlist
}
\makeatother

\usepackage[linesnumbered,commentsnumbered,ruled,vlined]{algorithm2e}
\usepackage{amssymb}
\usepackage{graphicx}
\usepackage{subfig}
\usepackage{makeidx}
\usepackage{multirow}
\usepackage{listings}
\usepackage[singlelinecheck=false]{caption}
\usepackage{xcolor}
\usepackage{array}
\usepackage{balance}
\usepackage{amsmath}
\usepackage{fancyvrb}
\usepackage{tabularx}
\usepackage{varwidth} 
\usepackage[inline]{enumitem}
\usepackage{lipsum}
\usepackage{enumitem}
\usepackage[T1]{fontenc}

\lstdefinelanguage[ARM]{Assembler}%
{morekeywords={adc,adcal,adcals,adccc,adcccs,adccs,adccss,adceq,adceqs,    %
		adcge,adcges,adcgt,adcgts,adchi,adchis,adchs,adchss,adcle,adcles,       %
		adclo,adclos,adcls,adclss,adclt,adclts,adcmi,adcmis,adcne,adcnes,       %
		adcpl,adcpls,adcs,adcvc,adcvcs,adcvs,adcvss,add,addal,addals,addcc,     %
		addccs,addcs,addcss,addeq,addeqs,addge,addges,addgt,addgts,addhi,       %
		addhis,addhs,addhss,addle,addles,addlo,addlos,addls,addlss,addlt,       %
		addlts,addmi,addmis,addne,addnes,addpl,addpls,adds,addvc,addvcs,addvs,  %
		addvss,and,andal,andals,andcc,andccs,andcs,andcss,andeq,andeqs,andge,   %
		andges,andgt,andgts,andhi,andhis,andhs,andhss,andle,andles,andlo,       %
		andlos,andls,andlss,andlt,andlts,andmi,andmis,andne,andnes,andpl,       %
		andpls,ands,andvc,andvcs,andvs,andvss,b,bal,bcc,bcs,beq,bge,bgt,bhi,    %
		bhs,bic,bical,bicals,biccc,bicccs,biccs,biccss,biceq,biceqs,bicge,      %
		bicges,bicgt,bicgts,bichi,bichis,bichs,bichss,bicle,bicles,biclo,       %
		biclos,bicls,biclss,biclt,biclts,bicmi,bicmis,bicne,bicnes,bicpl,       %
		bicpls,bics,bicvc,bicvcs,bicvs,bicvss,bkpt,bl,blal,blcc,blcs,ble,bleq,  %
		blge,blgt,blhi,blhs,blle,bllo,blls,bllt,blmi,blne,blo,blpl,bls,blt,     %
		blvc,blvs,blx,blxal,blxcc,blxcs,blxeq,blxge,blxgt,blxhi,blxhs,blxle,    %
		blxlo,blxls,blxlt,blxmi,blxne,blxpl,blxvc,blxvs,bmi,bne,bpl,bvc,bvs,    %
		bx,bxal,bxcc,bxcs,bxeq,bxge,bxgt,bxhi,bxhs,bxj,bxjal,bxjcc,bxjcs,       %
		bxjeq,bxjge,bxjgt,bxjhi,bxjhs,bxjle,bxjlo,bxjls,bxjlt,bxjmi,bxjne,      %
		bxjpl,bxjvc,bxjvs,bxle,bxlo,bxls,bxlt,bxmi,bxne,bxpl,bxvc,bxvs,cdp,     %
		cdp2,cdpal,cdpcc,cdpcs,cdpeq,cdpge,cdpgt,cdphi,cdphs,cdple,cdplo,       %
		cdpls,cdplt,cdpmi,cdpne,cdppl,cdpvc,cdpvs,clz,clzal,clzcc,clzcs,clzeq,  %
		clzge,clzgt,clzhi,clzhs,clzle,clzlo,clzls,clzlt,clzmi,clzne,clzpl,      %
		clzvc,clzvs,cmn,cmnal,cmncc,cmncs,cmneq,cmnge,cmngt,cmnhi,cmnhs,cmnle,  %
		cmnlo,cmnls,cmnlt,cmnmi,cmnne,cmnpl,cmnvc,cmnvs,cmp,cmpal,cmpcc,cmpcs,  %
		cmpeq,cmpge,cmpgt,cmphi,cmphs,cmple,cmplo,cmpls,cmplt,cmpmi,cmpne,      %
		cmppl,cmpvc,cmpvs,cps,cpsid,cpsie,cpy,cpyal,cpycc,cpycs,cpyeq,cpyge,    %
		cpygt,cpyhi,cpyhs,cpyle,cpylo,cpyls,cpylt,cpymi,cpyne,cpypl,cpyvc,      %
		cpyvs,eor,eoral,eorals,eorcc,eorccs,eorcs,eorcss,eoreq,eoreqs,eorge,    %
		eorges,eorgt,eorgts,eorhi,eorhis,eorhs,eorhss,eorle,eorles,eorlo,       %
		eorlos,eorls,eorlss,eorlt,eorlts,eormi,eormis,eorne,eornes,eorpl,       %
		eorpls,eors,eorvc,eorvcs,eorvs,eorvss,ldc,ldc2,ldcal,ldccc,ldccs,       %
		ldceq,ldcge,ldcgt,ldchi,ldchs,ldcle,ldclo,ldcls,ldclt,ldcmi,ldcne,      %
		ldcpl,ldcvc,ldcvs,ldmalda,ldmaldb,ldmalea,ldmaled,ldmalfa,ldmalfd,      %
		ldmalia,ldmalib,ldmccda,ldmccdb,ldmccea,ldmcced,ldmccfa,ldmccfd,        %
		ldmccia,ldmccib,ldmcsda,ldmcsdb,ldmcsea,ldmcsed,ldmcsfa,ldmcsfd,        %
		ldmcsia,ldmcsib,ldmda,ldmdb,ldmea,ldmed,ldmeqda,ldmeqdb,ldmeqea,        %
		ldmeqed,ldmeqfa,ldmeqfd,ldmeqia,ldmeqib,ldmfa,ldmfd,ldmgeda,ldmgedb,    %
		ldmgeea,ldmgeed,ldmgefa,ldmgefd,ldmgeia,ldmgeib,ldmgtda,ldmgtdb,        %
		ldmgtea,ldmgted,ldmgtfa,ldmgtfd,ldmgtia,ldmgtib,ldmhida,ldmhidb,        %
		ldmhiea,ldmhied,ldmhifa,ldmhifd,ldmhiia,ldmhiib,ldmhsda,ldmhsdb,        %
		ldmhsea,ldmhsed,ldmhsfa,ldmhsfd,ldmhsia,ldmhsib,ldmia,ldmib,ldmleda,    %
		ldmledb,ldmleea,ldmleed,ldmlefa,ldmlefd,ldmleia,ldmleib,ldmloda,        %
		ldmlodb,ldmloea,ldmloed,ldmlofa,ldmlofd,ldmloia,ldmloib,ldmlsda,        %
		ldmlsdb,ldmlsea,ldmlsed,ldmlsfa,ldmlsfd,ldmlsia,ldmlsib,ldmltda,        %
		ldmltdb,ldmltea,ldmlted,ldmltfa,ldmltfd,ldmltia,ldmltib,ldmmida,        %
		ldmmidb,ldmmiea,ldmmied,ldmmifa,ldmmifd,ldmmiia,ldmmiib,ldmneda,        %
		ldmnedb,ldmneea,ldmneed,ldmnefa,ldmnefd,ldmneia,ldmneib,ldmplda,        %
		ldmpldb,ldmplea,ldmpled,ldmplfa,ldmplfd,ldmplia,ldmplib,ldmvcda,        %
		ldmvcdb,ldmvcea,ldmvced,ldmvcfa,ldmvcfd,ldmvcia,ldmvcib,ldmvsda,        %
		ldmvsdb,ldmvsea,ldmvsed,ldmvsfa,ldmvsfd,ldmvsia,ldmvsib,ldr,ldral,      %
		ldralb,ldralbt,ldrald,ldralh,ldralsb,ldralsh,ldralt,ldrb,ldrbt,ldrcc,   %
		ldrccb,ldrccbt,ldrccd,ldrcch,ldrccsb,ldrccsh,ldrcct,ldrcs,ldrcsb,       %
		ldrcsbt,ldrcsd,ldrcsh,ldrcssb,ldrcssh,ldrcst,ldrd,ldreq,ldreqb,         %
		ldreqbt,ldreqd,ldreqh,ldreqsb,ldreqsh,ldreqt,ldrex,ldrexal,ldrexcc,     %
		ldrexcs,ldrexeq,ldrexge,ldrexgt,ldrexhi,ldrexhs,ldrexle,ldrexlo,        %
		ldrexls,ldrexlt,ldrexmi,ldrexne,ldrexpl,ldrexvc,ldrexvs,ldrge,ldrgeb,   %
		ldrgebt,ldrged,ldrgeh,ldrgesb,ldrgesh,ldrget,ldrgt,ldrgtb,ldrgtbt,      %
		ldrgtd,ldrgth,ldrgtsb,ldrgtsh,ldrgtt,ldrh,ldrhi,ldrhib,ldrhibt,ldrhid,  %
		ldrhih,ldrhisb,ldrhish,ldrhit,ldrhs,ldrhsb,ldrhsbt,ldrhsd,ldrhsh,       %
		ldrhssb,ldrhssh,ldrhst,ldrle,ldrleb,ldrlebt,ldrled,ldrleh,ldrlesb,      %
		ldrlesh,ldrlet,ldrlo,ldrlob,ldrlobt,ldrlod,ldrloh,ldrlosb,ldrlosh,      %
		ldrlot,ldrls,ldrlsb,ldrlsbt,ldrlsd,ldrlsh,ldrlssb,ldrlssh,ldrlst,       %
		ldrlt,ldrltb,ldrltbt,ldrltd,ldrlth,ldrltsb,ldrltsh,ldrltt,ldrmi,        %
		ldrmib,ldrmibt,ldrmid,ldrmih,ldrmisb,ldrmish,ldrmit,ldrne,ldrneb,       %
		ldrnebt,ldrned,ldrneh,ldrnesb,ldrnesh,ldrnet,ldrpl,ldrplb,ldrplbt,      %
		ldrpld,ldrplh,ldrplsb,ldrplsh,ldrplt,ldrsb,ldrsh,ldrt,ldrvc,ldrvcb,     %
		ldrvcbt,ldrvcd,ldrvch,ldrvcsb,ldrvcsh,ldrvct,ldrvs,ldrvsb,ldrvsbt,      %
		ldrvsd,ldrvsh,ldrvssb,ldrvssh,ldrvst,mar,maral,marcc,marcs,mareq,       %
		marge,margt,marhi,marhs,marle,marlo,marls,marlt,marmi,marne,marpl,      %
		marvc,marvs,mcr,mcr2,mcral,mcrcc,mcrcs,mcreq,mcrge,mcrgt,mcrhi,mcrhs,   %
		mcrle,mcrlo,mcrls,mcrlt,mcrmi,mcrne,mcrpl,mcrr,mcrr2,mcrral,mcrrcc,     %
		mcrrcs,mcrreq,mcrrge,mcrrgt,mcrrhi,mcrrhs,mcrrle,mcrrlo,mcrrls,mcrrlt,  %
		mcrrmi,mcrrne,mcrrpl,mcrrvc,mcrrvs,mcrvc,mcrvs,mia,miaal,miacc,miacs,   %
		miaeq,miage,miagt,miahi,miahs,miale,mialo,mials,mialt,miami,miane,      %
		miaph,miaphal,miaphcc,miaphcs,miapheq,miaphge,miaphgt,miaphhi,miaphhs,  %
		miaphle,miaphlo,miaphls,miaphlt,miaphmi,miaphne,miaphpl,miaphvc,        %
		miaphvs,miapl,miavc,miavs,miaxy,miaxyal,miaxycc,miaxycs,miaxyeq,        %
		miaxyge,miaxygt,miaxyhi,miaxyhs,miaxyle,miaxylo,miaxyls,miaxylt,        %
		miaxymi,miaxyne,miaxypl,miaxyvc,miaxyvs,mla,mlaal,mlaals,mlacc,mlaccs,  %
		mlacs,mlacss,mlaeq,mlaeqs,mlage,mlages,mlagt,mlagts,mlahi,mlahis,       %
		mlahs,mlahss,mlale,mlales,mlalo,mlalos,mlals,mlalss,mlalt,mlalts,       %
		mlami,mlamis,mlane,mlanes,mlapl,mlapls,mlas,mlavc,mlavcs,mlavs,mlavss,  %
		mov,moval,movals,movcc,movccs,movcs,movcss,moveq,moveqs,movge,movges,   %
		movgt,movgts,movhi,movhis,movhs,movhss,movle,movles,movlo,movlos,       %
		movls,movlss,movlt,movlts,movmi,movmis,movne,movnes,movpl,movpls,movs,  %
		movvc,movvcs,movvs,movvss,mra,mraal,mracc,mracs,mraeq,mrage,mragt,      %
		mrahi,mrahs,mrale,mralo,mrals,mralt,mrami,mrane,mrapl,mravc,mravs,mrc,  %
		mrc2,mrcal,mrccc,mrccs,mrceq,mrcge,mrcgt,mrchi,mrchs,mrcle,mrclo,       %
		mrcls,mrclt,mrcmi,mrcne,mrcpl,mrcvc,mrcvs,mrrc,mrrc2,mrrcal,mrrccc,     %
		mrrccs,mrrceq,mrrcge,mrrcgt,mrrchi,mrrchs,mrrcle,mrrclo,mrrcls,mrrclt,  %
		mrrcmi,mrrcne,mrrcpl,mrrcvc,mrrcvs,mrs,mrsal,mrscc,mrscs,mrseq,mrsge,   %
		mrsgt,mrshi,mrshs,mrsle,mrslo,mrsls,mrslt,mrsmi,mrsne,mrspl,mrsvc,      %
		mrsvs,msr,msral,msrcc,msrcs,msreq,msrge,msrgt,msrhi,msrhs,msrle,msrlo,  %
		msrls,msrlt,msrmi,msrne,msrpl,msrvc,msrvs,mul,mulal,mulals,mulcc,       %
		mulccs,mulcs,mulcss,muleq,muleqs,mulge,mulges,mulgt,mulgts,mulhi,       %
		mulhis,mulhs,mulhss,mulle,mulles,mullo,mullos,mulls,mullss,mullt,       %
		mullts,mulmi,mulmis,mulne,mulnes,mulpl,mulpls,muls,mulvc,mulvcs,mulvs,  %
		mulvss,mvn,mvnal,mvnals,mvncc,mvnccs,mvncs,mvncss,mvneq,mvneqs,mvnge,   %
		mvnges,mvngt,mvngts,mvnhi,mvnhis,mvnhs,mvnhss,mvnle,mvnles,mvnlo,       %
		mvnlos,mvnls,mvnlss,mvnlt,mvnlts,mvnmi,mvnmis,mvnne,mvnnes,mvnpl,       %
		mvnpls,mvns,mvnvc,mvnvcs,mvnvs,mvnvss,nop,orr,orral,orrals,orrcc,       %
		orrccs,orrcs,orrcss,orreq,orreqs,orrge,orrges,orrgt,orrgts,orrhi,       %
		orrhis,orrhs,orrhss,orrle,orrles,orrlo,orrlos,orrls,orrlss,orrlt,       %
		orrlts,orrmi,orrmis,orrne,orrnes,orrpl,orrpls,orrs,orrvc,orrvcs,orrvs,  %
		orrvss,pkhbt,pkhbtal,pkhbtcc,pkhbtcs,pkhbteq,pkhbtge,pkhbtgt,pkhbthi,   %
		pkhbths,pkhbtle,pkhbtlo,pkhbtls,pkhbtlt,pkhbtmi,pkhbtne,pkhbtpl,        %
		pkhbtvc,pkhbtvs,pkhtb,pkhtbal,pkhtbcc,pkhtbcs,pkhtbeq,pkhtbge,pkhtbgt,  %
		pkhtbhi,pkhtbhs,pkhtble,pkhtblo,pkhtbls,pkhtblt,pkhtbmi,pkhtbne,        %
		pkhtbpl,pkhtbvc,pkhtbvs,pld,pop,popal,popcc,popcs,popeq,popge,popgt,    %
		pophi,pophs,pople,poplo,popls,poplt,popmi,popne,poppl,popvc,popvs,      %
		push,pushal,pushcc,pushcs,pusheq,pushge,pushgt,pushhi,pushhs,pushle,    %
		pushlo,pushls,pushlt,pushmi,pushne,pushpl,pushvc,pushvs,qadd,qadd16,    %
		qadd16al,qadd16cc,qadd16cs,qadd16eq,qadd16ge,qadd16gt,qadd16hi,         %
		qadd16hs,qadd16le,qadd16lo,qadd16ls,qadd16lt,qadd16mi,qadd16ne,         %
		qadd16pl,qadd16vc,qadd16vs,qadd8,qadd8al,qadd8cc,qadd8cs,qadd8eq,       %
		qadd8ge,qadd8gt,qadd8hi,qadd8hs,qadd8le,qadd8lo,qadd8ls,qadd8lt,        %
		qadd8mi,qadd8ne,qadd8pl,qadd8vc,qadd8vs,qaddal,qaddcc,qaddcs,qaddeq,    %
		qaddge,qaddgt,qaddhi,qaddhs,qaddle,qaddlo,qaddls,qaddlt,qaddmi,qaddne,  %
		qaddpl,qaddsubx,qaddsubxal,qaddsubxcc,qaddsubxcs,qaddsubxeq,            %
		qaddsubxge,qaddsubxgt,qaddsubxhi,qaddsubxhs,qaddsubxle,qaddsubxlo,      %
		qaddsubxls,qaddsubxlt,qaddsubxmi,qaddsubxne,qaddsubxpl,qaddsubxvc,      %
		qaddsubxvs,qaddvc,qaddvs,qdadd,qdaddal,qdaddcc,qdaddcs,qdaddeq,         %
		qdaddge,qdaddgt,qdaddhi,qdaddhs,qdaddle,qdaddlo,qdaddls,qdaddlt,        %
		qdaddmi,qdaddne,qdaddpl,qdaddvc,qdaddvs,qdsub,qdsubal,qdsubcc,qdsubcs,  %
		qdsubeq,qdsubge,qdsubgt,qdsubhi,qdsubhs,qdsuble,qdsublo,qdsubls,        %
		qdsublt,qdsubmi,qdsubne,qdsubpl,qdsubvc,qdsubvs,qsub,qsub16,qsub16al,   %
		qsub16cc,qsub16cs,qsub16eq,qsub16ge,qsub16gt,qsub16hi,qsub16hs,         %
		qsub16le,qsub16lo,qsub16ls,qsub16lt,qsub16mi,qsub16ne,qsub16pl,         %
		qsub16vc,qsub16vs,qsub8,qsub8al,qsub8cc,qsub8cs,qsub8eq,qsub8ge,        %
		qsub8gt,qsub8hi,qsub8hs,qsub8le,qsub8lo,qsub8ls,qsub8lt,qsub8mi,        %
		qsub8ne,qsub8pl,qsub8vc,qsub8vs,qsubaddx,qsubaddxal,qsubaddxcc,         %
		qsubaddxcs,qsubaddxeq,qsubaddxge,qsubaddxgt,qsubaddxhi,qsubaddxhs,      %
		qsubaddxle,qsubaddxlo,qsubaddxls,qsubaddxlt,qsubaddxmi,qsubaddxne,      %
		qsubaddxpl,qsubaddxvc,qsubaddxvs,qsubal,qsubcc,qsubcs,qsubeq,qsubge,    %
		qsubgt,qsubhi,qsubhs,qsuble,qsublo,qsubls,qsublt,qsubmi,qsubne,qsubpl,  %
		qsubvc,qsubvs,rev,rev16,rev16al,rev16cc,rev16cs,rev16eq,rev16ge,        %
		rev16gt,rev16hi,rev16hs,rev16le,rev16lo,rev16ls,rev16lt,rev16mi,        %
		rev16ne,rev16pl,rev16vc,rev16vs,reval,revcc,revcs,reveq,revge,revgt,    %
		revhi,revhs,revle,revlo,revls,revlt,revmi,revne,revpl,revsh,revshal,    %
		revshcc,revshcs,revsheq,revshge,revshgt,revshhi,revshhs,revshle,        %
		revshlo,revshls,revshlt,revshmi,revshne,revshpl,revshvc,revshvs,revvc,  %
		revvs,rfeda,rfedb,rfeea,rfeed,rfefa,rfefd,rfeia,rfeib,rsb,rsbal,        %
		rsbals,rsbcc,rsbccs,rsbcs,rsbcss,rsbeq,rsbeqs,rsbge,rsbges,rsbgt,       %
		rsbgts,rsbhi,rsbhis,rsbhs,rsbhss,rsble,rsbles,rsblo,rsblos,rsbls,       %
		rsblss,rsblt,rsblts,rsbmi,rsbmis,rsbne,rsbnes,rsbpl,rsbpls,rsbs,rsbvc,  %
		rsbvcs,rsbvs,rsbvss,rsc,rscal,rscals,rsccc,rscccs,rsccs,rsccss,rsceq,   %
		rsceqs,rscge,rscges,rscgt,rscgts,rschi,rschis,rschs,rschss,rscle,       %
		rscles,rsclo,rsclos,rscls,rsclss,rsclt,rsclts,rscmi,rscmis,rscne,       %
		rscnes,rscpl,rscpls,rscs,rscvc,rscvcs,rscvs,rscvss,sadd16,sadd16al,     %
		sadd16cc,sadd16cs,sadd16eq,sadd16ge,sadd16gt,sadd16hi,sadd16hs,         %
		sadd16le,sadd16lo,sadd16ls,sadd16lt,sadd16mi,sadd16ne,sadd16pl,         %
		sadd16vc,sadd16vs,sadd8,sadd8al,sadd8cc,sadd8cs,sadd8eq,sadd8ge,        %
		sadd8gt,sadd8hi,sadd8hs,sadd8le,sadd8lo,sadd8ls,sadd8lt,sadd8mi,        %
		sadd8ne,sadd8pl,sadd8vc,sadd8vs,saddsubx,saddsubxal,saddsubxcc,         %
		saddsubxcs,saddsubxeq,saddsubxge,saddsubxgt,saddsubxhi,saddsubxhs,      %
		saddsubxle,saddsubxlo,saddsubxls,saddsubxlt,saddsubxmi,saddsubxne,      %
		saddsubxpl,saddsubxvc,saddsubxvs,sbc,sbcal,sbcals,sbccc,sbcccs,sbccs,   %
		sbccss,sbceq,sbceqs,sbcge,sbcges,sbcgt,sbcgts,sbchi,sbchis,sbchs,       %
		sbchss,sbcle,sbcles,sbclo,sbclos,sbcls,sbclss,sbclt,sbclts,sbcmi,       %
		sbcmis,sbcne,sbcnes,sbcpl,sbcpls,sbcs,sbcvc,sbcvcs,sbcvs,sbcvss,sel,    %
		selal,selcc,selcs,seleq,selge,selgt,selhi,selhs,selle,sello,sells,      %
		sellt,selmi,selne,selpl,selvc,selvs,setend,shadd16,shadd16al,           %
		shadd16cc,shadd16cs,shadd16eq,shadd16ge,shadd16gt,shadd16hi,shadd16hs,  %
		shadd16le,shadd16lo,shadd16ls,shadd16lt,shadd16mi,shadd16ne,shadd16pl,  %
		shadd16vc,shadd16vs,shadd8,shadd8al,shadd8cc,shadd8cs,shadd8eq,         %
		shadd8ge,shadd8gt,shadd8hi,shadd8hs,shadd8le,shadd8lo,shadd8ls,         %
		shadd8lt,shadd8mi,shadd8ne,shadd8pl,shadd8vc,shadd8vs,shaddsubx,        %
		shaddsubxal,shaddsubxcc,shaddsubxcs,shaddsubxeq,shaddsubxge,            %
		shaddsubxgt,shaddsubxhi,shaddsubxhs,shaddsubxle,shaddsubxlo,            %
		shaddsubxls,shaddsubxlt,shaddsubxmi,shaddsubxne,shaddsubxpl,            %
		shaddsubxvc,shaddsubxvs,shsub16,shsub16al,shsub16cc,shsub16cs,          %
		shsub16eq,shsub16ge,shsub16gt,shsub16hi,shsub16hs,shsub16le,shsub16lo,  %
		shsub16ls,shsub16lt,shsub16mi,shsub16ne,shsub16pl,shsub16vc,shsub16vs,  %
		shsub8,shsub8al,shsub8cc,shsub8cs,shsub8eq,shsub8ge,shsub8gt,shsub8hi,  %
		shsub8hs,shsub8le,shsub8lo,shsub8ls,shsub8lt,shsub8mi,shsub8ne,         %
		shsub8pl,shsub8vc,shsub8vs,shsubaddx,shsubaddxal,shsubaddxcc,           %
		shsubaddxcs,shsubaddxeq,shsubaddxge,shsubaddxgt,shsubaddxhi,            %
		shsubaddxhs,shsubaddxle,shsubaddxlo,shsubaddxls,shsubaddxlt,            %
		shsubaddxmi,shsubaddxne,shsubaddxpl,shsubaddxvc,shsubaddxvs,smlad,      %
		smladal,smladcc,smladcs,smladeq,smladge,smladgt,smladhi,smladhs,        %
		smladle,smladlo,smladls,smladlt,smladmi,smladne,smladpl,smladvc,        %
		smladvs,smladx,smladxal,smladxcc,smladxcs,smladxeq,smladxge,smladxgt,   %
		smladxhi,smladxhs,smladxle,smladxlo,smladxls,smladxlt,smladxmi,         %
		smladxne,smladxpl,smladxvc,smladxvs,smlal,smlalal,smlalals,smlalcc,     %
		smlalccs,smlalcs,smlalcss,smlald,smlaldal,smlaldcc,smlaldcs,smlaldeq,   %
		smlaldge,smlaldgt,smlaldhi,smlaldhs,smlaldle,smlaldlo,smlaldls,         %
		smlaldlt,smlaldmi,smlaldne,smlaldpl,smlaldvc,smlaldvs,smlaldx,          %
		smlaldxal,smlaldxcc,smlaldxcs,smlaldxeq,smlaldxge,smlaldxgt,smlaldxhi,  %
		smlaldxhs,smlaldxle,smlaldxlo,smlaldxls,smlaldxlt,smlaldxmi,smlaldxne,  %
		smlaldxpl,smlaldxvc,smlaldxvs,smlaleq,smlaleqs,smlalge,smlalges,        %
		smlalgt,smlalgts,smlalhi,smlalhis,smlalhs,smlalhss,smlalle,smlalles,    %
		smlallo,smlallos,smlalls,smlallss,smlallt,smlallts,smlalmi,smlalmis,    %
		smlalne,smlalnes,smlalpl,smlalpls,smlals,smlalvc,smlalvcs,smlalvs,      %
		smlalvss,smlalxy,smlalxyal,smlalxycc,smlalxycs,smlalxyeq,smlalxyge,     %
		smlalxygt,smlalxyhi,smlalxyhs,smlalxyle,smlalxylo,smlalxyls,smlalxylt,  %
		smlalxymi,smlalxyne,smlalxypl,smlalxyvc,smlalxyvs,smlawy,smlawyal,      %
		smlawycc,smlawycs,smlawyeq,smlawyge,smlawygt,smlawyhi,smlawyhs,         %
		smlawyle,smlawylo,smlawyls,smlawylt,smlawymi,smlawyne,smlawypl,         %
		smlawyvc,smlawyvs,smlaxy,smlaxyal,smlaxycc,smlaxycs,smlaxyeq,smlaxyge,  %
		smlaxygt,smlaxyhi,smlaxyhs,smlaxyle,smlaxylo,smlaxyls,smlaxylt,         %
		smlaxymi,smlaxyne,smlaxypl,smlaxyvc,smlaxyvs,smlsd,smlsdal,smlsdcc,     %
		smlsdcs,smlsdeq,smlsdge,smlsdgt,smlsdhi,smlsdhs,smlsdle,smlsdlo,        %
		smlsdls,smlsdlt,smlsdmi,smlsdne,smlsdpl,smlsdvc,smlsdvs,smlsdx,         %
		smlsdxal,smlsdxcc,smlsdxcs,smlsdxeq,smlsdxge,smlsdxgt,smlsdxhi,         %
		smlsdxhs,smlsdxle,smlsdxlo,smlsdxls,smlsdxlt,smlsdxmi,smlsdxne,         %
		smlsdxpl,smlsdxvc,smlsdxvs,smlsld,smlsldal,smlsldcc,smlsldcs,smlsldeq,  %
		smlsldge,smlsldgt,smlsldhi,smlsldhs,smlsldle,smlsldlo,smlsldls,         %
		smlsldlt,smlsldmi,smlsldne,smlsldpl,smlsldvc,smlsldvs,smlsldx,          %
		smlsldxal,smlsldxcc,smlsldxcs,smlsldxeq,smlsldxge,smlsldxgt,smlsldxhi,  %
		smlsldxhs,smlsldxle,smlsldxlo,smlsldxls,smlsldxlt,smlsldxmi,smlsldxne,  %
		smlsldxpl,smlsldxvc,smlsldxvs,smmla,smmlaal,smmlacc,smmlacs,smmlaeq,    %
		smmlage,smmlagt,smmlahi,smmlahs,smmlale,smmlalo,smmlals,smmlalt,        %
		smmlami,smmlane,smmlapl,smmlar,smmlaral,smmlarcc,smmlarcs,smmlareq,     %
		smmlarge,smmlargt,smmlarhi,smmlarhs,smmlarle,smmlarlo,smmlarls,         %
		smmlarlt,smmlarmi,smmlarne,smmlarpl,smmlarvc,smmlarvs,smmlavc,smmlavs,  %
		smmls,smmlsal,smmlscc,smmlscs,smmlseq,smmlsge,smmlsgt,smmlshi,smmlshs,  %
		smmlsle,smmlslo,smmlsls,smmlslt,smmlsmi,smmlsne,smmlspl,smmlsr,         %
		smmlsral,smmlsrcc,smmlsrcs,smmlsreq,smmlsrge,smmlsrgt,smmlsrhi,         %
		smmlsrhs,smmlsrle,smmlsrlo,smmlsrls,smmlsrlt,smmlsrmi,smmlsrne,         %
		smmlsrpl,smmlsrvc,smmlsrvs,smmlsvc,smmlsvs,smmul,smmulal,smmulcc,       %
		smmulcs,smmuleq,smmulge,smmulgt,smmulhi,smmulhs,smmulle,smmullo,        %
		smmulls,smmullt,smmulmi,smmulne,smmulpl,smmulr,smmulral,smmulrcc,       %
		smmulrcs,smmulreq,smmulrge,smmulrgt,smmulrhi,smmulrhs,smmulrle,         %
		smmulrlo,smmulrls,smmulrlt,smmulrmi,smmulrne,smmulrpl,smmulrvc,         %
		smmulrvs,smmulvc,smmulvs,smuad,smuadal,smuadcc,smuadcs,smuadeq,         %
		smuadge,smuadgt,smuadhi,smuadhs,smuadle,smuadlo,smuadls,smuadlt,        %
		smuadmi,smuadne,smuadpl,smuadvc,smuadvs,smuadx,smuadxal,smuadxcc,       %
		smuadxcs,smuadxeq,smuadxge,smuadxgt,smuadxhi,smuadxhs,smuadxle,         %
		smuadxlo,smuadxls,smuadxlt,smuadxmi,smuadxne,smuadxpl,smuadxvc,         %
		smuadxvs,smull,smullal,smullals,smullcc,smullccs,smullcs,smullcss,      %
		smulleq,smulleqs,smullge,smullges,smullgt,smullgts,smullhi,smullhis,    %
		smullhs,smullhss,smullle,smullles,smulllo,smulllos,smullls,smulllss,    %
		smulllt,smulllts,smullmi,smullmis,smullne,smullnes,smullpl,smullpls,    %
		smulls,smullvc,smullvcs,smullvs,smullvss,smulwy,smulwyal,smulwycc,      %
		smulwycs,smulwyeq,smulwyge,smulwygt,smulwyhi,smulwyhs,smulwyle,         %
		smulwylo,smulwyls,smulwylt,smulwymi,smulwyne,smulwypl,smulwyvc,         %
		smulwyvs,smulxy,smulxyal,smulxycc,smulxycs,smulxyeq,smulxyge,smulxygt,  %
		smulxyhi,smulxyhs,smulxyle,smulxylo,smulxyls,smulxylt,smulxymi,         %
		smulxyne,smulxypl,smulxyvc,smulxyvs,smusd,smusdal,smusdcc,smusdcs,      %
		smusdeq,smusdge,smusdgt,smusdhi,smusdhs,smusdle,smusdlo,smusdls,        %
		smusdlt,smusdmi,smusdne,smusdpl,smusdvc,smusdvs,smusdx,smusdxal,        %
		smusdxcc,smusdxcs,smusdxeq,smusdxge,smusdxgt,smusdxhi,smusdxhs,         %
		smusdxle,smusdxlo,smusdxls,smusdxlt,smusdxmi,smusdxne,smusdxpl,         %
		smusdxvc,smusdxvs,srsda,srsdb,srsea,srsed,srsfa,srsfd,srsia,srsib,      %
		ssat,ssat16,ssat16al,ssat16cc,ssat16cs,ssat16eq,ssat16ge,ssat16gt,      %
		ssat16hi,ssat16hs,ssat16le,ssat16lo,ssat16ls,ssat16lt,ssat16mi,         %
		ssat16ne,ssat16pl,ssat16vc,ssat16vs,ssatal,ssatcc,ssatcs,ssateq,        %
		ssatge,ssatgt,ssathi,ssaths,ssatle,ssatlo,ssatls,ssatlt,ssatmi,ssatne,  %
		ssatpl,ssatvc,ssatvs,ssub16,ssub16al,ssub16cc,ssub16cs,ssub16eq,        %
		ssub16ge,ssub16gt,ssub16hi,ssub16hs,ssub16le,ssub16lo,ssub16ls,         %
		ssub16lt,ssub16mi,ssub16ne,ssub16pl,ssub16vc,ssub16vs,ssub8,ssub8al,    %
		ssub8cc,ssub8cs,ssub8eq,ssub8ge,ssub8gt,ssub8hi,ssub8hs,ssub8le,        %
		ssub8lo,ssub8ls,ssub8lt,ssub8mi,ssub8ne,ssub8pl,ssub8vc,ssub8vs,        %
		ssubaddx,ssubaddxal,ssubaddxcc,ssubaddxcs,ssubaddxeq,ssubaddxge,        %
		ssubaddxgt,ssubaddxhi,ssubaddxhs,ssubaddxle,ssubaddxlo,ssubaddxls,      %
		ssubaddxlt,ssubaddxmi,ssubaddxne,ssubaddxpl,ssubaddxvc,ssubaddxvs,stc,  %
		stc2,stcal,stccc,stccs,stceq,stcge,stcgt,stchi,stchs,stcle,stclo,       %
		stcls,stclt,stcmi,stcne,stcpl,stcvc,stcvs,stmalda,stmaldb,stmalea,      %
		stmaled,stmalfa,stmalfd,stmalia,stmalib,stmccda,stmccdb,stmccea,        %
		stmcced,stmccfa,stmccfd,stmccia,stmccib,stmcsda,stmcsdb,stmcsea,        %
		stmcsed,stmcsfa,stmcsfd,stmcsia,stmcsib,stmda,stmdb,stmea,stmed,        %
		stmeqda,stmeqdb,stmeqea,stmeqed,stmeqfa,stmeqfd,stmeqia,stmeqib,stmfa,  %
		stmfd,stmgeda,stmgedb,stmgeea,stmgeed,stmgefa,stmgefd,stmgeia,stmgeib,  %
		stmgtda,stmgtdb,stmgtea,stmgted,stmgtfa,stmgtfd,stmgtia,stmgtib,        %
		stmhida,stmhidb,stmhiea,stmhied,stmhifa,stmhifd,stmhiia,stmhiib,        %
		stmhsda,stmhsdb,stmhsea,stmhsed,stmhsfa,stmhsfd,stmhsia,stmhsib,stmia,  %
		stmib,stmleda,stmledb,stmleea,stmleed,stmlefa,stmlefd,stmleia,stmleib,  %
		stmloda,stmlodb,stmloea,stmloed,stmlofa,stmlofd,stmloia,stmloib,        %
		stmlsda,stmlsdb,stmlsea,stmlsed,stmlsfa,stmlsfd,stmlsia,stmlsib,        %
		stmltda,stmltdb,stmltea,stmlted,stmltfa,stmltfd,stmltia,stmltib,        %
		stmmida,stmmidb,stmmiea,stmmied,stmmifa,stmmifd,stmmiia,stmmiib,        %
		stmneda,stmnedb,stmneea,stmneed,stmnefa,stmnefd,stmneia,stmneib,        %
		stmplda,stmpldb,stmplea,stmpled,stmplfa,stmplfd,stmplia,stmplib,        %
		stmvcda,stmvcdb,stmvcea,stmvced,stmvcfa,stmvcfd,stmvcia,stmvcib,        %
		stmvsda,stmvsdb,stmvsea,stmvsed,stmvsfa,stmvsfd,stmvsia,stmvsib,str,    %
		stral,stralb,stralbt,strald,stralh,stralt,strb,strbt,strcc,strccb,      %
		strccbt,strccd,strcch,strcct,strcs,strcsb,strcsbt,strcsd,strcsh,        %
		strcst,strd,streq,streqb,streqbt,streqd,streqh,streqt,strex,strexal,    %
		strexcc,strexcs,strexeq,strexge,strexgt,strexhi,strexhs,strexle,        %
		strexlo,strexls,strexlt,strexmi,strexne,strexpl,strexvc,strexvs,strge,  %
		strgeb,strgebt,strged,strgeh,strget,strgt,strgtb,strgtbt,strgtd,        %
		strgth,strgtt,strh,strhi,strhib,strhibt,strhid,strhih,strhit,strhs,     %
		strhsb,strhsbt,strhsd,strhsh,strhst,strle,strleb,strlebt,strled,        %
		strleh,strlet,strlo,strlob,strlobt,strlod,strloh,strlot,strls,strlsb,   %
		strlsbt,strlsd,strlsh,strlst,strlt,strltb,strltbt,strltd,strlth,        %
		strltt,strmi,strmib,strmibt,strmid,strmih,strmit,strne,strneb,strnebt,  %
		strned,strneh,strnet,strpl,strplb,strplbt,strpld,strplh,strplt,strt,    %
		strvc,strvcb,strvcbt,strvcd,strvch,strvct,strvs,strvsb,strvsbt,strvsd,  %
		strvsh,strvst,sub,subal,subals,subcc,subccs,subcs,subcss,subeq,subeqs,  %
		subge,subges,subgt,subgts,subhi,subhis,subhs,subhss,suble,subles,       %
		sublo,sublos,subls,sublss,sublt,sublts,submi,submis,subne,subnes,       %
		subpl,subpls,subs,subvc,subvcs,subvs,subvss,swi,swial,swicc,swics,      %
		swieq,swige,swigt,swihi,swihs,swile,swilo,swils,swilt,swimi,swine,      %
		swipl,swivc,swivs,swp,swpal,swpalb,swpb,swpcc,swpccb,swpcs,swpcsb,      %
		swpeq,swpeqb,swpge,swpgeb,swpgt,swpgtb,swphi,swphib,swphs,swphsb,       %
		swple,swpleb,swplo,swplob,swpls,swplsb,swplt,swpltb,swpmi,swpmib,       %
		swpne,swpneb,swppl,swpplb,swpvc,swpvcb,swpvs,swpvsb,sxtab,sxtab16,      %
		sxtab16al,sxtab16cc,sxtab16cs,sxtab16eq,sxtab16ge,sxtab16gt,sxtab16hi,  %
		sxtab16hs,sxtab16le,sxtab16lo,sxtab16ls,sxtab16lt,sxtab16mi,sxtab16ne,  %
		sxtab16pl,sxtab16vc,sxtab16vs,sxtabal,sxtabcc,sxtabcs,sxtabeq,sxtabge,  %
		sxtabgt,sxtabhi,sxtabhs,sxtable,sxtablo,sxtabls,sxtablt,sxtabmi,        %
		sxtabne,sxtabpl,sxtabvc,sxtabvs,sxtah,sxtahal,sxtahcc,sxtahcs,sxtaheq,  %
		sxtahge,sxtahgt,sxtahhi,sxtahhs,sxtahle,sxtahlo,sxtahls,sxtahlt,        %
		sxtahmi,sxtahne,sxtahpl,sxtahvc,sxtahvs,sxtb,sxtb16,sxtb16al,sxtb16cc,  %
		sxtb16cs,sxtb16eq,sxtb16ge,sxtb16gt,sxtb16hi,sxtb16hs,sxtb16le,         %
		sxtb16lo,sxtb16ls,sxtb16lt,sxtb16mi,sxtb16ne,sxtb16pl,sxtb16vc,         %
		sxtb16vs,sxtbal,sxtbcc,sxtbcs,sxtbeq,sxtbge,sxtbgt,sxtbhi,sxtbhs,       %
		sxtble,sxtblo,sxtbls,sxtblt,sxtbmi,sxtbne,sxtbpl,sxtbvc,sxtbvs,sxth,    %
		sxthal,sxthcc,sxthcs,sxtheq,sxthge,sxthgt,sxthhi,sxthhs,sxthle,sxthlo,  %
		sxthls,sxthlt,sxthmi,sxthne,sxthpl,sxthvc,sxthvs,teq,teqal,teqcc,       %
		teqcs,teqeq,teqge,teqgt,teqhi,teqhs,teqle,teqlo,teqls,teqlt,teqmi,      %
		teqne,teqpl,teqvc,teqvs,tst,tstal,tstcc,tstcs,tsteq,tstge,tstgt,tsthi,  %
		tsths,tstle,tstlo,tstls,tstlt,tstmi,tstne,tstpl,tstvc,tstvs,uadd16,     %
		uadd16al,uadd16cc,uadd16cs,uadd16eq,uadd16ge,uadd16gt,uadd16hi,         %
		uadd16hs,uadd16le,uadd16lo,uadd16ls,uadd16lt,uadd16mi,uadd16ne,         %
		uadd16pl,uadd16vc,uadd16vs,uadd8,uadd8al,uadd8cc,uadd8cs,uadd8eq,       %
		uadd8ge,uadd8gt,uadd8hi,uadd8hs,uadd8le,uadd8lo,uadd8ls,uadd8lt,        %
		uadd8mi,uadd8ne,uadd8pl,uadd8vc,uadd8vs,uaddsubx,uaddsubxal,            %
		uaddsubxcc,uaddsubxcs,uaddsubxeq,uaddsubxge,uaddsubxgt,uaddsubxhi,      %
		uaddsubxhs,uaddsubxle,uaddsubxlo,uaddsubxls,uaddsubxlt,uaddsubxmi,      %
		uaddsubxne,uaddsubxpl,uaddsubxvc,uaddsubxvs,uhadd16,uhadd16al,          %
		uhadd16cc,uhadd16cs,uhadd16eq,uhadd16ge,uhadd16gt,uhadd16hi,uhadd16hs,  %
		uhadd16le,uhadd16lo,uhadd16ls,uhadd16lt,uhadd16mi,uhadd16ne,uhadd16pl,  %
		uhadd16vc,uhadd16vs,uhadd8,uhadd8al,uhadd8cc,uhadd8cs,uhadd8eq,         %
		uhadd8ge,uhadd8gt,uhadd8hi,uhadd8hs,uhadd8le,uhadd8lo,uhadd8ls,         %
		uhadd8lt,uhadd8mi,uhadd8ne,uhadd8pl,uhadd8vc,uhadd8vs,uhaddsubx,        %
		uhaddsubxal,uhaddsubxcc,uhaddsubxcs,uhaddsubxeq,uhaddsubxge,            %
		uhaddsubxgt,uhaddsubxhi,uhaddsubxhs,uhaddsubxle,uhaddsubxlo,            %
		uhaddsubxls,uhaddsubxlt,uhaddsubxmi,uhaddsubxne,uhaddsubxpl,            %
		uhaddsubxvc,uhaddsubxvs,uhsub16,uhsub16al,uhsub16cc,uhsub16cs,          %
		uhsub16eq,uhsub16ge,uhsub16gt,uhsub16hi,uhsub16hs,uhsub16le,uhsub16lo,  %
		uhsub16ls,uhsub16lt,uhsub16mi,uhsub16ne,uhsub16pl,uhsub16vc,uhsub16vs,  %
		uhsub8,uhsub8al,uhsub8cc,uhsub8cs,uhsub8eq,uhsub8ge,uhsub8gt,uhsub8hi,  %
		uhsub8hs,uhsub8le,uhsub8lo,uhsub8ls,uhsub8lt,uhsub8mi,uhsub8ne,         %
		uhsub8pl,uhsub8vc,uhsub8vs,uhsubaddx,uhsubaddxal,uhsubaddxcc,           %
		uhsubaddxcs,uhsubaddxeq,uhsubaddxge,uhsubaddxgt,uhsubaddxhi,            %
		uhsubaddxhs,uhsubaddxle,uhsubaddxlo,uhsubaddxls,uhsubaddxlt,            %
		uhsubaddxmi,uhsubaddxne,uhsubaddxpl,uhsubaddxvc,uhsubaddxvs,umaal,      %
		umaalal,umaalcc,umaalcs,umaaleq,umaalge,umaalgt,umaalhi,umaalhs,        %
		umaalle,umaallo,umaalls,umaallt,umaalmi,umaalne,umaalpl,umaalvc,        %
		umaalvs,umlal,umlalal,umlalals,umlalcc,umlalccs,umlalcs,umlalcss,       %
		umlaleq,umlaleqs,umlalge,umlalges,umlalgt,umlalgts,umlalhi,umlalhis,    %
		umlalhs,umlalhss,umlalle,umlalles,umlallo,umlallos,umlalls,umlallss,    %
		umlallt,umlallts,umlalmi,umlalmis,umlalne,umlalnes,umlalpl,umlalpls,    %
		umlals,umlalvc,umlalvcs,umlalvs,umlalvss,umull,umullal,umullals,        %
		umullcc,umullccs,umullcs,umullcss,umulleq,umulleqs,umullge,umullges,    %
		umullgt,umullgts,umullhi,umullhis,umullhs,umullhss,umullle,umullles,    %
		umulllo,umulllos,umullls,umulllss,umulllt,umulllts,umullmi,umullmis,    %
		umullne,umullnes,umullpl,umullpls,umulls,umullvc,umullvcs,umullvs,      %
		umullvss,uqadd16,uqadd16al,uqadd16cc,uqadd16cs,uqadd16eq,uqadd16ge,     %
		uqadd16gt,uqadd16hi,uqadd16hs,uqadd16le,uqadd16lo,uqadd16ls,uqadd16lt,  %
		uqadd16mi,uqadd16ne,uqadd16pl,uqadd16vc,uqadd16vs,uqadd8,uqadd8al,      %
		uqadd8cc,uqadd8cs,uqadd8eq,uqadd8ge,uqadd8gt,uqadd8hi,uqadd8hs,         %
		uqadd8le,uqadd8lo,uqadd8ls,uqadd8lt,uqadd8mi,uqadd8ne,uqadd8pl,         %
		uqadd8vc,uqadd8vs,uqaddsubx,uqaddsubxal,uqaddsubxcc,uqaddsubxcs,        %
		uqaddsubxeq,uqaddsubxge,uqaddsubxgt,uqaddsubxhi,uqaddsubxhs,            %
		uqaddsubxle,uqaddsubxlo,uqaddsubxls,uqaddsubxlt,uqaddsubxmi,            %
		uqaddsubxne,uqaddsubxpl,uqaddsubxvc,uqaddsubxvs,uqsub16,uqsub16al,      %
		uqsub16cc,uqsub16cs,uqsub16eq,uqsub16ge,uqsub16gt,uqsub16hi,uqsub16hs,  %
		uqsub16le,uqsub16lo,uqsub16ls,uqsub16lt,uqsub16mi,uqsub16ne,uqsub16pl,  %
		uqsub16vc,uqsub16vs,uqsub8,uqsub8al,uqsub8cc,uqsub8cs,uqsub8eq,         %
		uqsub8ge,uqsub8gt,uqsub8hi,uqsub8hs,uqsub8le,uqsub8lo,uqsub8ls,         %
		uqsub8lt,uqsub8mi,uqsub8ne,uqsub8pl,uqsub8vc,uqsub8vs,uqsubaddx,        %
		uqsubaddxal,uqsubaddxcc,uqsubaddxcs,uqsubaddxeq,uqsubaddxge,            %
		uqsubaddxgt,uqsubaddxhi,uqsubaddxhs,uqsubaddxle,uqsubaddxlo,            %
		uqsubaddxls,uqsubaddxlt,uqsubaddxmi,uqsubaddxne,uqsubaddxpl,            %
		uqsubaddxvc,uqsubaddxvs,usad8,usad8al,usad8cc,usad8cs,usad8eq,usad8ge,  %
		usad8gt,usad8hi,usad8hs,usad8le,usad8lo,usad8ls,usad8lt,usad8mi,        %
		usad8ne,usad8pl,usad8vc,usad8vs,usada8,usada8al,usada8cc,usada8cs,      %
		usada8eq,usada8ge,usada8gt,usada8hi,usada8hs,usada8le,usada8lo,         %
		usada8ls,usada8lt,usada8mi,usada8ne,usada8pl,usada8vc,usada8vs,usat,    %
		usat16,usat16al,usat16cc,usat16cs,usat16eq,usat16ge,usat16gt,usat16hi,  %
		usat16hs,usat16le,usat16lo,usat16ls,usat16lt,usat16mi,usat16ne,         %
		usat16pl,usat16vc,usat16vs,usatal,usatcc,usatcs,usateq,usatge,usatgt,   %
		usathi,usaths,usatle,usatlo,usatls,usatlt,usatmi,usatne,usatpl,usatvc,  %
		usatvs,usub16,usub16al,usub16cc,usub16cs,usub16eq,usub16ge,usub16gt,    %
		usub16hi,usub16hs,usub16le,usub16lo,usub16ls,usub16lt,usub16mi,         %
		usub16ne,usub16pl,usub16vc,usub16vs,usub8,usub8al,usub8cc,usub8cs,      %
		usub8eq,usub8ge,usub8gt,usub8hi,usub8hs,usub8le,usub8lo,usub8ls,        %
		usub8lt,usub8mi,usub8ne,usub8pl,usub8vc,usub8vs,usubaddx,usubaddxal,    %
		usubaddxcc,usubaddxcs,usubaddxeq,usubaddxge,usubaddxgt,usubaddxhi,      %
		usubaddxhs,usubaddxle,usubaddxlo,usubaddxls,usubaddxlt,usubaddxmi,      %
		usubaddxne,usubaddxpl,usubaddxvc,usubaddxvs,uxtab,uxtab16,uxtab16al,    %
		uxtab16cc,uxtab16cs,uxtab16eq,uxtab16ge,uxtab16gt,uxtab16hi,uxtab16hs,  %
		uxtab16le,uxtab16lo,uxtab16ls,uxtab16lt,uxtab16mi,uxtab16ne,uxtab16pl,  %
		uxtab16vc,uxtab16vs,uxtabal,uxtabcc,uxtabcs,uxtabeq,uxtabge,uxtabgt,    %
		uxtabhi,uxtabhs,uxtable,uxtablo,uxtabls,uxtablt,uxtabmi,uxtabne,        %
		uxtabpl,uxtabvc,uxtabvs,uxtah,uxtahal,uxtahcc,uxtahcs,uxtaheq,uxtahge,  %
		uxtahgt,uxtahhi,uxtahhs,uxtahle,uxtahlo,uxtahls,uxtahlt,uxtahmi,        %
		uxtahne,uxtahpl,uxtahvc,uxtahvs,uxtb,uxtb16,uxtb16al,uxtb16cc,          %
		uxtb16cs,uxtb16eq,uxtb16ge,uxtb16gt,uxtb16hi,uxtb16hs,uxtb16le,         %
		uxtb16lo,uxtb16ls,uxtb16lt,uxtb16mi,uxtb16ne,uxtb16pl,uxtb16vc,         %
		uxtb16vs,uxtbal,uxtbcc,uxtbcs,uxtbeq,uxtbge,uxtbgt,uxtbhi,uxtbhs,       %
		uxtble,uxtblo,uxtbls,uxtblt,uxtbmi,uxtbne,uxtbpl,uxtbvc,uxtbvs,uxth,    %
		uxthal,uxthcc,uxthcs,uxtheq,uxthge,uxthgt,uxthhi,uxthhs,uxthle,uxthlo,  %
		uxthls,uxthlt,uxthmi,uxthne,uxthpl,uxthvc,uxthvs},%
	morekeywords=[2]{.2byte,.4byte,.8byte,.abort,.abort,.align,.altmacro,     %
		.arch,.arch_extension,.arm,.ascii,.asciz,.balign,.bss,                  %
		.bundle_align_mode,.bundle_lock,,.bundle_unlock,.byte,.cantunwind,      %
		.cfi_endproc,,.cfi_startproc,.code,.comm,.cpu,.data,.def,.desc,.dim,    %
		.dn,.double,.eabi_attribute,.eject,.else,.elseif,.end,.endef,.endfunc,  %
		.endif,.equ,.equiv,.eqv,.err,.error,.even,.exitm,.extend,.extend.,      %
		.extern,.fail,.file,.fill,.float,.fnend,.fnstart,.force_thumb,.fpu,     %
		.func,.global,.globl,.gnu_attribute,.handlerdata,.hidden,.hword,        %
		.ident,.if,.incbin,.include,.inst,.inst.n,.inst.w,.int,.internal,.irp,  %
		.irpc,.lcomm,.ldouble,.lflags,.line,.linkonce,.list,.ln,.loc,           %
		.loc_mark_labels,.local,.long,.ltorg,.ltorg.,.macro,.movsp,.mri,        %
		.noaltmacro,.nolist,.object_arch,.octa,.offset,.org,.p2align,.packed,   %
		.pad,.personality,.personalityindex,.pool,.popsection,.previous,        %
		.print,.protected,.psize,.purgem,.pushsection,.qn,.quad,.reloc,.rept,   %
		.req,.save,.sbttl,.scl,.secrel32,.section,.set,.setfp,.short,.single,   %
		.size,.skip,.sleb128,.space,.stabd,,.stabn,,.stabs,.string,.string16,   %
		.string32,.string64,.string8,.struct,.subsection,.symver,.syntax,.tag,  %
		.text,.thumb,.thumb_func,.thumb_set,.title,.tlsdescseq,.type,.uleb128,  %
		.unreq,.unwind_raw,.val,.version,.vsave,.vtable_entry,.vtable_inherit,  %
		.warning,.weak,.weakref,.word},%
	alsoletter={.,0,1,2,3,4,5,6,7,8,9},%
	alsodigit={?},%
	sensitive=false,%
	morestring=[b]",%
	morecomment=[s]{/*}{*/},%
	morecomment=[l]@,%
	morecomment=[l]//,%
}[keywords,comments,strings]

\usepackage[referable]{threeparttablex}
\renewlist{tablenotes}{enumerate}{1}
\setlist[tablenotes]{label=\tnote{\alph*},ref=\alph*,labelindent=\tabcolsep,labelsep=.2em,leftmargin=*,align=left,before={\footnotesize}}

\graphicspath{{figure/}}

\usepackage[justification=centering]{caption}

\SetKwProg{Al}{Algorithm}{}{}
\SetAlFnt{\small\rmfamily}
\SetKw{Continue}{continue}
\SetKw{And}{and}
\SetKw{Or}{or}
\SetKw{Not}{not}
\SetKw{In}{in}
\SetKw{Null}{NULL}

\definecolor{darkspringgreen}{rgb}{0.09,0.45,0.27}

\SetCommentSty{mycommfont}

\begin{document}

\setcounter{secnumdepth}{3}
\newcommand{\tabincell}[2]{\begin{tabular}{@{}#1@{}}#2\end{tabular}}
\newcommand{\binCMP}{\mbox{\textsc{BinMatch}}}
\newcommand{\Bindiff}{Bindiff}
\newcommand{\pngfunc}{\emph{png\_set\_unknown\_chunks}}
\def\tabularxcolumn#1{m{#1}}

\title{A Semantics-Based Hybrid Approach on Binary Code Similarity Comparison}

\author{Yikun Hu, Hui Wang, Yuanyuan Zhang, Bodong Li and Dawu Gu* \thanks{* Corresponding author. E-mail: dwgu@sjtu.edu.cn}% <-this % stops a space
\IEEEcompsocitemizethanks{\IEEEcompsocthanksitem Y. Hu is with the Department
of Computer Science and Technology, SEIEE, Shanghai Jiao Tong University, Shanghai, China.%\protect\\
\IEEEcompsocthanksitem H. Wang is with the Department
of Computer Science and Technology, SEIEE, Shanghai Jiao Tong University, Shanghai, China.%\protect\\
\IEEEcompsocthanksitem Y. Zhang is with the Department
of Computer Science and Technology, SEIEE, Shanghai Jiao Tong University, Shanghai, China.%\protect\\
\IEEEcompsocthanksitem B. Li is with the Department
of Computer Science and Technology, SEIEE, Shanghai Jiao Tong University, Shanghai, China.%\protect\\
\IEEEcompsocthanksitem D. Gu is with the Department
of Computer Science and Technology, SEIEE, Shanghai Jiao Tong University, Shanghai, China.%\protect\\
}
\thanks{A preliminary version of this paper~\cite{hu2018binmatch} appeared in the {\normalfont Proceedings of the 34th IEEE International Conference on Software Maintenance and Evolution~(ICSME'18)}, Madrid, Spain, September 23-29, 2018.}}

\markboth{Journal of \LaTeX\ Class Files,~Vol.~14, No.~8, August~2015}%
{Hu \MakeLowercase{\textit{et al.}}: Semantics-Based Hybrid Binary Code Similarity Comparison across Architectures}

\IEEEtitleabstractindextext{%
\begin{abstract}
Binary code similarity comparison is a methodology for identifying similar or identical code fragments in binary programs.
It is indispensable in fields of software engineering and security, which has many important applications~(e.g.,~plagiarism detection, bug detection).
With the widespread of smart and IoT~(Internet of Things) devices, an increasing number of programs are ported to multiple architectures~(e.g. ARM, MIPS).
It becomes necessary to detect similar binary code across architectures as well.
The main challenge of this topic lies in the semantics-equivalent code transformation resulting from different compilation settings, code obfuscation, and varied instruction set architectures.
Another challenge is the trade-off between comparison accuracy and coverage.
Unfortunately, existing methods still heavily rely on semantics-less code features which are susceptible to the code transformation.
Additionally, they perform the comparison merely either in a static or in a dynamic manner, which cannot achieve high accuracy and coverage simultaneously.
In this paper, we propose a semantics-based hybrid method to compare binary function similarity.
We execute the reference function with test cases, then emulate the execution of every target function with the runtime information migrated from the reference function.
Semantic signatures are extracted during the execution as well as the emulation.
Lastly, similarity scores are calculated from the signatures to measure the likeness of functions.
We have implemented the method in a prototype system designated as \binCMP\ which performs binary code similarity comparison across architectures of x86, ARM and MIPS on the Linux platform.
We evaluate \binCMP\ with nine real-word projects compiled with different compilation settings, on variant architectures, and with commonly-used obfuscation methods, totally performing over 100 million pairs of function comparison.
The experimental results show that \binCMP\ is resilient to the semantics-equivalent code transformation.
Besides, it not only covers all target functions for similarity comparison, but also improves the accuracy comparing to the state-of-the-art solutions.
\end{abstract}

}

\maketitle

\IEEEdisplaynontitleabstractindextext

% \fi
%\IEEEpeerreviewmaketitle

\newsavebox{\globalreference}
\begin{lrbox}{\globalreference}
	\begin{minipage}{.45\linewidth}
		\begin{lstlisting}[basicstyle=\ttfamily\scriptsize, language={[x86masm]Assembler}, numberstyle=\tiny, tabsize=4, keywordstyle=\color{blue!70}, commentstyle=\color{red!50!green!50!blue!50}, rulesepcolor=\color{red!20!green!20!blue!20}, escapechar=!, captionpos=b, numberstyle=\tiny, numbers=left, numbersep=-11pt, belowcaptionskip=-10pt]
      mov     ecx, gvar1
      test    ecx, ecx
      mov     eax, gvar2
      add     ecx, eax
		\end{lstlisting}
	\end{minipage}
\end{lrbox}

\newsavebox{\globaltarget}
\begin{lrbox}{\globaltarget}
	\begin{minipage}{.45\linewidth}
		\begin{lstlisting}[basicstyle=\ttfamily\scriptsize, language={[x86masm]Assembler}, numberstyle=\tiny, tabsize=4, keywordstyle=\color{blue!70}, commentstyle=\color{red!50!green!50!blue!50}, rulesepcolor=\color{red!20!green!20!blue!20}, escapechar=!, captionpos=b, numberstyle=\tiny, numbers=left, numbersep=-11pt, belowcaptionskip=-10pt]
      mov     ecx, gvar1!'!
      mov     ebp, gvar2!'!
      test    ebp, ebp
      add     ebp, ecx
		\end{lstlisting}
	\end{minipage}
\end{lrbox}

\newsavebox{\switchjmp}
\begin{lrbox}{\switchjmp}
	\begin{lstlisting}[basicstyle=\ttfamily\scriptsize, language={[x86masm]Assembler}, numberstyle=\tiny, tabsize=4, keywordstyle=\color{blue!70}, commentstyle=\color{red!50!green!50!blue!50}, rulesepcolor=\color{red!20!green!20!blue!20}, escapechar=!, captionpos=b, numberstyle=\tiny, numbers=left, numbersep=-11pt, belowcaptionskip=-10pt]
      mov     edx, [ebp-0E4h] ; load a local variable
      lea     eax, [edx-0Ah] ; get the index
      cmp     eax, 2Ah
      ja      loc_8052880 ; the default case
      jmp     dword ptr [eax*4+808F630h]; indirect jump
	\end{lstlisting}
\end{lrbox}

\newsavebox{\armswitchjmp}
\begin{lrbox}{\armswitchjmp}
	\begin{lstlisting}[basicstyle=\ttfamily\scriptsize, language={[ARM]Assembler}, numberstyle=\tiny, tabsize=4, keywordstyle=\color{blue!70}, commentstyle=\color{red!50!green!50!blue!50}, rulesepcolor=\color{red!20!green!20!blue!20}, escapechar=!, captionpos=b, numberstyle=\tiny, numbers=left, numbersep=-11pt, belowcaptionskip=-10pt]
      LDR     R1, [SP, #0x88+arg_0] !\textcolor{red!50!green!50!blue!50}{; load the index}!
      CMP     R1, #8
      LDRLS   PC, [PC, R1, LSL#2] !\textcolor{red!50!green!50!blue!50}{; indirect jump}!
      B       loc_410F4 !\textcolor{red!50!green!50!blue!50}{; the default case}!
    !\textcolor{red!50!green!50!blue!50}{--------------------------------------------------------------------------}!
      DCD loc_410F8
      DCD loc_40FCC
      ...
	\end{lstlisting}
\end{lrbox}

\newsavebox{\inputindirectcall}
\begin{lrbox}{\inputindirectcall}
	\begin{lstlisting}[basicstyle=\ttfamily\scriptsize, language={[x86masm]Assembler}, numberstyle=\tiny, tabsize=4, keywordstyle=\color{blue!70}, commentstyle=\color{red!50!green!50!blue!50}, rulesepcolor=\color{red!20!green!20!blue!20}, escapechar=!, captionpos=b, numberstyle=\tiny, numbers=left, numbersep=-11pt, belowcaptionskip=-10pt]
      mov     eax, [ebp+arg_0] ; load the first argument
      mov     eax, [eax] 
      ...
      call    eax ; indirect call
	\end{lstlisting}
\end{lrbox}

\newsavebox{\cfgindirectcall}
\begin{lrbox}{\cfgindirectcall}
	\begin{lstlisting}[basicstyle=\ttfamily\scriptsize, language={[x86masm]Assembler}, numberstyle=\tiny, tabsize=4, keywordstyle=\color{blue!70}, commentstyle=\color{red!50!green!50!blue!50}, rulesepcolor=\color{red!20!green!20!blue!20}, escapechar=!, captionpos=b, numberstyle=\tiny, numbers=left, numbersep=-11pt, belowcaptionskip=-10pt]
      test    eax, eax ; input realted
      jnz     short loc_806E0D4
      mov     ds:dword_808C810, 808157Bh ; the false branch
      jmp     loc_806E0E9
    loc_806E0D4:
      mov     ds:dword_808C810, 80815FCh ; the true branch
    loc_806E0E9:
      mov     eax, ds:dword_808C810
      ...
      call    eax ; indirect call
	\end{lstlisting}
\end{lrbox}

\IEEEraisesectionheading{\section{Introduction}\label{sec:introduction}}

\IEEEPARstart{B}inary code similarity comparison is a fundamental methodology which identifies similar or identical code fragments in target binary programs with the reference code.
It has numerous important applications in software engineering as well as security, for example, plagiarism detection~\cite{jhi2011value, zhang2012first, zhang2014program}, code searching~\cite{Khoo2013rendezvous, david2014tracelet, chandramohan2016bingo}, program comprehension~\cite{hu2016cross}, malware lineage inference~\cite{walenstein2007software, lindorfer2012lines, ming2017binsim}, patch code analysis~\cite{brumley2008automatic, zhang2018precise}, known vulnerability detection~\cite{pewny2014leveraging, pewny2015cross, eschweiler2016discovre, feng2016scalable},~etc.
In addition, with the development of smart and IoT~(Internet of Things) devices, binary code similarity comparison is also required to be performed across multiple architectures considering above applications.
Therefore, to improve the productivity and ensure the security of the software, it is necessary to effectively compare binary code similarity across architectures.

To fulfill the target, there exist two challenges.
The first one is the \emph{semantics-equivalent code transformation}~(\textbf{C1}).
It results from different compilation settings~\cite{egele2014blanket}~(i.e., different compilers or optimization options), code obfuscation~\cite{luo2014semantics,ming2017binsim}, and varied instruction set architectures~(ISAs)~\cite{pewny2015cross}.
Because of the code transformation, even though two pieces of binary code are compiled from the same code base~(i.e.,~semantically equivalent), they would differ significantly on the syntax and structure level, such as variant instruction sequences and control flow graphs,~etc.
The other challenge lies in the \emph{trade-off} between comparison accuracy and coverage~(\textbf{C2})~\cite{wang2017memory}.
Dynamic methods procure rich semantics from code execution to guarantee the high accuracy of comparison, yet 
they analyze merely the executed code, leading to low code coverage.
In contrast, 
static methods 
are able to cover all program components, while they heavily rely on syntax or structure-based code features which lacks semantics and thus produce less accurate results.

In the literature, it has drawn much attention to compare the similarity of binary code.
However, existing solutions adopt either static methods which depend on semantics-less code features or dynamic methods which merely care about executed code.
They cannot reach the compromise between comparison accuracy, which corresponds to \textbf{C1}, and coverage.
Typically,
static methods discovRE~\cite{eschweiler2016discovre}, Genius~\cite{feng2016scalable}, and Kam1n0~\cite{ding2016kam1n0} extract code features from control flow graphs, and measure the binary function similarity basing on graph isomorphism.
\mbox{Multi-MH}~\cite{pewny2015cross}, BinGo~\cite{chandramohan2016bingo}, and IMF-sim~\cite{wang2017memory} capture behaviors of a binary function by sampling it with random values.
Since the random input lacks semantics and is commonly illegal for the function, it could hardly trigger the core semantics of that function.
Besides, Asm2Vec~\cite{ding2019asm2vec} leverages machine learning techniques to extract code features from the lexical relationships of assembly code tokens,
while it is still syntax-based and suffers from \textbf{C1}.
For dynamic methods, although \mbox{Ming~et al.}~\cite{ming2017binsim}, \mbox{Jhi~et al.}~\cite{jhi2011value}, and \mbox{Zhang~et al.}~\cite{zhang2012first} adopt semantics-based code features, i.e.,~system calls and invariant values during execution, they perform detection merely on executed code.
BLEX~\cite{egele2014blanket} pursues high code coverage at the cost of breaking normal execution of binary functions, distorting the semantics inferred from the collected features.
Thus, it is necessary to propose a method which only depends on semantics and takes advantages of both static and dynamic techniques so as to achieve high accuracy and coverage for binary code similarity comparison.

In this paper,
we propose \binCMP, a semantics-based hybrid method, to fulfill the target.
Given the reference function, \binCMP\ aims to identify its match of similar semantics in the target binary program.
\binCMP\ firstly instruments the reference function, and executes it with available test cases to record its runtime information.
It then migrates the runtime information to each function of the target program, and emulates the execution of that function.
During the execution of the reference function and the emulation of the target functions, the semantic signature of each function is extracted simultaneously.
Finally, \binCMP\ compares the signature of the reference function with that of each target function in pairs to measure their similarity.
Semantics describes the processes a computer follows when executing a program, which could be shown by describing the relationship between the input and output of a program~\cite{harper2016practical}.
To overcome \textbf{C1} of semantics-equivalent code transformation,
\binCMP\ only relies on signatures generated from the input/output and intermediate processing data collected during the~(emulated) execution of the whole reference or target function.
To address \textbf{C2} of the trade-off between comparison accuracy and coverage, \binCMP\ adopts the hybrid method which captures signatures either in a static or in a dynamic manner.
By executing the reference function and emulating the target functions, \binCMP\ is able to extract their semantics-based signatures from the~(emulated) executions.
Because of the emulation, it is not necessary to really run the target program. \binCMP\ emulates the target functions with the runtime information migrated from the reference function. Thus, it could cover all functions of the target program.

We have implemented a prototype of \binCMP\ using the above method.
We evaluate it with nine real-world projects compiled with various compilation settings, obfuscation configurations, and ISAs on the 32-bit Linux platform, totally performing over \emph{100 million} pairs of function comparisons.
The experimental results indicated that \binCMP\ not only is robust to semantics-equivalent code transformation, but also outperforms the state-of-the-art solutions of binary code similarity comparison.

The paper makes the following contributions.
\begin{itemize}
	\item We propose \binCMP, a semantics-based hybrid method, to compare binary code similarity.
	It captures the semantic signature of a binary function either in a dynamic~(execution) or in a static~(emulation) manner.
	Thus, it could not only detect similar functions accurately with signatures of rich semantics, but also cover all target functions under analysis.
	
	\item \binCMP\ emulates the execution of a function by migrating existing runtime information.
	To smooth the process of migration,
	we propose novel strategies to handle global variable reading, indirect jumping/calling, and library function invocation.
	
	\item We implement \binCMP\ in a prototype system which supports cross-architecture binary code similarity comparison on the 32-bit Linux platform.
	\binCMP\ is evaluated with nine real-world projects which are compiled with different compilation settings, obfuscation configurations, and instruction set architectures.
	The experimental results show that \binCMP\ is robust to the semantics-equivalent code transformation.
	Besides, it covers all candidate target functions for similarity comparison, and outperforms the state-of-the-art solutions.
\end{itemize}

As this work is an extended version of our conference paper~\cite{hu2018binmatch}, we list below, the contributions of this extension:
\begin{enumerate}[wide, labelindent=0pt, label=\textbf{(\arabic*)}]
	\item \textbf{Effectiveness:} We adopt Intel C++ Compiler~(ICC) to compile the object projects, and leverage \binCMP\ to compare the similarity of the resulting binary functions to those compiled by GCC and Clang as well as the obfuscated ones.
	Besides, we conduct experiments to evaluate the capacity of \binCMP\ in comparing similar binary function across the mainstream architectures, i.e.,~x86, ARM and MIPS.
	In addition to Kam1n0~\cite{ding2016kam1n0} and BinDiff~\cite{flake2004structural}, we compare the results of \binCMP\ to those of Asm2Vec~\cite{ding2019asm2vec} and CACompare~\cite{hu2017binary} as well.
	The experimental results further show the effectiveness of \binCMP\ in handling semantics-equivalent code transformation which exists in the binary code.
	
	\item \textbf{Practicability:} Despite effective, \binCMP\ is inefficient. To make the method more practical, we propose a strategy to prune the process of signature comparison.
	Additionally, we adopt a hash-based technology to efficiently estimate the similarity of function signatures.
	
	\item \textbf{Investigation into thresholds:} 
	To reach the compromise between comparison accuracy and efficiency,
	we investigate the thresholds for applying the pruning strategy and the hash-based technology, and find suitable values for \binCMP.
	The results indicate that it thus fulfills the comparison efficiently.
	Besides, it also outperforms the existing solutions from the perspective of accuracy.
\end{enumerate}

The rest of this paper is organized as follows.
Section~\ref{sec:example} introduces a motivating example and presents the system overview of \binCMP.
Section~\ref{sec:approach} introduces how \binCMP\ extracts semantics signatures of binary functions and compares their similarity.
Section~\ref{sec:implementation} presents several aspects to implement \binCMP.
The experimental results are shown and analyzed in Section~\ref{sec:evaluation}.
Some related issues are discussed in Section~\ref{sec:discussion}.
Section~\ref{sec:relate} discusses the related work and the conclusion follows in Section~\ref{sec:conclusion}.

\section{Motivation and Overview}\label{sec:example}
In this section, we firstly present a typical application of binary code similarity comparison, and illustrate the challenges of the topic with an example.
Then, we explain the basic idea of \binCMP\ and show the overview of its system.

\begin{figure}[t]
	\subfloat[GCC -O0\label{fig:cfg_libpng_gcc_O0}]{\includegraphics[scale=.1]{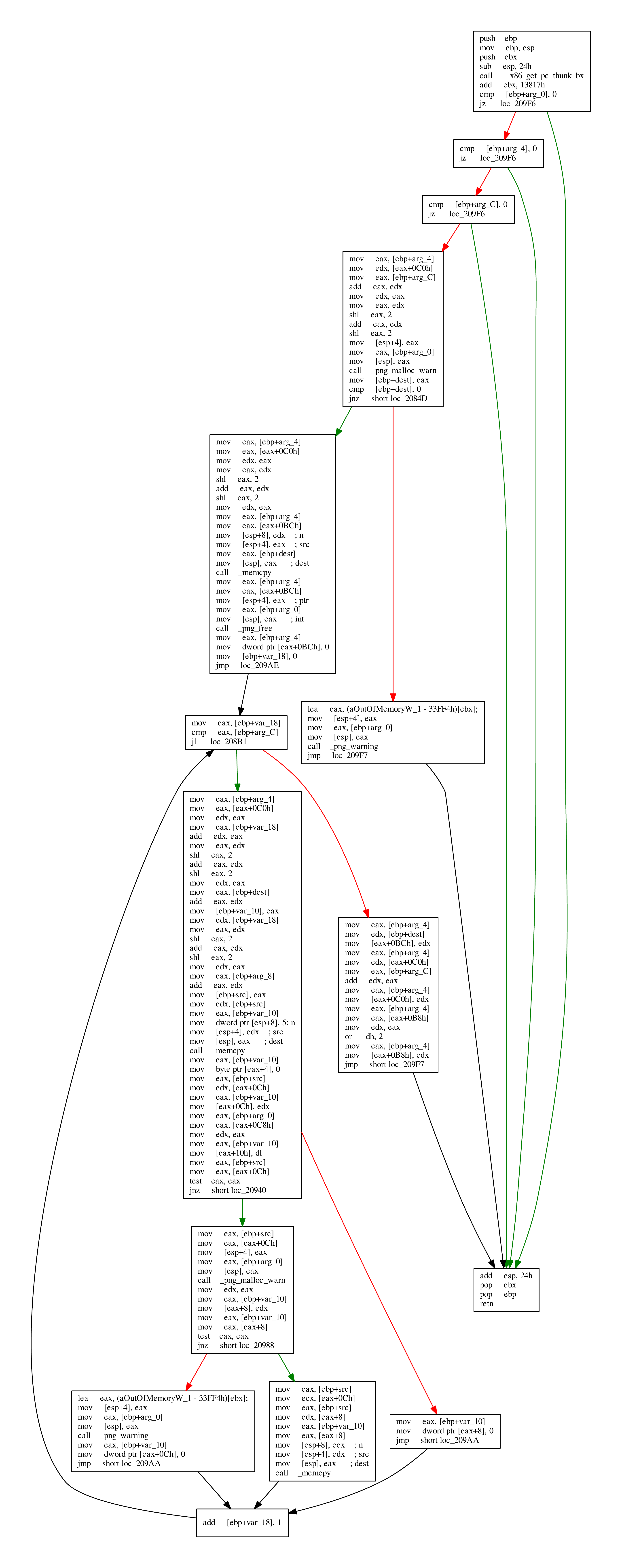}}
	\hfil
	\subfloat[Clang -O2\label{fig:cfg_libpng_clang_O2}]{\includegraphics[scale=.1]{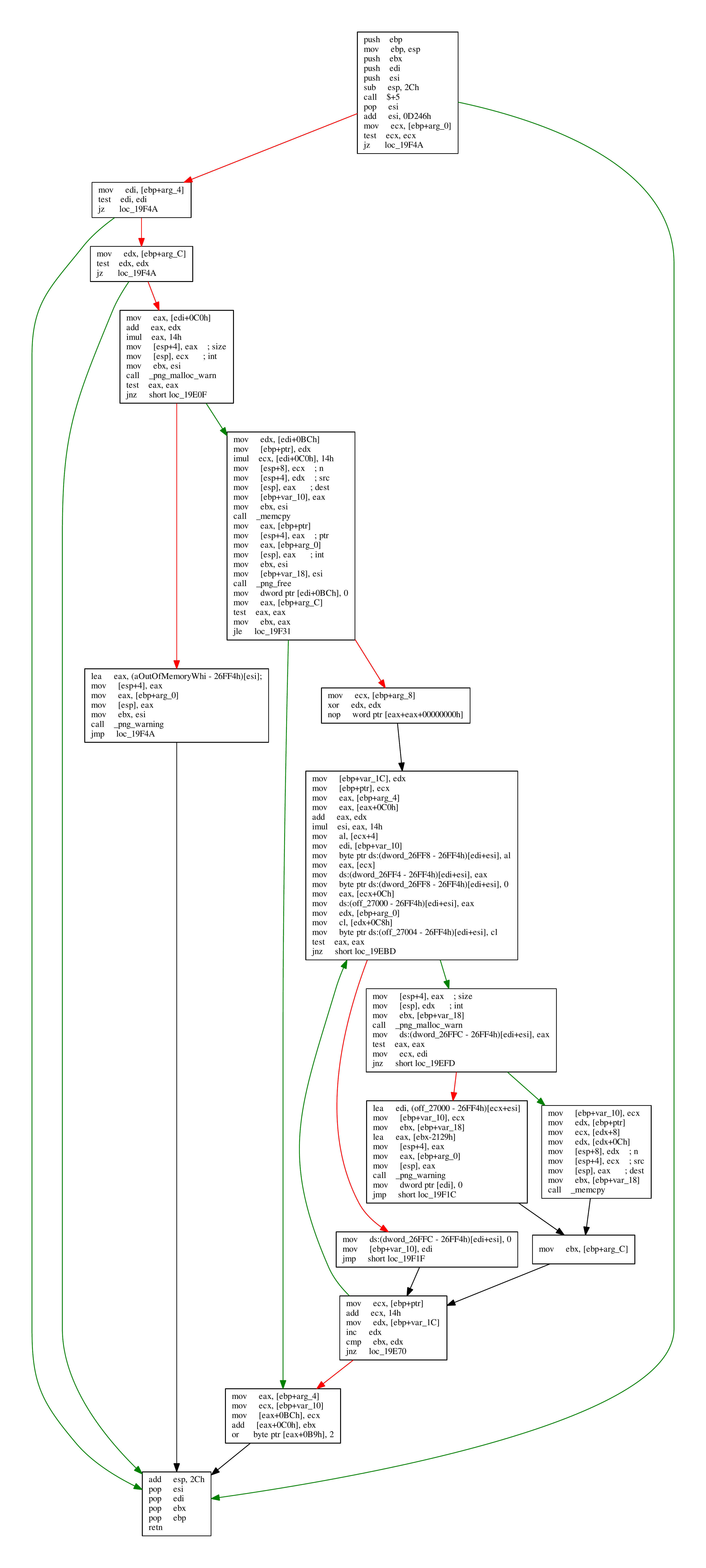}}
	\hfil
	\subfloat[NConvert v6.17\label{fig:cfg_libpng_nconvert}]{\includegraphics[scale=.1]{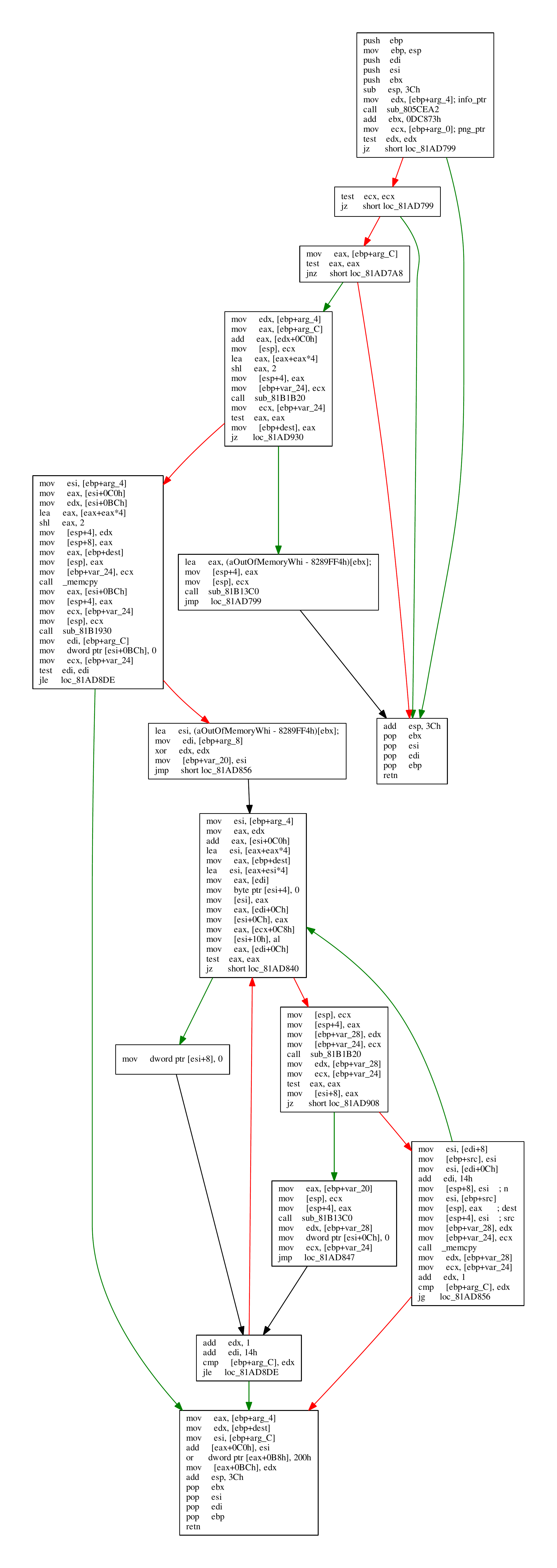}}
	\caption{Control Flow Graphs of png\_set\_unknown\_chunks}
	\label{fig:cfg_libpng}
\end{figure}

\subsection{The Motivating Example}\label{sec:example:movtivating}
Known vulnerability detection is a typical application of binary code similarity comparison~\cite{chandramohan2016bingo,pewny2014leveraging,pewny2015cross,eschweiler2016discovre, feng2016scalable}.
Given a piece of code which contains a known vulnerability, it is possible to locate its similar~(or identical) match in other programs so as to check whether they are vulnerable or not.

\texttt{NConvert}\footnote{https://www.xnview.com/en/nconvert/} is a closed-source image processor which supports multiple image formats.
It handles files of the PNG~(Portable Network Graphics) format with the statically-linked open-source library \texttt{libpng}\footnote{http://www.libpng.org/pub/png/libpng.html}.
Unfortunately, \texttt{libpng} is found to suffer from an integer overflow vulnerability in the function \emph{png\_set\_unknown\_chunks} before the version of 1.5.14~(CVE-2013-7353\footnote{https://www.cvedetails.com/cve/CVE-2013-7353/}).
The vulnerability allows attackers to cause a denial of service via a crafted image.
To ensure whether \texttt{NConvert} suffers from the vulnerability, analyzers firstly need to locate the potential vulnerable function in it.

Since the source code of \texttt{libpng} is available, it is reasonable to locate the target function via code similarity comparison.
\texttt{NConvert} is closed-source that only its executable is accessible, and the compilation setting of the executable is unknown.
Even though executables are compiled from the same code base, different compilation settings would lead to \emph{semantics-equivalent code transformation}, generating syntax and structure-variant binary code of equal semantics~(\textbf{C1}).
Figure~\ref{fig:cfg_libpng} presents the CFGs~(Control Flow Graphs) of \pngfunc.
Functions in Figure~\ref{fig:cfg_libpng_gcc_O0} and Figure~\ref{fig:cfg_libpng_clang_O2} are compiled from the source code of \mbox{\texttt{libpng} v1.5.12} with the setting of \mbox{gcc -O0} and \mbox{clang -O2} separately, while Figure~\ref{fig:cfg_libpng_nconvert} is extracted from the executable of \mbox{\texttt{NConvert} v6.17} via manual reverse engineering.
Because of the code transformation, despite the same semantics, the three functions differ in instruction sequences and CFGs.
Thus, methods relying on syntax or structure code features~(e.g.,~CFG isomorphism, binary code hashing) become ineffective.

Another problem is the \emph{trade-off} between comparison accuracy and coverage~(\textbf{C2}).
Existing dynamic analysis-based methods only handle the executed code.
However, \pngfunc\ is statically-linked, mixing with the user-defined functions in the executable of \texttt{NConvert}.
It requires huge extra work for dynamic methods to generate test cases in order to cover the target function, which is still an issue of binary code dynamic analysis~\cite{kargen2015turning}.
In contrast, static analysis-based methods could cover all functions of \texttt{NConvert}.
Nevertheless, they depend on semantics-less code features because they perform without actually executing the code.
Therefore, the static methods cannot handle the semantics-equivalent code transformation.

\begin{figure*}[t]
	\centering
	\includegraphics[scale=.37]{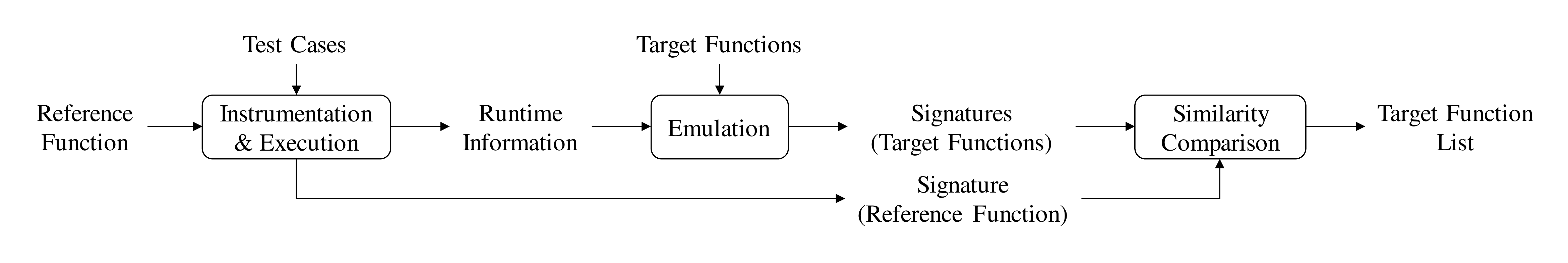}
	\caption{System Architecture of \binCMP}
	\label{fig:sys_ovv}
\end{figure*}

\subsection{System Overview of \binCMP}
We propose \binCMP\ to compare the similarity of binary functions.
Given a \emph{reference} function,
\binCMP\ finds its match of similar semantics in the target binary program, returning a list of functions~(the \emph{target} functions) of the target program, which is ranked basing on the similarity of semantics.

Figure~\ref{fig:sys_ovv} presents the work flow of \binCMP.
Provided the reference function has been well analyzed or understood~(\pngfunc), \binCMP\ dynamically instruments and executes it with available test cases, capturing its semantic signature~(\S\ref{sec:approach:signature}).
Meanwhile, runtime information is recorded during the execution~(\S\ref{sec:approach:execution}).
Then, \binCMP\ emulates every function of the target program~(\texttt{NConvert}) with the runtime information.
During the emulation, signature of each target function is extracted as well.
Afterward, \binCMP\ compares the signature of the reference function to that of each target function in pairs, and computes their similarity score~(\S\ref{sec:approach:similarity}).
Finally, \binCMP\ generates a list of target functions ranked by the similarity scores in descending order.

In Summary, to overcome \textbf{C1}, 
\binCMP\ completely depends on the semantics-based signature which is generated from the input/output and the intermediate processing data during the~(emulated) execution of a function.
To address \textbf{C2}, \binCMP\ captures function signatures in a hybrid manner.
It extracts the signature of the reference function via \emph{dynamically executing} its test cases.
We assume that the reference function has been well studied and its test cases are available.
In above example, the integer overflow of \pngfunc\ has been known, and its test case could be found in the \texttt{libpng} project or the vulnerability database.
Then, with the runtime information of the reference function, \binCMP\ extracts the signature of each function of the target program~(\texttt{NConvert}) via \emph{static emulation}.
Therefore, \binCMP\ is able to cover all target functions and detect similar function with signatures of rich semantics.

\section{Methodology}\label{sec:approach}
In this section, we firstly introduce the semantic signatures adopted by \binCMP.
Then, we discuss how it captures the signatures of binary functions and measures their similarity.

\subsection{Semantic Signature}\label{sec:approach:signature}
The semantics describes the processes a computer follows when executing a program.
It could be shown by describing the relationship between the input and output of the program~\cite{harper2016practical}.
Thus, given a specific input, we focus on two points to reveal the semantics of a binary function:
\begin{enumerate*}[label=\roman*)]
	\item \emph{what} is the corresponding output after the function processing the input, and
	\item \emph{how} the function processes the input to generate the output.
\end{enumerate*}
The signature adopted by \binCMP\ consists of the following features:
\begin{itemize}
	\item \emph{Output Values:}
	For a binary function,
	the feature consists of the return value and the global variable values written to the memory.
	It covers the output of a function. When given the specific input, the feature directly shows the semantics of a function.
	
	\item \emph{Comparison Operand Values:}
	The feature is composed of values for comparison operations whose results decide the following control flow of an~(emulated) execution.
	A function might have numerous paths, while only one is triggered by the input to generate the output.
	The feature describes how an input chooses the path of a function to produce the corresponding output, indicating the relationship between the input and output.
	Therefor, it reflects the semantics of a function.
	
	\item \emph{Invoked Standard Library functions:}
	Standard library functions provide fundamental operations for implementing user-defined functions. They have complete functionality, such as \texttt{malloc} meaning dynamic memory allocation, \texttt{fread} representing file reading, etc.
	Then the invocations of those library functions indicate the semantics of the~(emulated) execution.
	Besides, the feature has been show to be effective for binary code similarity comparison~\cite{wang2009begavior, wang2012can}.
	Thus, it is adopted as complement to the semantic signature of \binCMP.
\end{itemize}

During the~(emulated) execution of a binary function, \binCMP\ captures the sequence of above features, and considers it as the signature of that function for latter similarity comparison.

\begin{algorithm}[t]
	\SetKwInput{Input}{Input}
	\SetKwInOut{Output}{Output}
	\DontPrintSemicolon
	
	\caption{Instrumentation}
	\label{algthm:instrumentation}
	
	\Input{Instruction under Analysis $\mathcal{I}$}
	\Output{Instruction after Instrumentation $\mathcal{I}_r$}
	
	\Al{Instrumentation ($\mathcal{I}$)}{
		
		$\mathcal{I}_r \leftarrow \mathcal{I}$\\
		\tcp{capture features for the signature}
		\If{$\mathcal{I}$ outputs data}{
			$\mathcal{I}_r \leftarrow$ {\ttfamily\bfseries record\_data\_val} ($\mathcal{I}_r$)
		}
		\If{$\mathcal{I}$ performs comparison}{
			$\mathcal{I}_r \leftarrow$ {\ttfamily\bfseries record\_oprd\_val} ($\mathcal{I}_r$)
		}
		\If{$\mathcal{I}$ calls a standard library function}{
			$\mathcal{I}_r \leftarrow$ {\ttfamily\bfseries record\_libc\_name} ($\mathcal{I}_r$)
		}
		
		\tcp{record runtime information}
		\If{$\mathcal{I}$ reads an argument of the function}{
			$\mathcal{I}_r \leftarrow$ {\ttfamily\bfseries record\_arg\_val} ($\mathcal{I}_r$) 
		}
		\ElseIf{$\mathcal{I}$ uses global variable}{
			$\mathcal{I} \leftarrow$ {\ttfamily\bfseries record\_var\_val} ($\mathcal{I}_r$)
		}
		\ElseIf{$\mathcal{I}$ calls a function indirectly}{
			$\mathcal{I}_r \leftarrow$ {\ttfamily\bfseries record\_func\_addr} ($\mathcal{I}_r$)
		}
		\ElseIf{a subroutine returns}{
			$\mathcal{I}_r \leftarrow$ {\ttfamily\bfseries record\_ret\_val} ($\mathcal{I}_r$)
		}

		\KwRet{$\mathcal{I}_r$}
	}
\end{algorithm}

\begin{figure}[t]
	\centering
	\includegraphics[scale=.21]{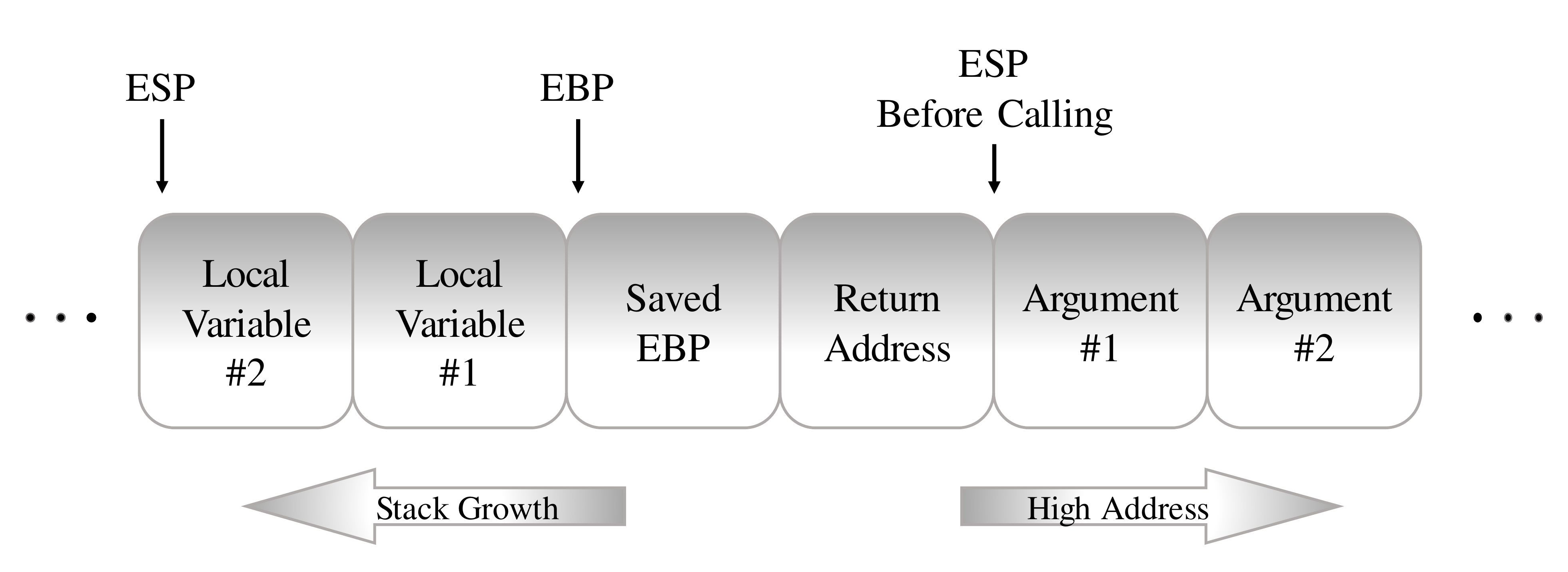}
	\caption{Calling Stack of \emph{cdecl}}
	\label{fig:cdecl_call_stack}
\end{figure}

\subsection{Instrumentation and Execution}\label{sec:approach:execution}
In this step, \binCMP\ dynamically instruments the reference function \emph{R} to extract its signature by executing the available test cases.
Meanwhile, runtime information for Emulation~(\S\ref{sec:approach:emulation}) is recorded as well.

Algorithm~\ref{algthm:instrumentation} presents the pseudo-code of instrumentation.
For the instruction $\mathcal{I}$ of \emph{R}, if it outputs data, performs comparison operations, or calls a standard library function, \binCMP\ injects code before it to capture corresponding features, then generates the signature of \emph{R}~(\mbox{Line 4-9}).

\mbox{Line 11-18} present the code for recording runtime information of \emph{R}'s execution.
Similar functions should behave similarly if they are executed with the same input~\cite{chandramohan2016bingo,pewny2015cross,egele2014blanket,hu2017binary}.
Therefore, \binCMP\ records the input of \emph{R}'s execution, which is provided for the Emulation in the next step.
For a binary function, the input consists of the argument values, global values, and return values of its subroutines~\cite{zhang2018precise}.
According to \emph{cdecl}, the default calling convention of x86, function arguments are prepared by callers and passed through the stack, as shown in Figure~\ref{fig:cdecl_call_stack}.
In contrast, 32-bit ARM and MIPS have specific argument registers~(i.e.,~R0-R3 for ARM, \$a0-\$a3 for MIPS).
When the function has arguments more than the registers, the surplus ones are passed by the stack, which is similar to \emph{cdecl}.
Thus, if $\mathcal{I}$ reads a variable from argument registers or stack with the address higher than the stack pointer~(ESP in Figure~\ref{fig:cdecl_call_stack}) before calling, \binCMP\ considers the variable as a function argument and records its value~(\mbox{Line 11-12}).

For ELF~(Executable and Linkable Format) files, global data is placed in specific data sections, e.g.,~\texttt{.bss} for uninitialized global data.
Therefore, if $\mathcal{I}$ uses data within those sections, \binCMP\ considers the values as global data and records them along with the accessing addresses~(\mbox{Line 13-14}). After this step, a record sequence of accessed global variables is generated. If a variable is accessed for multiple times, there would be the same number of records in the sequence as well.
Besides, \binCMP\ records the target addresses of subroutines indirectly invoked by \emph{R}~(\mbox{Line 15-16}).
The return values of all subroutines are recorded as well, including user-defined functions and library functions~(\mbox{Line 17-18}).

\begin{algorithm}[t]
	\SetKwInput{Input}{Input}
	\SetKwInOut{Output}{Output}
	\DontPrintSemicolon
	
	\caption{Emulation}
	\label{algthm:emulation}
	
	\Input{Emulated Memory Space of the Target Function $\mathcal{M}$}
	\Input{Runtime Value Set of the Reference Function $\mathcal{S}$}
	
	\Al{Emulation ($\mathcal{M}$, $\mathcal{S}$)}{
		{\ttfamily\bfseries init\_func\_stack} ($\mathcal{M}$)\\
		{\ttfamily\bfseries assign\_func\_arg} ($\mathcal{M}$, $\mathcal{S}$)\\
		\ForEach{instruction $I$ to be emulated}{
			\If{$I$ uses a global variable}{
				\If{the variable is accessed for the first time}{
					{\ttfamily\bfseries migrate\_var\_val} ($\mathcal{M}$, $\mathcal{S}$)
				}
			}
			\If{$I$ calls a function indirectly}{
				$addr \leftarrow$ {\ttfamily\bfseries get\_tar\_addr} ($I$)\\
				\If{$addr \in \mathcal{S}$}{
					{\ttfamily\bfseries migrate\_ret\_val} ($\mathcal{M}$, $addr$, $\mathcal{S}$)
				}
				\ElseIf{$addr$ is an illegal function address}{
					{\ttfamily\bfseries exit\_emulation}()
				}
			}
			\If{$I$ invokes a standard library function}{
				$libf \leftarrow$ {\ttfamily\bfseries get\_func\_name} ($I$)\\
				\If{$libf$ needs system supports}{
					{\ttfamily\bfseries migrate\_ret\_val} ($\mathcal{M}$, $libf$, $\mathcal{S}$)\\
					{\ttfamily\bfseries record\_feat\_val} ($\mathcal{M}$, $I$)\\
					\Continue
				}
			}
			\tcp{capture features for the signature}
			\If{$I$ contains features}{
				{\ttfamily\bfseries record\_feat\_val} ($\mathcal{M}$, $I$)
			}
			{\ttfamily\bfseries emulate\_inst} ($\mathcal{M}$, $I$, $\mathcal{S}$)
		}
	}
\end{algorithm}

\subsection{Emulation}\label{sec:approach:emulation}
For every target function \emph{T} to be compared with the reference function \emph{R},
\binCMP\ emulates its execution with runtime information recorded in the last step.
The semantic signature of \emph{T} is captured simultaneously.
The basic idea of the process is to emulate \emph{T} with the same input~(i.e.,~runtime information) of \emph{R} as if it was executed in the memory space of \emph{R}.
If \emph{T} is the match of \emph{R}, then their generated signatures should be similar.

To fulfill the emulation, \binCMP\ firstly needs to prepare the stack frame for \emph{T} which is similar to dynamic execution~(\S\ref{sec:approach:emulation:init}).
Then, it provides \emph{T} with the input of \emph{R} to perform the emulation.
From a function's perspective, the input consists of arguments, global variables, and return values of subroutines~\cite{zhang2018precise}.
Therefore, we need to handle function argument assignment~(\S\ref{sec:approach:emulation:argument}) and global variable reading~(\S\ref{sec:approach:emulation:gread}).
Since the targets of direct user-defined function calls are explicit, we then just focus on indirect calls~(\S\ref{sec:approach:emulation:indirec_call}) and standard library function calls which might require the system support~(\S\ref{sec:approach:emulation:libc}).
It is also necessary to consider indirect jumps whose target addresses are implicit for emulation~(\S\ref{sec:approach:emulation:indirect_jump}).

Algorithm~\ref{algthm:emulation} presents the pseudo-code of the process.
\binCMP\ firstly prepares the stack frame for the function emulation, including initializing the stack pointer values~(Line~2) and providing \emph{T} with the arguments of \emph{R}~(Line~3).
Before emulating the instructions of \emph{T} with the runtime intermediate data of \emph{R}~(\mbox{Line 23}), \binCMP\ needs to handle global variable reading~(\mbox{Line 5-7}), indirect function calling~(\mbox{Line 8-13}), and standard library function invocation~(\mbox{Line 14-19}) if necessary.
If \emph{T} is not the match of \emph{R}, 
the emulation might access illegal memory addresses which have never been recorded in the last step.
\binCMP\ then stops the emulation.
Additionally, \binCMP\ records the features of \emph{T} to generate its signature~(\mbox{Line 21-22}).
Next, we discuss the algorithm of emulation in more details.

\subsubsection{Stack Frame Pointer Initialization}\label{sec:approach:emulation:init}
Similar to execution, every \emph{T} for emulation has its own stack frame which is accessed by the stack pointer or the base pointer~(e.g.,~ESP or EBP) with relative offsets.
Before emulating, \binCMP\ assigns the stack and base pointers with those initial values of the reference function.
After assigning the argument values~(\S\ref{sec:approach:emulation:argument}), the arrangement of the stack frame is decided by the code of \emph{T}, such as pushing or popping values, allocating memory for local variables,~etc.

\subsubsection{Function Argument Assignment}\label{sec:approach:emulation:argument}
In our scenario, functions for similarity comparison are compiled from the same code base, i.e.,~they have identical interface with the same number and order of arguments.
According the calling convention, \binCMP\ recognizes the argument list of \emph{T}.
If the argument number of \emph{T} equals to that of \emph{R}, \binCMP\ assigns the argument values of \emph{R} to those of \emph{T} in order.
Otherwise, \emph{T} cannot be the match of \emph{R}, then \binCMP\ skips the emulation.
For example, \emph{R} and \emph{T} have the following argument lists:
\begin{center}
	\ttfamily
	\emph{R}(rarg\_0, rarg\_1, rarg\_2)\\
	\emph{T}(targ\_0, targ\_1, targ\_2)
\end{center}
Provided \binCMP\ has the values of \texttt{rarg\_0} and \texttt{rarg\_2}~(\emph{R} only accesses the two arguments in the execution), it assigns their values to \texttt{targ\_0} and \texttt{targ\_2} separately.
To make the emulation smooth, arguments without corresponding values~(\texttt{targ\_1}) are assigned with a predefined value~(e.g.,~\texttt{0xDEADBEEF}).

\subsubsection{Global Variable Reading}\label{sec:approach:emulation:gread}

In the execution of the reference function \emph{R}, it might read global~(or static) variables whose values have been modified by former executed code.
For example, \emph{R} accesses a global variable \texttt{gvar} whose initial value is~\emph{0}.
During the execution, before \emph{R} is invoked, its caller modifies \texttt{gvar} with the value~\emph{1}.
Then \emph{R} processes with \texttt{gvar} of value~\emph{1}.
To ensure the target function \emph{T} is emulated with the same input as \emph{R}, the modified global values should be assigned to the corresponding addresses which \emph{T} reads from.

Global variables are stored in specific data sections of an executable file~(e.g.,~\texttt{.data} section).
The size of each variable is decided by the source code. The location of the variable is determined during the process of compilation and not changed afterward.
Therefore, if the addresses of \emph{T}'s global variable accessing are input-unrelated, their explicit values could inferred during the emulation.
Recall that \binCMP\ has generated a record sequence of accessed global variables during the execution of \emph{R}~(\S\ref{sec:approach:execution}).
Then, it migrates the unassigned values in the record sequence to the inferred addresses according to the usage order.
If the address originates from the input, which is actually the one processed by \emph{R}~(e.g.,~pointer assigned as an argument of \emph{T}), \binCMP\ then directly reuses the first unassigned value of that address in the record sequence for the emulation.
In addition, if the address is illegal for both \emph{R} and \emph{T}, \binCMP\ stops and exits the emulation.

\begin{figure}[t]
	\subfloat[Reference Function\label{fig:global_reference}]{\usebox{\globalreference}}
	\hfil
	\subfloat[Target Function\label{fig:global_target}]{\usebox{\globaltarget}}
	\caption{Global Variable Value Migration}
	\label{fig:global_migration}
\end{figure}

Figure~\ref{fig:global_migration} shows an example of two functions for global variable value migration which bases on the usage order.
During the execution of \emph{R}, two global variables \texttt{gvar1} and \texttt{gvar2} are read at Line~1 and Line~3 separately in Figure~\ref{fig:global_reference}.
\texttt{gvar1} is used to test its value at Line~2, and \texttt{gvar2} is used for the addition operation at Line~4.
So the usage order of the two variable is \mbox{\texttt{[gvar1}, \texttt{gvar2]}}.
When emulating \emph{T} in Figure~\ref{fig:global_target}, \binCMP\ identifies \texttt{ecx} and \texttt{ebp} are loaded with global variables \texttt{gvar1'} and \texttt{gvar2'} at Line~1 and Line~2.
Then, it finds \texttt{ebp} is used for testing at Line~3, and \texttt{ecx} is used for the addition at Line~4 afterward.
The usage order of the global variables in Figure~\ref{fig:global_target} is \mbox{\texttt{[gvar2'}, \texttt{gvar1']}}.
Therefore, \binCMP\ assigns the value of \texttt{gvar1} to \texttt{gvar2'}, and \texttt{gvar2} to \texttt{gvar1'} accordingly.
If there are no enough global values to assign~(e.g.,~\emph{T} reads two global variables but \emph{R} reads only one), \binCMP\ stops and exits the emulation.

\begin{figure}
	%\centering
	\usebox{\switchjmp}
	\caption{Indirect Jump of a Switch on x86}
	\label{fig:swt_jmp}
\end{figure}

\subsubsection{Indirect Jumping}\label{sec:approach:emulation:indirect_jump}
An indirect jump~(or branch) is implemented with a jump table which contains an ordered list of target addresses.
For x86 and MIPS, jump tables are stored in \texttt{.rodata}, the read-only data section of an executable.
Therefore, similar to reading a global data structure, a jump table entry is accessed by adding the offset to the base address of the jump table.
The base address is a constant value, and the offset is computed from the input.
Figure~\ref{fig:swt_jmp} shows an indirect jump of a switch structure on x86.
At Line~2, the index value is computed with \texttt{edx}, a value of an input-related local variable, and stored in \texttt{eax}.
If the index value is not above \texttt{0x2A}, which represents the default case, an indirect jump is performed according to the jump table whose base address is \texttt{0x808F630}~(Line~5).

\begin{figure}[t]
	%\centering
	\usebox{\armswitchjmp}
	\caption{Indirect Jump of a Switch on ARM}
	\label{fig:arm_swt_jmp}
\end{figure}

On ARM, jump tables are inlined into the code.
They directly follow the code which accesses the tables.
Figure~\ref{fig:arm_swt_jmp} presents the indirect jump and jump table of a switch structure on ARM.
It loads the index from the first function argument~(\texttt{arg\_0}), storing it in \texttt{R1}~(Line~1).
The index is compared with \texttt{0x8}~(Line~2).
If it is larger than \texttt{0x8}, the program directly jumps to the default case~(Line~4).
Otherwise, the program refers to the jump table and gets the corresponding target address~(Line~3).
Since the jump table is attached to the jumping code, \texttt{PC}~(or \texttt{R15}, the Program Counter) is used as the base address.
Note that the code in Figure~\ref{fig:arm_swt_jmp} is compiled with the A32 instruction set of ARM, which has the fixed instruction length of \mbox{32 bits}~(\mbox{4 bytes}).
Because of the processor's pipeline, the \texttt{PC} value is always 8-byte ahead the current executed instruction.
When executing the loading instruction at Line~3, the \texttt{PC} in the right operand is pointing to the first entry of the jump table at Line~6.
Therefore, on ARM, indirect jumps also access jump tables with the decided value as the base address.

During the compilation, entries of jump tables are sorted and placed in the resulting binary code.
With the same input, code of identical semantics would jump to the same path to process the input.
Thus, \binCMP\ just follows the emulated control flow and has no need to do extra work for indirect jumps.

\begin{figure}[t]
	\usebox{\inputindirectcall}
	\caption{Indirect Call Decided by the Input on x86}
	\label{fig:ind_call_input}
\end{figure}

\subsubsection{Indirect Calling}\label{sec:approach:emulation:indirec_call}
Similar to indirect jumping, targets of indirect calls are decided by the input at runtime as well.
In some cases, target addresses directly come from the input, as shown in Figure~\ref{fig:ind_call_input}.
At Line~1, the first function argument~(\texttt{arg\_0}), which is the pointer of a data structure, is loaded to \texttt{eax}.
After the first member is fetched, which is a function address, the code calls the function indirectly.
Since the target function \emph{T} is emulated with the memory space of the reference function \emph{R}, if \emph{T} is the match of \emph{R}, targets of the indirect calls in above cases should be those invoked during the execution of \emph{R}.
\binCMP\ then migrates the return values of those calls to the corresponding ones of \emph{T}~(\mbox{Line 10-11} in Algorithm~\ref{algthm:emulation}).

\begin{figure}[t]
	%\centering
	\usebox{\cfgindirectcall}
	\caption{Indirect Call Affected by the Control Flow on x86}
	\label{fig:ind_call_cfg}
\end{figure}

In some cases, varied execution paths would generate different target addresses for a function call. That is decided by the input.
Figure~\ref{fig:ind_call_cfg} presents an example of the case.
At Line~1, the input-related value stored in \texttt{eax} is tested.
If it is not zero, the branch is taken at Line~2~(\texttt{jnz}: jump if not zero), jumping to \texttt{0x806E0D4}.
Then a function address \texttt{0x80815FC} is stored into data section at \texttt{0x808C810}~(Line~6).
Otherwise, another function address \texttt{0x808157B} is stored~(Line~3).
Afterward, the program jumps to the stored address indirectly~(Line~10).
Thus, the whole process merely depends on the input.
It is not necessary to do other work for indirect calls in such case.
In other cases, indirect calls are implemented with jump tables as well, such as virtual function tables.
\binCMP\ handles such indirect calls the same way as that for indirect jumps.

When the address is not a legal function address of either \emph{R} or \emph{T}, \emph{T} cannot be the match of \emph{R}.
\binCMP\ just stops the process and exits~(\mbox{Line 12-13} in Algorithm~\ref{algthm:emulation}).

\subsubsection{Standard Library Function Invocation}\label{sec:approach:emulation:libc}
If the target function \emph{T} calls a standard library function which requests the system support~(e.g., \texttt{malloc}), \binCMP\ skips its emulation and assigns it with the result of the corresponding one invoked by the reference function \emph{R}~(\mbox{Line 16-19} in Algorithm~\ref{algthm:emulation}).
For example, \emph{R} and \emph{T} calls following library functions in order:
\begin{center}
	\ttfamily
	R: malloc\_0, memcpy, malloc\_1\\
	T: malloc\_0', memset, malloc\_1'
\end{center}
\binCMP\ assigns return values of \texttt{malloc\_0}, \texttt{malloc\_1} to \texttt{malloc\_0'}, \texttt{malloc\_1'} separately, and skips the emulation.
In contrast, \texttt{memset} is emulated normally, because it has no need for the system support.
Afterward, when \emph{T} accesses the memory values on the heap, i.e.,~via the return values of \texttt{malloc\_0'} or \texttt{malloc\_1'},
it would be assigned with those of \emph{R} for the emulation.

\begin{algorithm}[t]
	\SetKwInput{Input}{Input}
	\SetKwInOut{Output}{Output}
	\DontPrintSemicolon
	
	\caption{Function Similarity Comparison}
	\label{algthm:similarity}
	
	\Input{Signature of the Reference Function $\mathcal{S}_r$}
	\Input{Signature of the Target Functions $\mathcal{S}_t$}
	\Input{Length Threshold $\mathcal{L}$}
	\Output{Similarity Score $\mathcal{S}$}
	
	\Al{Comparison ($\mathcal{S}_r$, $\mathcal{S}_t$, $\mathcal{L}$)}{
		$L_r \leftarrow$ {\ttfamily\bfseries length}($S_r$)\\
		\lIf{$L_r < \mathcal{L}$}{
			$F \leftarrow$ {\ttfamily\bfseries jaccard\_with\_lcs}
		}
		\lElse{
			$F \leftarrow$ {\ttfamily\bfseries hd\_with\_simhash}
		}
		$\mathcal{S} \leftarrow F(\mathcal{S}_r, \mathcal{S}_t)$\\
		\KwRet{$\mathcal{S}$}
	}
\end{algorithm}

\subsection{Similarity Comparison}\label{sec:approach:similarity}
\binCMP\ has captured the semantic signature~(the feature sequence) of the reference function via execution, and those of target functions via emulation.
In this step, 
it compares the signature of the reference function to that of each target function in pairs, and calculates their similarity score,
as shown in Algorithm~\ref{algthm:similarity}.
\binCMP\ adopts two solutions to measure the signature similarity.
One is the Jaccard Index~\cite{hamers1989similarity} with Longest Common Subsequence~(LCS)~\cite{bergroth2000survey}, the other is Hamming Distance~(HD)~\cite{hamming1950error} with SimHash~\cite{charikar2002similarity}.
The former solution is relatively more accurate but slow, while the latter one is fast but less accurate.
Thus, a length threshold~($\mathcal{L}$) is specified to select the suitable method.
When the lengths of the reference signatures are short, i.e.,~less than $\mathcal{L}$, \binCMP\ performs the comparison with the accurate matching~(\mbox{Line 3}).
Otherwise, it leverages the fuzzy matching to fulfill the target~(Line~4).
In such way, we aim to reach a compromise between comparison accuracy and efficiency.
We will discuss the value of $\mathcal{L}$ in Section~\ref{sec:eva:setup:parameter:length}.

After the comparison, \binCMP\ generates a list of target functions along with similarity scores, which is ranked by the scores in descending order.
Next, we discuss the details of the solutions adopted by \binCMP\ for similarity comparison.

\subsubsection{Jaccard Index with Longest Common Subsequence}\label{sec:approach:similarity:lcs}
Jaccard Index is a statistic used for measuring the similarity of sets.
Given two sets $S_r$ and $S_t$, the Jaccard Index is calculated as followed:
\begin{equation}\label{eq:jaccard}
	J(S_r, S_t) = \frac{|S_r \cap S_t|}{|S_r \cup S_t|} = \frac{|S_r \cap S_t|}{|S_r|+|S_t|-|S_r \cap S_t|}
\end{equation}
$J(S_r, S_t)$ ranges from 0 to 1, which is closer to 1 when $S_r$ and $S_t$ are considered to be more similar.

To better adapt to the scenario of \binCMP, we utilize the Longest Common Subsequence~(LCS) algorithm to the Jaccard Index.
On one hand, a signature is captured from the~(emulated) execution of a function. The appearance order of each entry in the signature is a latent feature as well.
The order reflects how the input is processed to generate the output, thus it is semantics-related.
On the other hand, the signature might be captured from an optimized or obfuscated binary function that it would contain diverse or noisy entries in the sequence.
LCS not only considers the element order of two sequences for comparison, but also allows skipping non-matching elements, which tolerates semantics-equivalent code transformation.
Hence, the LCS algorithm is suitable for signature similarity comparison of \binCMP.
With LCS, in Equation~\ref{eq:jaccard}, $S_r$ and $S_t$ represent the signatures of the reference and target functions.
$|S_r|$ and $|S_t|$ are their lengths, and $|S_r \cap S_t|$ is the LCS length of the two signatures.

\subsubsection{Hamming Distance with SimHash}\label{sec:approach:similarity:simhash}
SimHash is a solution which quickly estimates the similarity of two sets.
The basic idea of SimHash is similar items are hashed to similar hash values, i.e.,~with small bitwise hamming distances.
Assuming that the hash values have the size of \emph{F}, the similarity of two sets $S_r$ and $S_t$ is computed as followed:
\begin{equation}\label{eq:simhash}
	Sim(S_r, S_t) = 1 - \frac{HD[SH_F(S_r), SH_F(S_t)]}{F}
\end{equation}
Here, $SH_F(S)$ means the SimHash value of set $S$, ranging from $0$ to $2^F$.
Then, the hamming distance~(HD) of the two SimHash values ranges from $0$ to $F$.
As a result, $Sim(S_r, S_t)$ ranges from $0$ to $1$ as well.
The larger the value is, the more similar the two sets are considered to be.

In such way, comparing to the high time complexity $O(n^2)$ of the LCS algorithm, SimHash only has the time complexity of $O(n)$.
However, SimHash treats the feature sequences~(signatures) as sets.
It ignores the order of sequence elements, which is considered to be semantics-related.
Thus, it is less accurate in handling signatures extracted from optimized or obfuscated functions.
Consequently, hamming distance with SimHash computes the similarity of signatures more efficiently, while Jaccard Index with LCS has higher accuracy.

\section{Implementation}\label{sec:implementation}
Currently, \binCMP\ supports binary function similarity comparison of ELF~(Executable and Linkable Format) files.
It performs analysis on 32-bit Linux platform of three mainstream ISAs, i.e.,~x86, ARM, and MIPS.
Next, we discuss the key aspects of the implementation.

\subsection{Binary Function Boundary Identification}\label{sec:implementation:disassembling}
\binCMP\ performs comparison on the function level.
It requires the address and length information of each binary function under analysis.
Given an ELF file, \mbox{\texttt{IDA Pro} v6.6}\footnote{https://www.hex-rays.com/products/ida/} is adopted to disassemble it and identify the boundaries of its binary functions.
The plugin of \mbox{\texttt{IDA Pro}}, \texttt{IDAPython}, provides interfaces to obtain addresses of functions. For example, \mbox{\texttt{Functions(start, end)}} returns a list of function start addresses between address \texttt{start} and \texttt{end}.
As a result, we develop a script with IDAPython to acquire function addresses of binary programs automatically.
Besides, function arguments and switch structures are identified as well to assist in assigning argument values~(\S\ref{sec:approach:emulation:argument}) and emulating indirect jumps~(\S\ref{sec:approach:emulation:indirect_jump}).
Although the resulting disassembly of \mbox{\texttt{IDA Pro}} is not perfect~\cite{andriesse2016an,meng2016binary}, it is sufficient for the scenarios of \binCMP.

\subsection{Instrumentation and Emulation}\label{sec:implementation:instr_emult}
We implement the instrumentation module of \binCMP\ with \texttt{Valgrind}~\cite{nethercote2007valgrind}, a dynamic instrumentation framework.
\texttt{Valgrind} unifies binary code under analysis into VEX-IR, a RISC-like intermediate representation~(IR), and injects instrumentation code into the IR code.
Then, it translates the instrumented IR code into binaries for execution.
IR translation unifies the operations of binary code and facilitates the process of signature extraction.
For example, memory reading and writing instructions are all unified with \texttt{Load} and \texttt{Store}, the opcodes defined by VEX-IR.
Hence, we just concentrate on the specific operations of IR and ignore the complex instruction sets of different architectures.

The step of emulation is implemented basing on \texttt{angr}~\cite{shoshitaishvili2016sok}, a static binary analysis framework.
\texttt{angr} borrows VEX-IR from \texttt{Valgrind}, and translates binary code to be analyzed into IR statically.
Given a user-defined initial state, it provides a module named \texttt{SimProcedure} to emulate the execution of IR code.
SimProcedure allows injecting extra code to monitor the emulation of the IR code.
It actually emulates the process of instrumentation.
Besides, \texttt{angr} maintains a database of standard library functions to ease the emulation of those functions~(\S\ref{sec:approach:emulation:libc}).
Thus, we develop a script of monitoring code, which is similar to the instrumentation code developed with \texttt{Valgrind}, to capture semantic signatures during the emulation with \texttt{angr}.

\subsection{Function Inlining and Signature Inlining}
Function inlining is an operation which expands a callee to its caller.
It eliminates the calling and returning of the callee, improving the efficiency of code execution.
Then, it is adopted as a strategy for code optimization~\cite{chang1992profile}.
Function inlining also might be used as an obfuscation technique that modifies boundaries of functions~\cite{balakrishnan2005code}, posing difficulties to reverse engineering.

Since \binCMP\ works on the function level, function inlining would affect its accuracy of comparison.
For example, the reference function $M_r$ invokes the subroutine $N_r$ during the execution.
The corresponding function $N_t$ is inlined into $M_t$ in the target binary program, becoming $M_tN_t$.
Because the signature of $M_tN_t$ is actually extracted from two functions, while that of $M_r$ only contains one, \binCMP\ might miss the match of $[M_r, M_tN_t]$ finally.
To alleviate the side effects of function inlining, \binCMP\ inlines the signature of a callee to its caller, which is similar to the process of function inlining.
In above example, the signature of $N_r$ is then expanded in that of  $M_r$, becoming the signature of $M_rN_r$ for the similarity comparison.
Note that \binCMP\ only inlines the signatures of user-defined functions, not counting those of standard library functions.

\begin{algorithm}[t]
	\SetKwInput{Input}{Input}
	\SetKwInOut{Output}{Output}
	\DontPrintSemicolon
	
	\caption{Pruning Similarity Comparison}
	\label{algthm:prune_similarity}
	
	\Input{Signature of the Reference Function $\mathcal{S}_r$}
	\Input{Signature of the Target Functions $\mathcal{S}_t$}
	\Input{Length Threshold $\mathcal{L}$}
	\Input{Pruning Threshold $\mathcal{P}$}
	\Output{Similarity Score $\mathcal{S}$}
	
	\Al{pruningComparison ($\mathcal{S}_r$, $\mathcal{S}_t$, $\mathcal{L}$, $\mathcal{P}$)}{
		$L_r \leftarrow$ {\ttfamily\bfseries length}($\mathcal{S}_r$)\\
		$L_t \leftarrow$ {\ttfamily\bfseries length}($\mathcal{S}_t$)\\
		\tcp{pruning strategy}
		\lIf{{\ttfamily\bfseries max}($L_r$, $L_t$) / {\ttfamily\bfseries min}($L_r$, $L_t$) > $\mathcal{P}$}{$\mathcal{S} \leftarrow -1$}
		\tcp{Algorithm \ref{algthm:similarity}}
		\lElse{ $\mathcal{S \leftarrow}$ {\ttfamily\bfseries comparison}($\mathcal{S}_r$, $\mathcal{S}_t$, $\mathcal{L}$)}
		\KwRet{$\mathcal{S}$}
	}
\end{algorithm}

\subsection{Pruning Strategy of Similarity Comparison}\label{sec:implementation:pruning}

The code features adopted by \binCMP\ are semantics-related.
Intuitively, because of signature inline, signature lengths of similar functions should be close.
Thus, we propose a signature length-based pruning strategy to improve the efficiency of similarity comparison.
As presented at \mbox{Line 5} in Algorithm~\ref{algthm:prune_similarity}, given a pre-defined pruning threshold $\mathcal{P}$~(>1), \binCMP\ skips the comparison when the difference between two signature lengths is sufficiently large, i.e.,~the division of their lengths is lager than $\mathcal{P}$ or less than $\frac{1}{\mathcal{P}}$.
The two functions are considered to be dissimilar under that condition.
We will discuss the value of $\mathcal{P}$ in Section~\ref{sec:eva:setup:parameter:pruning}.

\section{Evaluation}\label{sec:evaluation}

We conduct empirical experiments to evaluate the effectiveness and capacity of \binCMP.
We firstly discuss thresholds~($\mathcal{L}$ in \S\ref{sec:approach:similarity} and $\mathcal{P}$ in \S\ref{sec:implementation:pruning}) adopted by \binCMP, which balance the accuracy and efficiency of comparison~(\S\ref{sec:eva:setup:parameter}).
Then, \binCMP\ is evaluated with binaries compiled with different compilation settings, including variant optimization options and compilers~(\S\ref{sec:eva:cc}).
We also evaluate the effectiveness of \binCMP\ in handling obfuscation by comparing binary functions with their obfuscated versions~(\S\ref{sec:eva:obf}).
Lastly, we leverage \binCMP\ to compare the similarity of binary code compiled with different ISAs, across x86, ARM and MIPS~(\S\ref{sec:eva:arch}).
The results of above experiments are all compared to those of existing solutions.

\subsection{Experiment Setup}
The evaluation is performed in the system of Ubuntu 16.04 which is running on an Intel Core \mbox{i7 @ 2.8GHz} CPU with 16G DDR3-RAM.

\begin{table*}[t]
	\renewcommand{\arraystretch}{1.5}
	\centering
	\caption{Object Projects of Evaluation}
	\label{tab:obj}
	\begin{tabularx}{.98\textwidth}{|c|c|X|l|}
		\hline
		Program & Version & \multicolumn{1}{c|}{Description} & \multicolumn{1}{c|}{Test Command} \\
		\hline
		\hline
		convert & 6.9.2 & Command-line interface to the ImageMagick image editor/converter & \texttt{convert} \emph{sample.png} -background black -alpha remove \emph{sample.jpg} \\
		\hline
		curl & 7.39 & Command-line tool for transferring data using various protocols & \texttt{curl} -O \emph{http://ftp.gnu.org/gnu/wget/wget-1.13.tar.xz} \\
		\hline
		ffmpeg & 2.7.2 & Program for transcoding multimedia files & \texttt{ffmpeg} -f image2 -i \emph{sample.png} \emph{sample.gif} \\
		\hline
		gzip & 1.6 & Program for file compression and decompression with the DEFLATE algorithm & \texttt{gzip} -\phantom{}-best -\phantom{}-recursive -\phantom{}-force \emph{sample\_directory} \\
		\hline
		lua & 5.2.3 & Scripting parser for Lua, a lightweight, multi-paradigm programming language & \texttt{lua} \emph{sample.lua} \\
		\hline
		mutt & 1.5.24 & Text-based email client for Unix-like systems & \texttt{mutt} -s \emph{"hello"} \emph{user@domain.com} < \emph{sample.txt} \\
		\hline
		openssl & 1.0.1p & Toolkit implementing the TLS/SSL protocols and a cryptography library & \texttt{openssl} s\_server -key \emph{key.pem} -cert \emph{cert.pem} -accept \emph{44330} -www \\
		\hline
		puttygen & 0.70 & Part of PUTTYGEN suit, a tool to generate and manipulate SSH public and private key pairs & \texttt{puttygen} -P \emph{sample.pem} -o \emph{key.pem}\\
		\hline
		wget & 1.15 & Program retrieving content from web servers via multiple protocols & \texttt{wget} \emph{http://ftp.gnu.org/gnu/wget/wget-1.13.tar.xz} -\phantom{}-no-cookies \\
		\hline
	\end{tabularx}
\end{table*}

\subsubsection{Dataset}\label{sec:eva:dataset}
We adopt programs of nine real-world projects as objects of the evaluation, as listed in Table~\ref{tab:obj}.
The object programs have various functionalities, including data transformation~(\texttt{convert}, \texttt{ffmpeg}), data compression~(\texttt{gzip}), code parsing~(\texttt{lua}), email posting~(\texttt{mutt}), etc.
With those objects, the effectiveness of \binCMP\ is shown to be not limited by the types of programs and functions under analysis.

For cross-compilation-setting comparison~(\S\ref{sec:eva:cc}), the objects are compiled with different compilers~(i.e.,~\mbox{GCC v4.9.3}, \mbox{Clang v4.0.0}, and \mbox{ICC v16.0.4}) and variant optimizations~(i.e.,~\texttt{-O3} and \texttt{-O0}).
For comparison with obfuscated code~(\S\ref{sec:eva:obf}), we adopt Obfuscator-LLVM~(\texttt{OLLVM})~\cite{ieeespro2015-JunodRWM} to obfuscate the object programs.
OLLVM provides three widely used techniques for obfuscation, including \emph{Instruction Substitution}, \emph{Bogus Control Flow}, and \emph{Control Flow Flattening}.
We use the three techniques to handle the object programs optimized with \texttt{-O3} and \texttt{-O0} respectively.
Then, for cross-architecture comparison~(\S\ref{sec:eva:arch}), the object programs are compiled for three architectures, i.e.,~x86, ARM, and MIPS, separately, with the compiler GCC and optimization option \texttt{-O3}.
As a result, we totally compile \emph{142} unique executables for the evaluation.

For each experiment, we select two executables of an object program, i.e.,~$E_{r}$~(the reference executable) and $E_t$~(the target executable).
\binCMP\ executes $E_r$ with the test command presented in Table~\ref{tab:obj}, and considers each executed function as a reference function.
Then, it compares every reference function to all target functions of $E_t$ in pairs to compute similarity scores.
Consequently, \binCMP\ performs over \emph{100 million} pairs of function comparisons in all the experiments.

\subsubsection{Ground Truth}\label{sec:eva:gtruth}
All the executables for the evaluation are stripped that their debug and symbol information is discarded.
To verify the correctness of the experimental results, we compile their \emph{extra unstripped copies}, and establish the ground truth with the symbol information.

For each reference function, \binCMP\ generates a list of target functions ranked by the similarity scores in descending order~(\S\ref{sec:approach:similarity}).
According to the ground truth, if the reference function name exists in the Top~\emph{K} entries of the resulting target function list, we consider the match of the reference function could be \emph{found} by \binCMP\ in $E_t$. 
Given the reference function, \binCMP\ is designed for assisting analyzers in looking for similar matches in target binaries. Thus, it is reasonable to assume the analyzers could further identify the correct match with acceptable amount of effort when provided with \emph{K} candidates.
In this paper, we assign \emph{K} with values of \emph{1}, \emph{5}, and \emph{10} respectively.

\subsubsection{Evaluation Metrics}
Similar to previous research~\cite{hu2016cross,egele2014blanket}, we measure the performance of \binCMP\ with $Accuracy$, the ratio of executed reference functions which could be \emph{found} in the Top~\emph{K} entries of the resulting target function lists. The formula is as followed:
\begin{equation}
Accuracy = \frac{|\text{Found Matches}|}{|\text{Reference Functions}|}
\end{equation}

\subsection{Parameter Settings}\label{sec:eva:setup:parameter}
We leverage either \emph{Jaccard Index with LCS}~(accurate but less efficient) or \emph{Hamming Distance with SimHash}~(efficient but less accurate) to measure the similarity of function signatures.
We propose the signature length threshold $\mathcal{L}$ to select suitable method for the measurement~(\S\ref{sec:approach:similarity}).
Besides, 
we introduce the ratio threshold $\mathcal{P}$ to prune unnecessary comparison to improve the efficiency of similarity measurement~(\S\ref{sec:implementation:pruning}).
In this section, we attempt to find the best values of $\mathcal{L}$ and $\mathcal{P}$ for the following experiments, which balance the accuracy and efficiency of \binCMP.

\begin{figure}[t]
	\hspace{-15pt}
	\includegraphics[scale=.47]{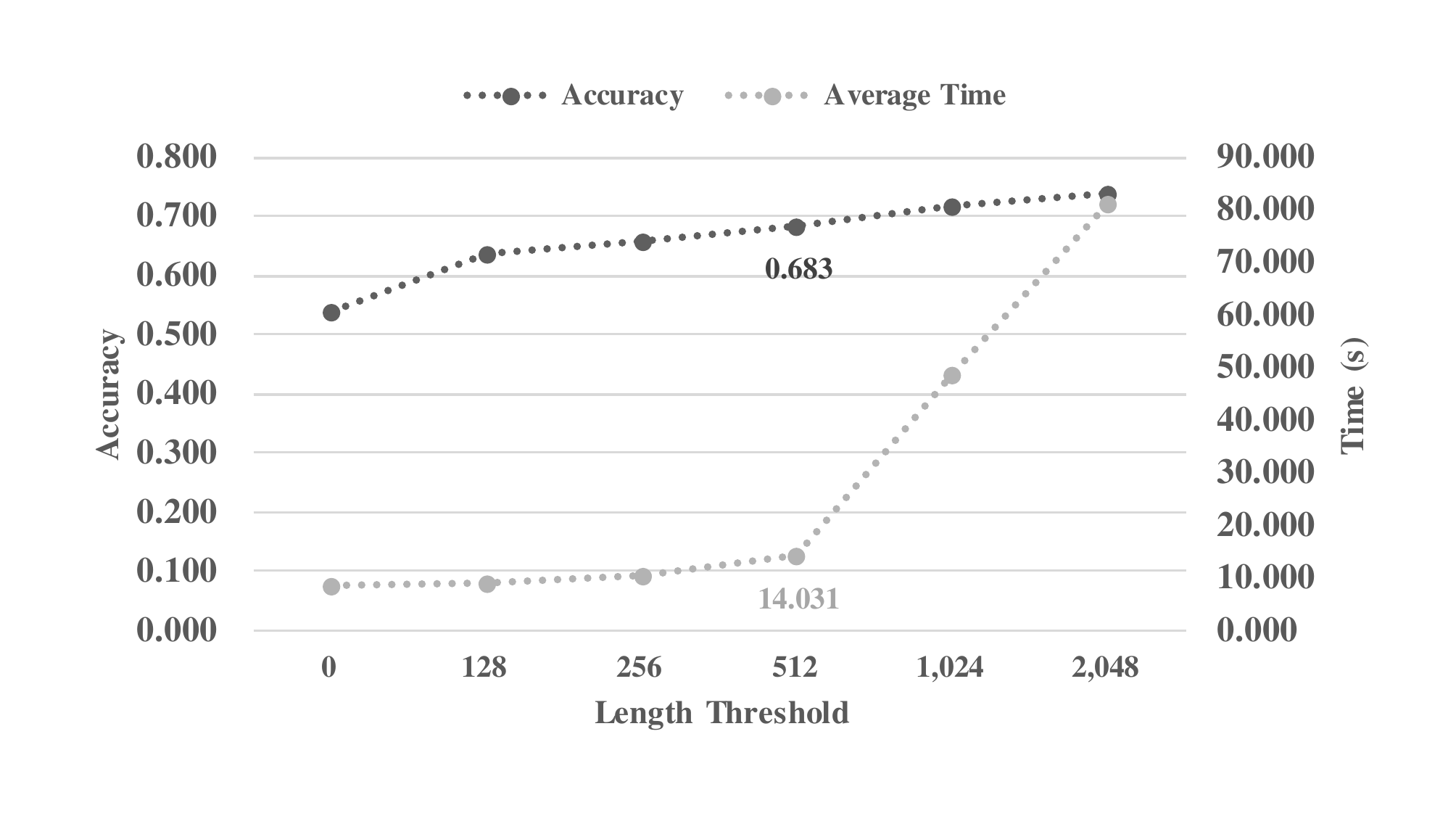}
	\caption{Accuracy and Time versus Length Threshold}
	\label{fig:length_threshold}
\end{figure}

\subsubsection{Length Threshold for Similarity Comparison}\label{sec:eva:setup:parameter:length}
We execute the reference executables, totally obtaining \emph{14,207} reference functions.
We randomly select \emph{4,000} of them, and investigate the performance of \binCMP\ with variant values of the length threshold $\mathcal{L}$.
The pruning strategy is disabled in this part of the experiments, i.e.,~$\mathcal{P}=+\infty$.

The results are shown in Figure~\ref{fig:length_threshold}.
The black line represents the Top~\emph{1} accuracy, and the gray line is the average time for processing each reference function.
As $\mathcal{L}$ increases exponentially, the corresponding time rises in a similar manner, while the accuracy grows much slower in a linear-like form.
When \binCMP\ performs the similarity comparison only with the Jaccard Index, i.e.,~$\mathcal{L}=+\infty$, it achieves the Top~\emph{1} accuracy of \emph{85.4\%}. That could be considered as the upper bound capacity of \binCMP. Nevertheless, the time consumption of each function reaches \emph{262.637} seconds. 
Considering the average time, note that $(512, 14.031)$ is the turning point of the line.
When $\mathcal{L}=512$, although the Top~\emph{1} accuracy is only \emph{68.3\%}, the Top~\emph{5} accuracy rises to \emph{86.4\%} which is comparable to the Top~\emph{1} accuracy when $\mathcal{L}=+\infty$.
Therefore, in the following experiments, \binCMP\ adopts:
\begin{equation}
\mathcal{L}=512.
\end{equation}

\begin{figure}[t]
	%\centering
	\hspace{-15pt}
	\includegraphics[scale=.45]{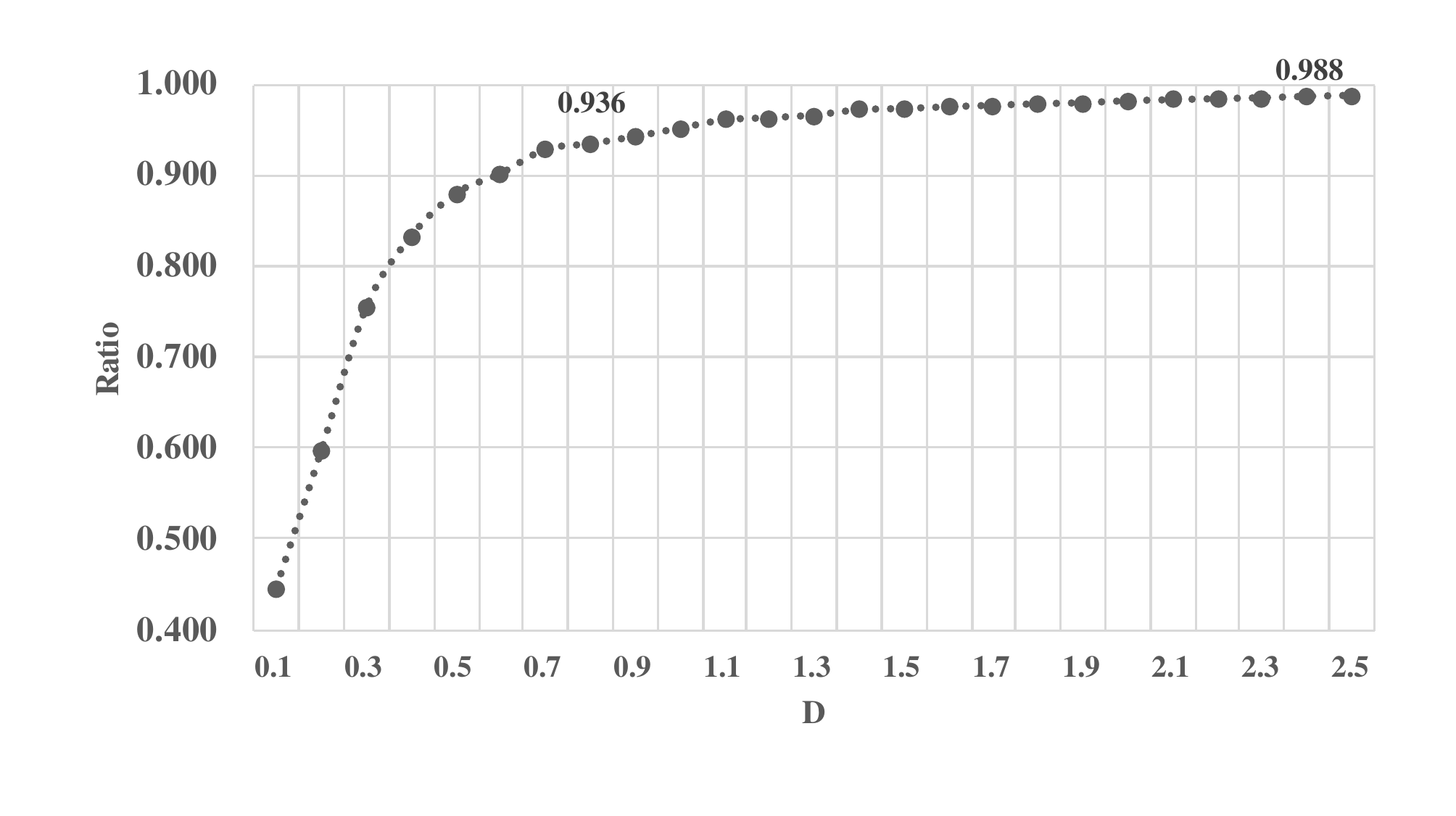}
	\caption{Accumulative Function Ratio versus Pruning Threshold}
	\label{fig:ratio_threshold}
\end{figure}

\subsubsection{Ratio Threshold for Pruning}\label{sec:eva:setup:parameter:pruning}

Considering the reference and target signatures $S_r$ and $S_t$ with the lengths of $L_r$ and $L_t$, where $L_r \neq L_t$, we define
\begin{equation}\label{eq:P:D}
D = \ln\frac{max(L_r, L_t)}{min(L_r, L_t)} = |\ln L_r - \ln L_t|
\end{equation}
According to Equation~\ref{eq:jaccard}, the possible maximum similarity score of $(S_r, S_t)$ computed by Jaccard Index is
\begin{equation}\label{eq:P:J_M}
J_M(S_r, S_t) = \frac{min(L_r, L_t)}{max(L_r, L_t)},\text{ if }S_r \subset S_t\text{ or }S_r \supset S_t
\end{equation}
As presented at \mbox{Line 5} in Algorithm~\ref{algthm:prune_similarity}, when
\begin{equation}\label{eq:P:ineq}
\frac{min(L_r, L_t)}{max(L_r, L_t)} < \frac{1}{\mathcal{P}},
\end{equation}
the pair of comparison $(S_r, S_t)$ is pruned.
Combining Equation~\ref{eq:P:ineq} with Equation~\ref{eq:P:D} and~\ref{eq:P:J_M}, we have
\begin{equation}
J(S_r, S_t) \le \frac{min(L_r, L_t)}{max(L_r, L_t)} = \frac{1}{e^D} < \frac{1}{\mathcal{P}}
\end{equation}
Therefore, in this section, we aim to find the acceptable maximum Jaccard Index of similarity comparison, i.e,~the minimum value of $D$, to fulfill the pruning.

We randomly select \emph{1,000} reference functions, and leverage \binCMP\ to perform similarity comparison with the Jaccard Index, i.e,~$\mathcal{L}=+\infty$.
Since we merely consider at most Top~\emph{10} candidates of the results, we investigate the 10th similarity scores for all the target functions in the resulting lists.
Then, we find the average value of all the 10th similarity scores is \emph{0.450}, and the minimum value is \emph{0.094}.
As a result, we obtain the candidate values of $D$ denoted as $D_{avg}=\ln\frac{1}{0.450}=0.798$ and $D_{min}=\ln\frac{1}{0.094}=2.364$.

We further select another \emph{3,000} reference functions randomly to test $D_{avg}$ and $D_{min}$.
According to the ground truth, we find the corresponding matches of the reference functions in target executables.
Then, we compare the signature length of each reference function to that of its corresponding target function.
We set $D \in (0, 2.5]$ with the step of $0.1$, obtaining the accumulative ratio of function pairs versus $D$ as presented in Figure~\ref{fig:ratio_threshold}.
When $D=0.8$, 93.6\% of function pairs are correctly covered, and there exist \emph{193} samples which are incorrectly pruned.
We observe the following reasons leading to the incorrectness:
\begin{itemize}
	\item \textbf{Duplicated Functions.}
	Compilers might create several copies of a function for the resulting executable.
	They ensure the jumping distance from a caller to its callee is less than the memory page size~(e.g.,~\texttt{0x1000} bytes for x86), avoiding page faults when calling functions and improving execution efficiency.
	
	\binCMP\ collects function signatures from~(emulated) executions.
	The signature of the original function would be divided into parts for its copies.
	Then the signature length of a duplicated function might be much less than that of the original one.
	
	\item \textbf{Compiler-created Functions.}
	A Compiler would replace standard library functions with its own efficient ones during compilation.
	Typically, we find ICC generates binaries inlined with \texttt{\_\_intel\_fast\_*} functions, e.g,~\texttt{\_\_intel\_fast\_memcmp}.
	After stripping symbol names, it is difficult to distinguish those functions from user-defined ones.
	\binCMP\ then records their code features as well which enlarge the signature lengths of their callers.
	
	\item \textbf{Transformation between \texttt{switch} and \texttt{if} structures.}
	In some cases, the \texttt{switch} and \texttt{if} structures could be transformed between each other equivalently.
	It is also an optimization strategy of compilers.
	Comparing to a \texttt{switch}, an \texttt{if} structure contains more condition comparisons for branches, which corresponds to the code feature of Comparison Operand Values~(\S\ref{sec:approach:signature}), and generates longer signatures.
\end{itemize}
\begin{figure*}[t]
	\centering
	\subfloat[GCC -O3 vs. -O0\label{fig:cross_opt:gcc}]{\includegraphics[scale=.29]{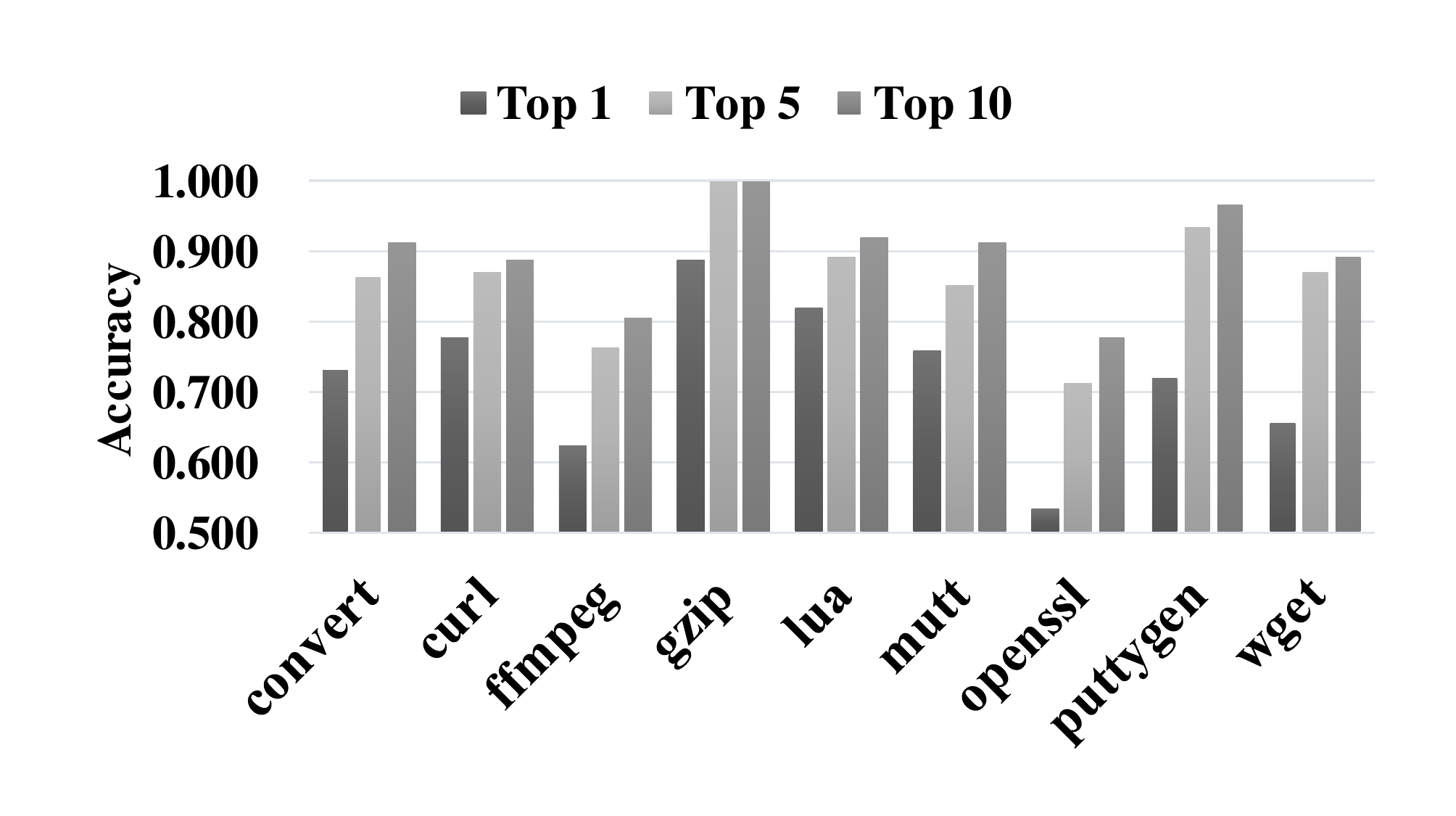}}
	\hfil
	\subfloat[Clang -O3 vs. -O0\label{fig:cross_opt:clang}]{\includegraphics[scale=.29]{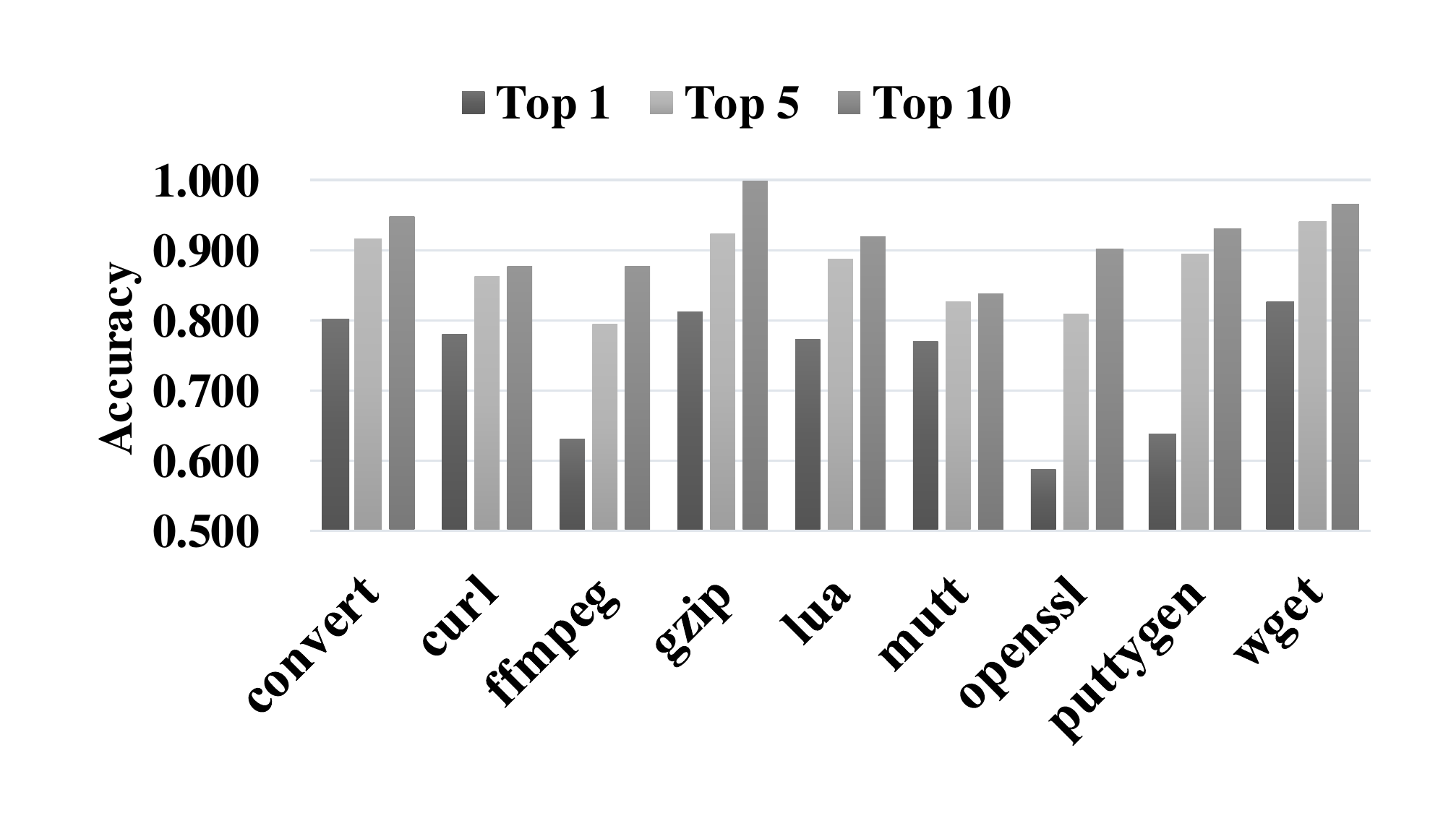}}
	\hfil
	\subfloat[ICC -O3 vs. -O0\label{fig:cross_opt:icc}]{\includegraphics[scale=.29]{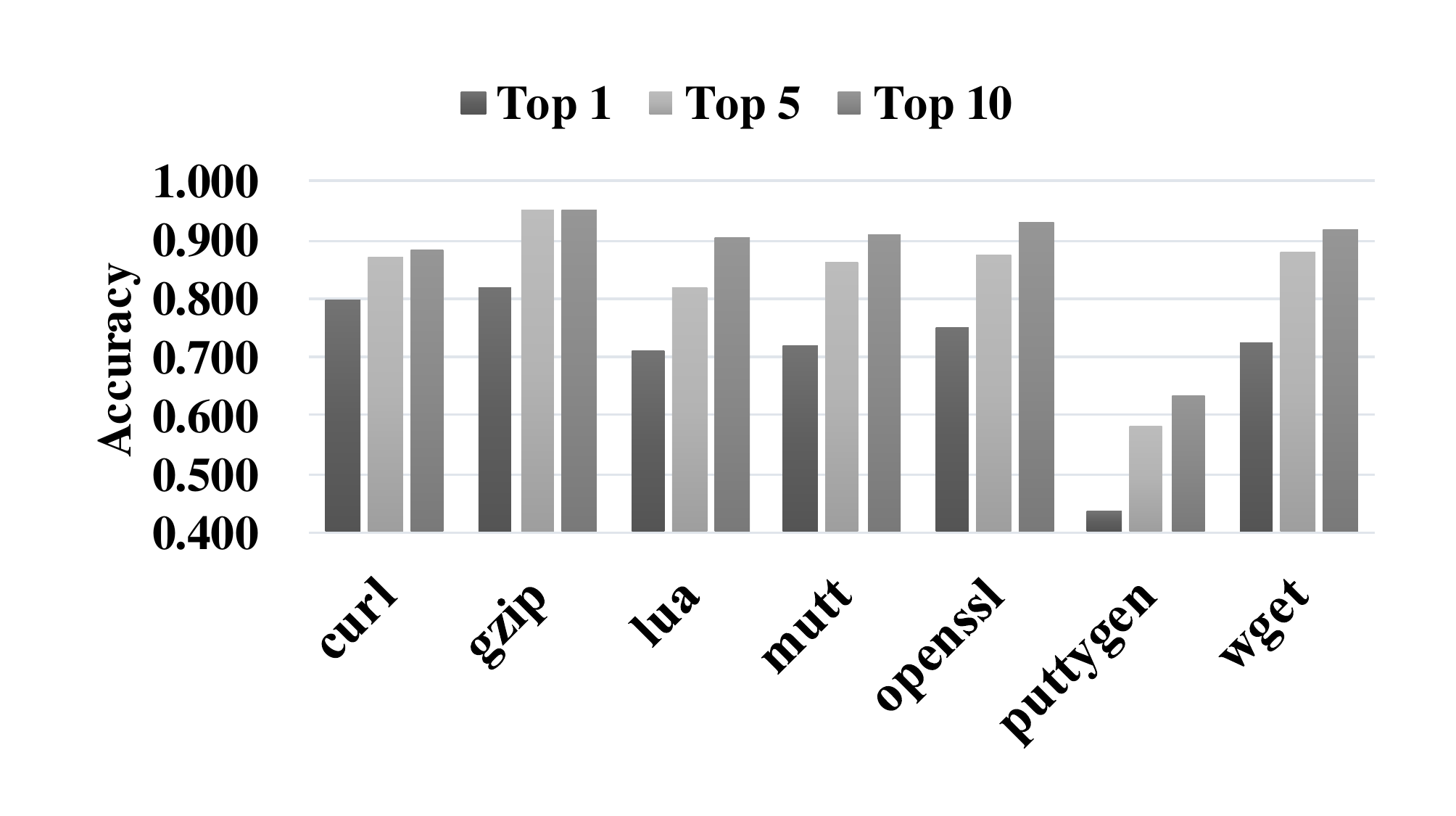}}
	\caption{Accuracy of Cross-optimization Comparison}
	\label{fig:cross_opt}
\end{figure*}
\begin{figure*}[t]
	\centering
	\subfloat[GCC -O3 vs. Clang -O0\label{fig:cross_com:gcc_clang}]{\includegraphics[scale=.29]{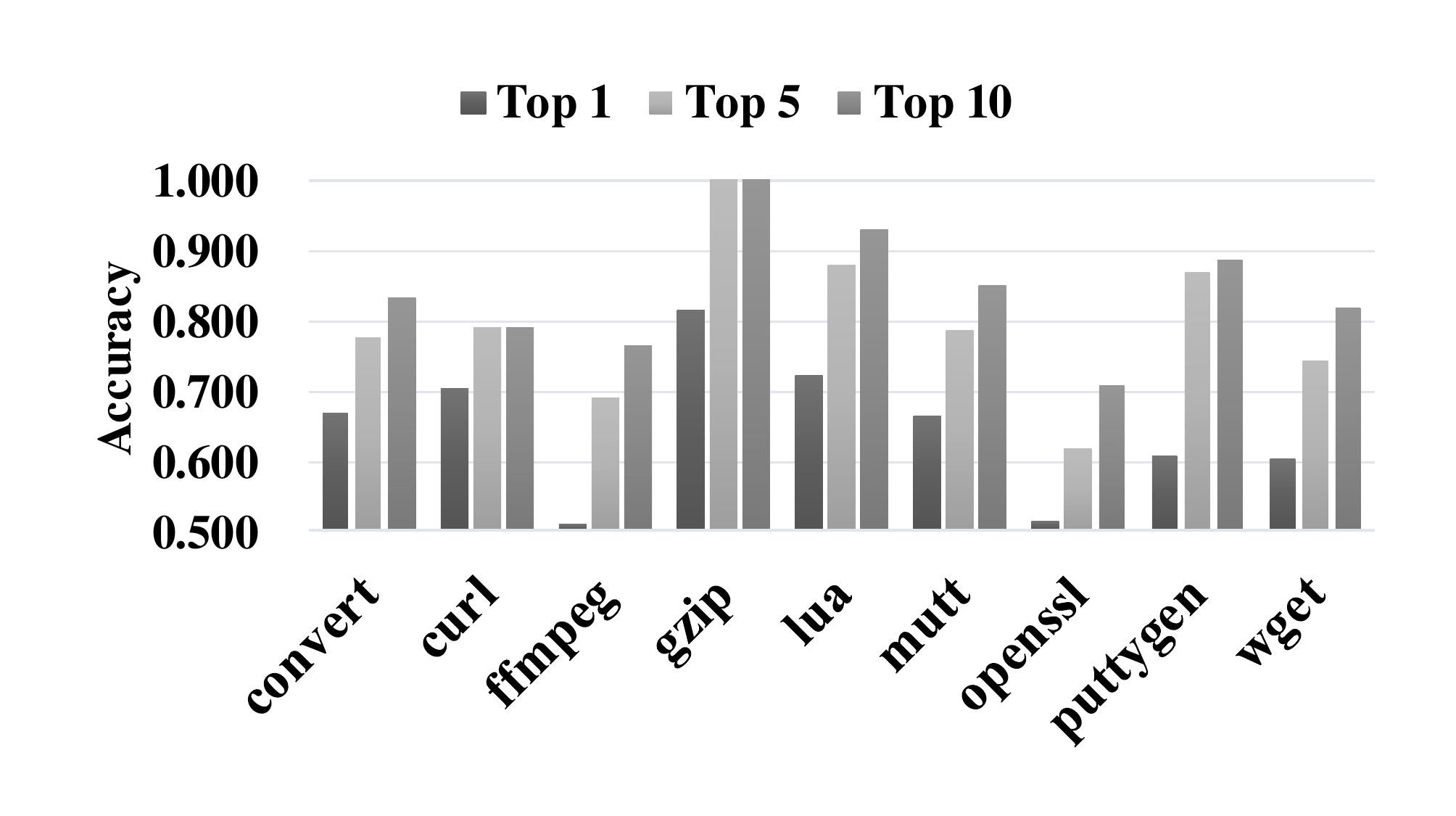}}
	\hfil
	\subfloat[Clang -O3 vs. GCC -O0\label{fig:cross_com:clang_gcc}]{\includegraphics[scale=.29]{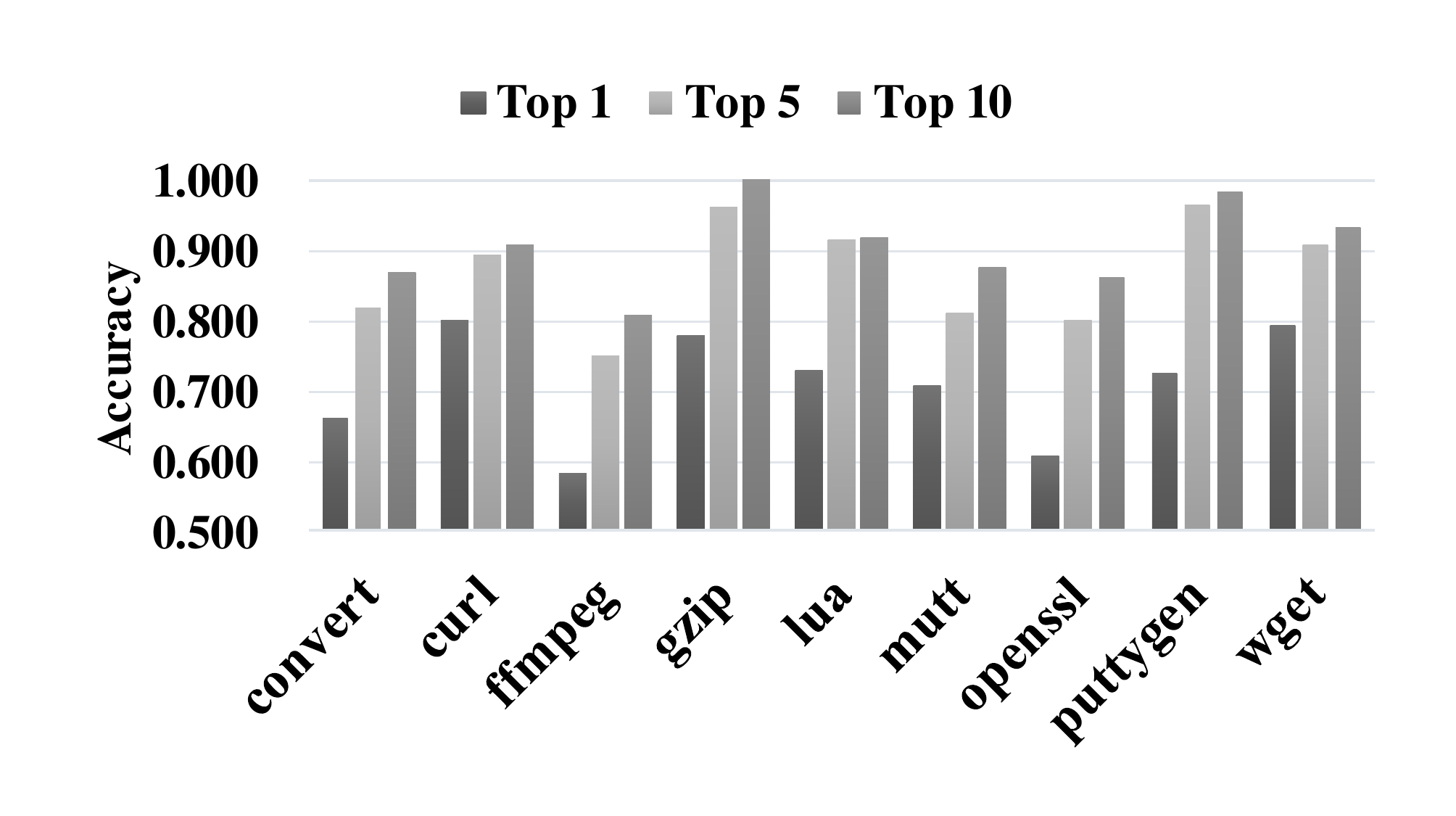}}
	\hfil
	\subfloat[ICC -O3 vs. GCC -O0\label{fig:cross_com:icc_gcc}]{\includegraphics[scale=.29]{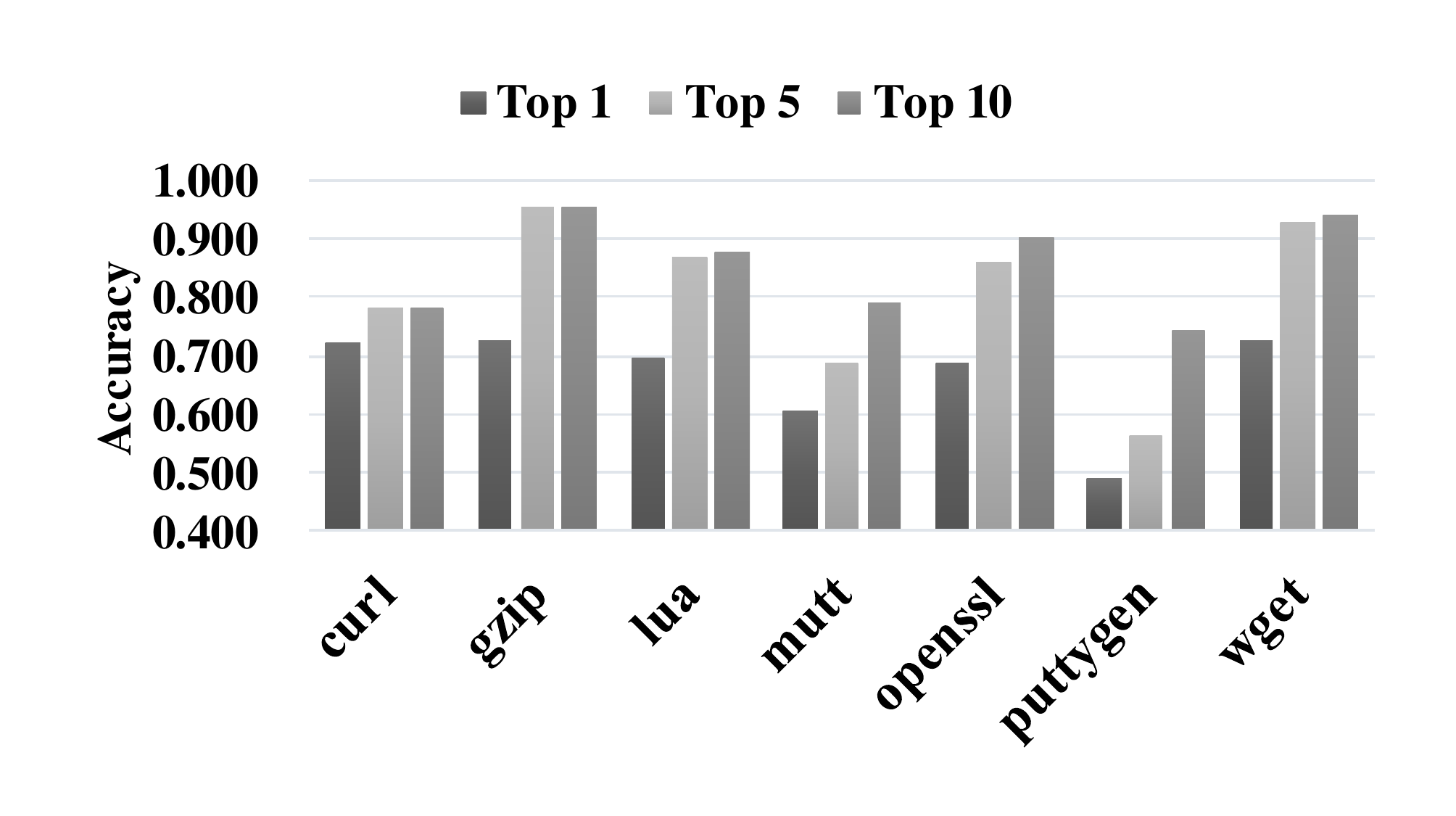}}
	\hfil
	\subfloat[GCC -O3 vs. ICC -O0\label{fig:cross_com:gcc_icc}]{\includegraphics[scale=.29]{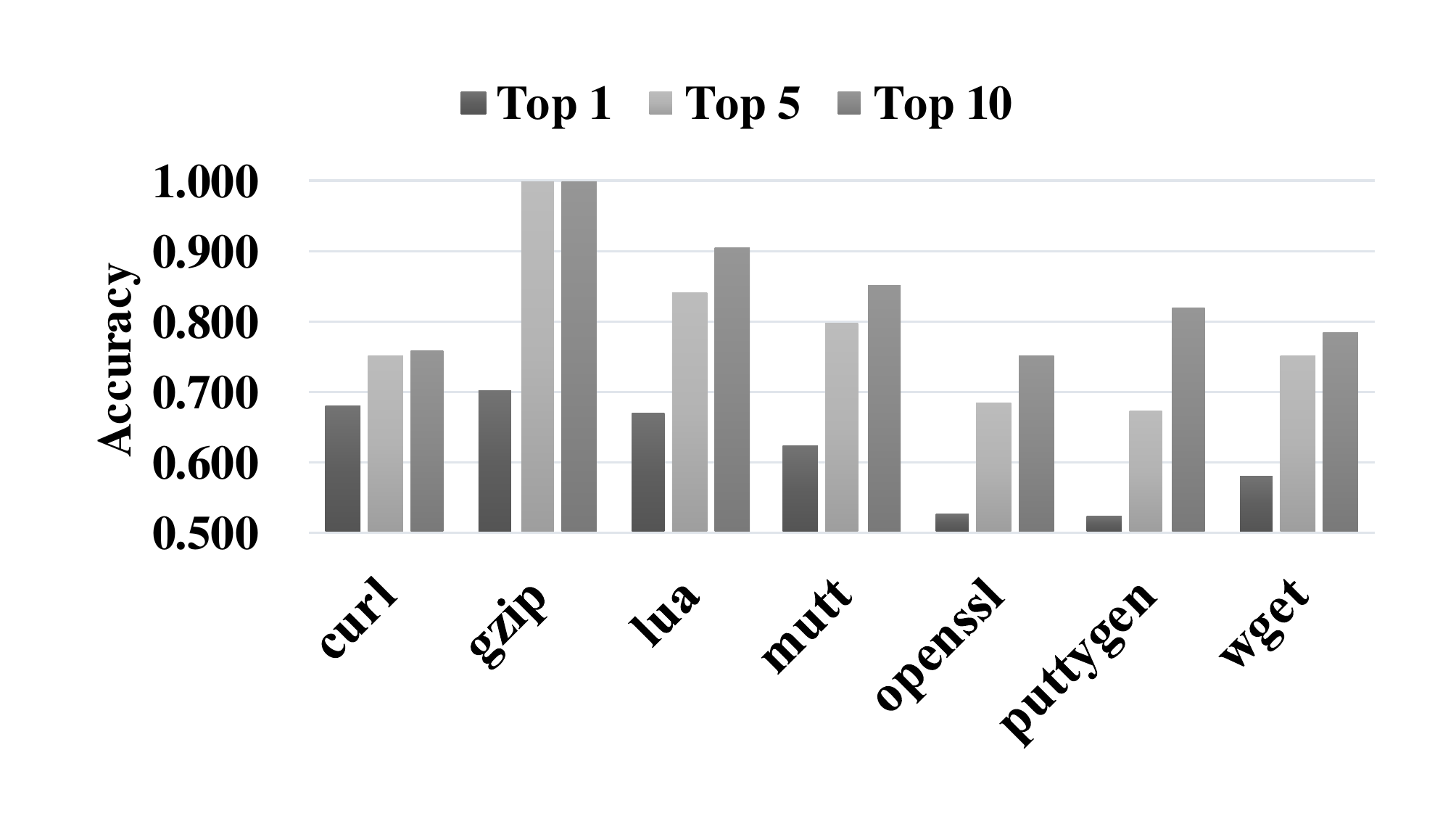}}
	\hfil
	\subfloat[Clang -O3 vs. ICC -O0\label{fig:cross_com:clang_icc}]{\includegraphics[scale=.29]{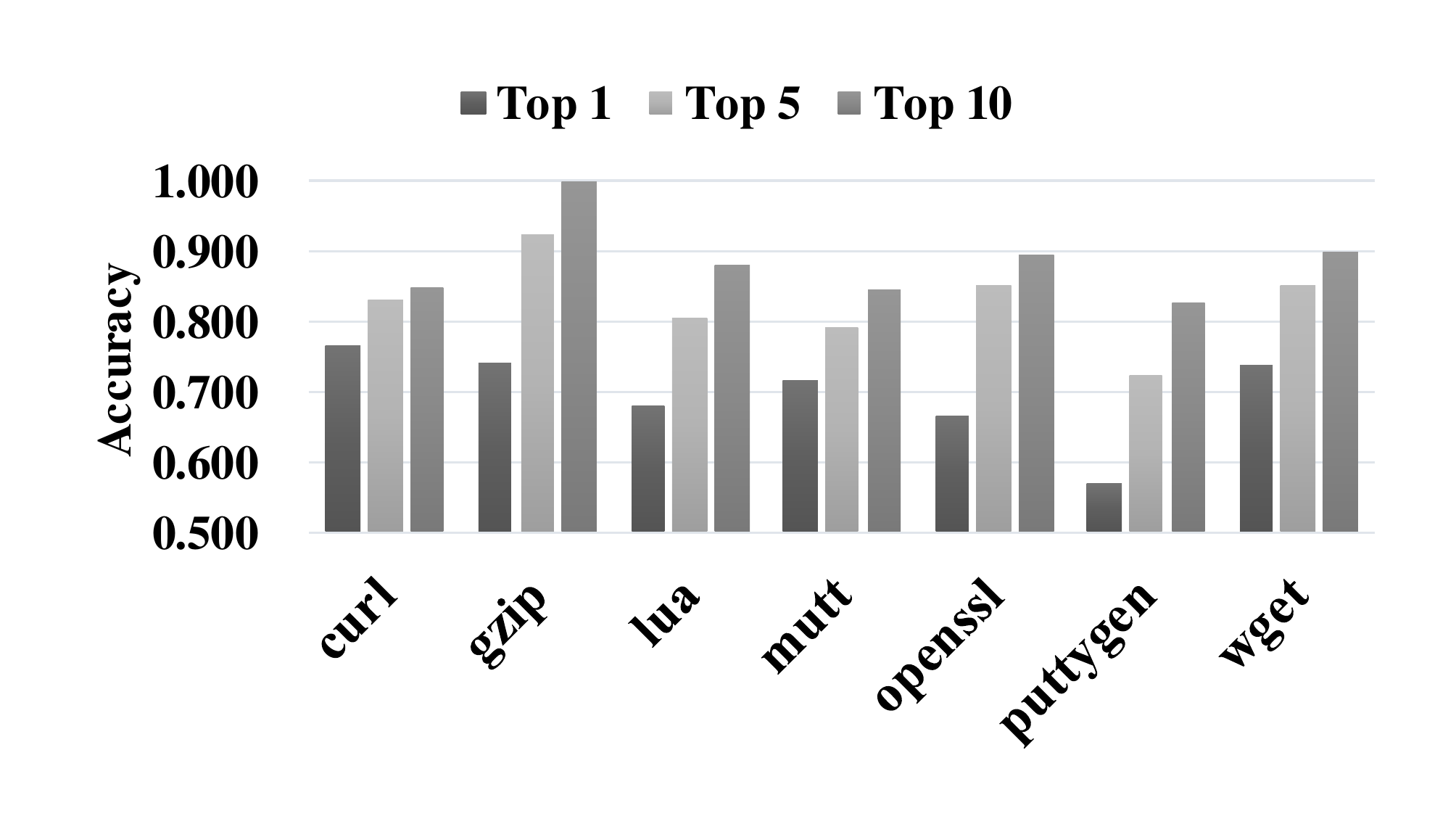}}
	\hfil
	\subfloat[ICC -O3 vs. Clang -O0\label{fig:cross_com:icc_clang}]{\includegraphics[scale=.29]{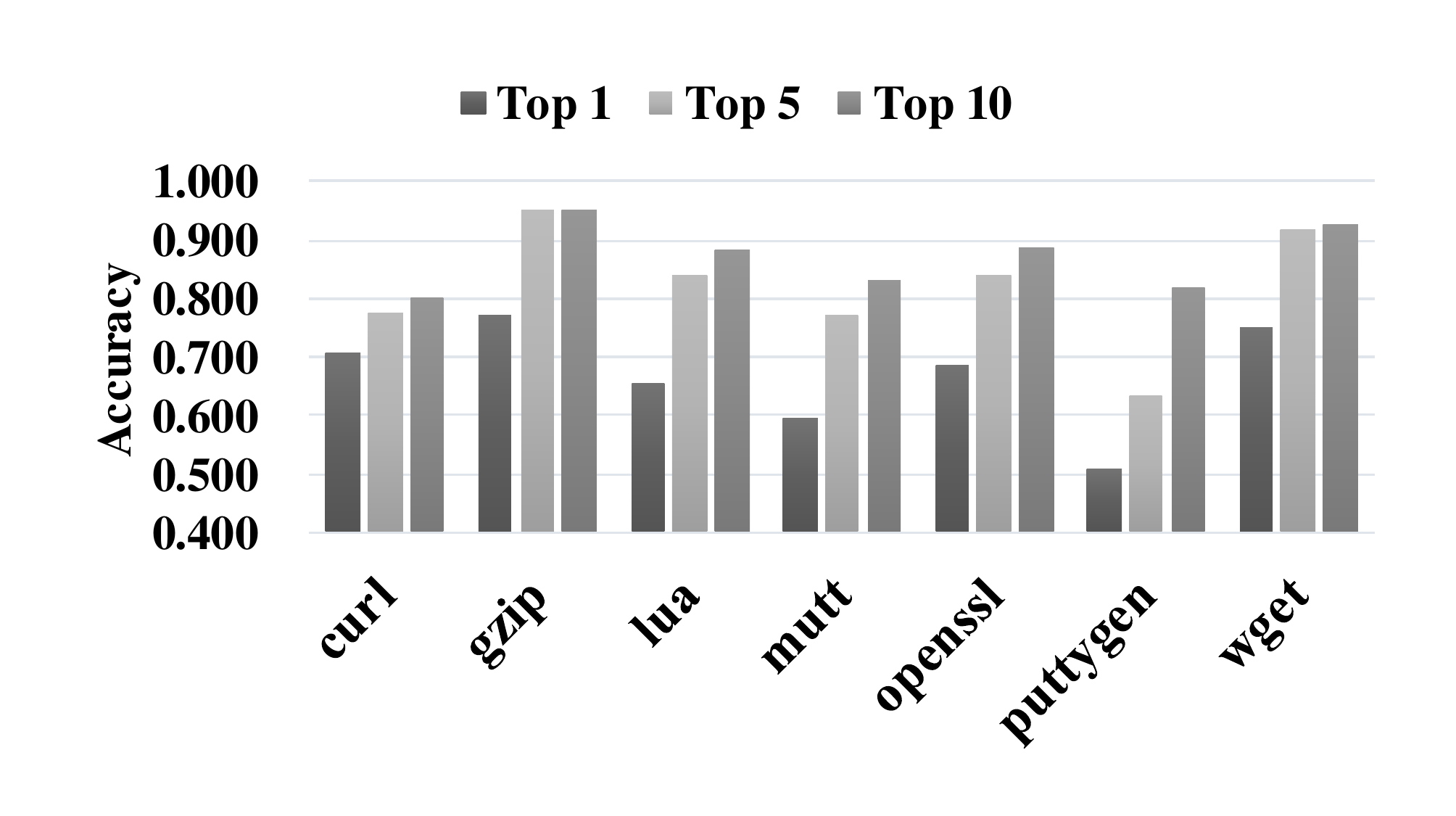}}
	\caption{Accuracy of Cross-compiler Comparison}
	\label{fig:cross_com}
\end{figure*}
When $D=2.4$, 98.8\% of samples are correctly handled.
We then use \binCMP\ to perform similarity comparison for the left \emph{37} samples~($\mathcal{L}=512$), and find that the differences between the reference and target signatures are so large that none of their scores could be ranked within Top~\emph{10} in the resulting lists.

Among the above reasons, \emph{duplicated functions} could be handled by combining their signatures.
Since they are exactly identical, it is possible to find all the duplicated functions with static analysis, such as binary function hashing.
Then, \binCMP\ records code features of duplicated functions as one function signature which could be considered as that of the original one.
However, it is difficult to decide whether a function is the compiler-created one in a stripped executable. It is also challenging to unify the representations of \texttt{switch} and \texttt{if} structures of binary programs.
Although we cannot perform pruning correctly for all cases, the above experiments indicate that the possibility is low to make the mistakes when $D=2.4$.
Therefore, in following experiments, \binCMP\ adopts
\begin{equation}
\mathcal{P} = e^{2.4} \approx 11.023.
\end{equation}

\subsection{Analysis across Compilation Settings}\label{sec:eva:cc}

\subsubsection{Cross-optimization Analysis}
In this section, we leverage \binCMP\ to match binary functions compiled with different optimizations.
For a compiler, higher optimization options contain all strategies specified by lower ones.
Taking \mbox{GCC v4.9.3} as an example\footnote{https://gcc.gnu.org/onlinedocs/gcc-4.9.3/gcc/Optimize-Options.html}, the option \texttt{-O3} enables all the \emph{68} optimizations of \texttt{-O2},
and turns on another \emph{9} optimization flags in addition.
\texttt{-O2} also covers all the \emph{32} strategies specified by \texttt{-O1}.
Thus, we only discuss the case of \texttt{-O3}~($E_{r}$) versus \texttt{-O0}~($E_{t}$), which has larger differences than any other pair of cross-optimization analysis.

Figure~\ref{fig:cross_opt} shows the accuracy of cross-optimization comparisons between each object program compiled by GCC, Clang, and ICC separately.
In Figure~\ref{fig:cross_opt:gcc}, the average accuracy of Top~\emph{1},~\emph{5}, and~\emph{10} is \emph{68.9\%}, \emph{82.5\%}, and \emph{87.0\%}.
The performance of \binCMP\ increases notably regrading the Top~\emph{5} and~\emph{10} target functions in the resulting lists.
Thus, the possibility is high for analyzers to find the real match by further considering the first five or ten candidates in a target function list.

\begin{table*}[t]
	\renewcommand{\arraystretch}{1.3}
	\captionsetup{width=.75\textwidth}
	\centering
	\caption{Comparison with the state-of-the-art methods \texttt{Asm2Vec}, \texttt{Kam1n0}, and the industrial tool \texttt{BinDiff}. @\emph{K} represents the Top~\emph{K} accuracy.}
	\label{tab:tool_cmp}
	\begin{tabular}{|c|c|c|c|c|c|c|c|c|c|}
		\hline
		\multirow{2}{*}{\tabincell{c}{Reference}} & \multirow{2}{*}{\tabincell{c}{Target}} & \multicolumn{3}{c|}{\textbf{\binCMP}} & \multicolumn{3}{c|}{Asm2Vec} & \multirow{2}{*}{\tabincell{c}{Kam1n0}} & \multirow{2}{*}{\tabincell{c}{BinDiff}} \\
		\cline{3-8}
		& & @1 & @5 & @10 & @1 & @5 & @10 & & \\
		\hline
		\hline
		\multirow{3}{*}{\tabincell{c}{GCC -O3}} & GCC -O0 & 0.689 & 0.825 & 0.870 & 0.444 & 0.623 & 0.674 & 0.288 & 0.338 \\
		& Clang -O0 & 0.614 & 0.748 & 0.808 & 0.417 & 0.580 & 0.629 & 0.212 & 0.273 \\
		& ICC -O0 & 0.603 & 0.756 & 0.811 & 0.370 & 0.553 & 0.619 & 0.209 & 0.277 \\
		\hline
		\multirow{3}{*}{\tabincell{c}{Clang -O3}} & Clang -O0 & 0.718 & 0.858 & 0.909 & 0.425 & 0.567 & 0.620 & 0.251 & 0.461 \\
		& GCC -O0 & 0.667 & 0.826 & 0.871 & 0.412 & 0.577 & 0.627 & 0.271 & 0.457 \\
		& ICC -O0 & 0.696 & 0.825 & 0.877 & 0.386 & 0.622 & 0.694 & 0.224 & 0.400 \\
		\hline
		\multirow{3}{*}{\tabincell{c}{ICC -O3}} & ICC -O0 & 0.726 & 0.846 & 0.895 & 0.323 & 0.546 & 0.627 & 0.276 & 0.332 \\
		& GCC -O0 & 0.666 & 0.816 & 0.865 & 0.343 & 0.399 & 0.465 & 0.182 & 0.219 \\
		& Clang -O0 & 0.673 & 0.808 & 0.853 & 0.315 & 0.533 & 0.628 & 0.191 & 0.212 \\
		\hline
		\hline
		\multicolumn{2}{|c|}{Average Accuracy} & 0.672 & 0.813 & 0.863 & 0.395 & 0.574 & 0.635 & 0.240 & 0.328 \\
		\hline
		\multicolumn{2}{|c|}{Time (s) / Function} & \multicolumn{3}{c|}{4.151} & \multicolumn{3}{c|}{1.255} & 1.332 & 0.210 \\
		\hline
	\end{tabular}
	
\end{table*}

The results of Clang- and ICC-compiled programs are similar.
The average accuracy of Top~\emph{1},~\emph{5},~\emph{10} is \emph{71.8\%}, \emph{85.8\%}, \emph{90.9\%} in Figure~\ref{fig:cross_opt:clang}, and \emph{72.6\%}, \emph{84.6\%}, \emph{89.5\%} in Figure~\ref{fig:cross_opt:icc}.
Since ICC fails to generate executables for \texttt{convert} and \texttt{ffmpeg} with the corresponding compilation settings, we only conduct experiments with the left seven object programs for ICC.
In Figure~\ref{fig:cross_opt:icc}, the results of \texttt{puttygen} are much worse than those of other object programs.
The Top~\emph{1} accuracy of \texttt{puttygen} is \emph{43.6\%}, while that of every other object exceeds \emph{70.0\%}.
When generating the executable of \texttt{puttygen} optimized with -O3, ICC inlines its own library functions to replace the standard ones, while such optimization is not applied to the -O0 version.
The test command of \texttt{puttygen}~(as presented in Table~\ref{tab:obj}) triggers the reference functions which frequently invoke those inlined by ICC, leading to huge differences in signatures for comparison.
\binCMP\ then produces the relative low accuracy.

\subsubsection{Cross-compiler Analysis}\label{sec:eva:cc:compiler}

In this section, \binCMP\ is evaluated with binaries compiled by different compilers.
Similar to the cross-optimization analysis, only the case of \texttt{-O3}~($E_r$) versus \texttt{-O0}~($E_t$) is considered.
The results are presented in Figure~\ref{fig:cross_com}.
\binCMP\ performs well in most cases.
The average Top~\emph{1},~\emph{5} and~\emph{10} accuracy of all experiments is \emph{65.0\%}, \emph{79.4\%}, and \emph{84.6\%} separately.

\emph{Compiler-created functions} of ICC still constitute the reason that affects the performance of \binCMP.
Specifically, the average Top~\emph{1} accuracy displayed in Figure~\ref{fig:cross_com:icc_gcc} and~\ref{fig:cross_com:icc_clang} is \emph{67.0\%}, while that of \mbox{ICC -O3 vs. -O0} in Figure~\ref{fig:cross_opt:icc} is \emph{72.6\%}.
Additionally, we find \emph{floating-point number} is another reason decreasing the accuracy.
GCC leverages x87 floating-point instructions to implement corresponding operations, while Clang and ICC uses the SSE~(Streaming SIMD Extensions) instruction set.
x87 adopts the FPU~(floating point unit) stack to assist in processing floating-point numbers. The operations deciding whether the stack is full or empty insert redundant entries to the semantic signature with comparison operand values~(\S\ref{sec:approach:signature}).
In contrast, SSE directly operates on a specific register set~(i.e.,~XMM registers) and has no extra operations.
Besides, x87 could handle single precision, double precision, and even 80-bit double-extended precision floating-point calculation, while SSE mainly processes single-precision data.
Due to the different precision of representations,
even though the floating-point numbers are the same, their values generated by different instruction sets are not equal, therefore affecting the accuracy.
As a result, when processing executables compiled by GCC~(\mbox{Figure~\ref{fig:cross_com:gcc_clang}-\ref{fig:cross_com:gcc_icc}}), the average Top~\emph{1} accuracy is \emph{63.9\%}.
In contrast, when the comparisons are performed between Clang and ICC~(Figure~\ref{fig:cross_com:clang_icc} and~\ref{fig:cross_com:icc_clang}), the corresponding accuracy is \emph{68.5\%}.

\subsubsection{Comparison with Existing Work}\label{sec:eva:cc:comp}

In this section, we compare \binCMP\ to the state-of-the-art methods Asm2Vec~\cite{ding2019asm2vec}, Kam1n0~\cite{ding2016kam1n0}, and the industrial tool BinDiff~\cite{flake2004structural} supported by Google, which are all open for public use.
Thus, we could use them to detect similar binary functions with the same settings as \binCMP.
Since \binCMP\ is evaluated with the executed reference functions, to make fair comparison, we investigate the performance of the three solutions with those reference functions as well.
We configure the three solutions with their default settings, and the results are displayed in Table~\ref{tab:tool_cmp}.
Similar to \binCMP, Asm2Vec returns a list of target functions for each reference function. Thus, we also present its accuracy of Top~\emph{1},~\emph{5},~\emph{10} respectively.
The last two rows show the average accuracy for all the experiments and the processing time of each function on average.
Obviously, \binCMP\ performs much better than other three from the perspective of accuracy.
Specifically, on average, its Top~\emph{1} accuracy even outperforms the Top~\emph{10} accuracy of Asm2Vec. 
Besides, benefiting from the adoption of SimHash and the pruning strategy, its average processing time of each function is around \emph{4} seconds.
Although \binCMP\ is still slower than the other three, considering the accuracy, it deserves the time.

Asm2Vec adopts machine learning techniques for binary function similarity comparison.
It treats each path of a function as a document, and leverage the PV-DM model~\cite{le2014distributed} to encode the function into a feature vector.
Asm2Vec explores the co-occurrence relationships among assembly code tokens, aiming to describe a binary function with the most representative~(or the unique) instructions.
However, binary code might be implemented with semantics-equivalent but different kinds of instructions, especially when the code is generated with variant compilation settings. 

Kam1n0 and BinDiff are typical solutions which rely on syntax and structure features to detect binary similar functions.
Kam1n0 captures features of a function from its control flow graph~(CFG), and encodes the features as a vector for indexing. Thus, essentially, it detects similar functions by analyzing graph isomorphism of CFG.
The relatively low accuracy of Kam1n0 indicates that compilation settings indeed affect representations of binaries, even though two pieces of code are compiled from the same code base.
In addition to measuring the similarity of CFG, BinDiff considers other features to compare similar functions, such as function hashing which compares the hash values of raw function bytes, call graph edges which matches functions basing on the dependencies in the call graphs,~etc. 
By carefully choosing suitable features to measure the similarity of functions, BinDiff becomes resilient towards code transformation resulting from different compilers or optimization options to an extent.
Therefore, it performs better than Kam1n0, but is still at a disadvantage comparing to \binCMP.

\subsection{Analysis on Obfuscated Code}\label{sec:eva:obf}

\begin{table*}[t]
	\renewcommand{\arraystretch}{1.3}
	\centering
	\caption{Accuracy of comparing with obfuscated code. The target executables are obfuscated with OLLVM~(BCF: Bogus Control Flow, FLA: Control Flow Flattening, SUB: Instructions Substitution). @\emph{K} represents the Top~\emph{K} accuracy.}
	\label{tab:obf}
	\begin{tabular}{|c|c|c|c|c|c|c|c|c|c|}
		\hline
		\multirow{2}{*}{\tabincell{c}{Reference}} & \multirow{2}{*}{\tabincell{c}{Target}} & \multirow{2}{*}{\tabincell{c}{Obf.}} & \multicolumn{3}{c|}{\textbf{\binCMP}} & \multicolumn{3}{c|}{Asm2Vec} & \multirow{2}{*}{BinDiff} \\
		\cline{4-9}
		& & & @1 & @5 & @10 & @1 & @5 & @10 & \\
		\hline
		\hline
		\multirow{6}{*}{\tabincell{c}{GCC -O3}} & \multirow{3}{*}{\tabincell{c}{OLLVM\\-O3}} & SUB & 0.755 & 0.873 & 0.912 & 0.530 & 0.710 & 0.760  & 0.678 \\
		& & BCF & 0.615 & 0.722 & 0.773 & 0.509 & 0.676 & 0.718 & 0.311 \\
		& & FLA & 0.521 & 0.629 & 0.705 & 0.358 & 0.529 & 0.565 & 0.430 \\
		\cline{2-10}
		& \multirow{3}{*}{\tabincell{c}{OLLVM\\-O0}} & SUB & 0.692 & 0.821 & 0.859 & 0.336 & 0.489 & 0.568 & 0.385 \\
		& & BCF & 0.504 & 0.561 & 0.580 & 0.302 & 0.473 & 0.535 & 0.211 \\
		& & FLA & 0.452 & 0.495 & 0.551 & 0.166 & 0.280 & 0.323 & 0.312 \\
		\hline
		\multirow{6}{*}{\tabincell{c}{Clang -O3}} & \multirow{3}{*}{\tabincell{c}{OLLVM\\-O3}} & SUB & 0.890 & 0.977 & 0.985 & 0.766 & 0.874 & 0.897 & 0.805 \\
		& & BCF & 0.647 & 0.752 & 0.842 & 0.624 & 0.777 & 0.814 & 0.334 \\
		& & FLA & 0.560 & 0.616 & 0.682 & 0.470 & 0.633 & 0.680 & 0.541 \\
		\cline{2-10}
		& \multirow{3}{*}{\tabincell{c}{OLLVM\\-O0}} & SUB & 0.758 & 0.901 & 0.942 & 0.351 & 0.525 & 0.590 & 0.497 \\
		& & BCF & 0.532 & 0.592 & 0.616 & 0.356 & 0.524 & 0.567 & 0.053 \\
		& & FLA & 0.485 & 0.546 & 0.611 & 0.241 & 0.398 & 0.441 & 0.280 \\
		\hline
		\multirow{6}{*}{\tabincell{c}{ICC -O3}} & \multirow{3}{*}{\tabincell{c}{OLLVM\\-O3}} & SUB & 0.621 & 0.750 & 0.805 & 0.482 & 0.632 & 0.691 & 0.609 \\
		& & BCF & 0.718 & 0.848 & 0.896 & 0.398 & 0.558 & 0.622 & 0.208 \\
		& & FLA & 0.473 & 0.523 & 0.609 & 0.292 & 0.414 & 0.469 & 0.272 \\
		\cline{2-10}
		& \multirow{3}{*}{\tabincell{c}{OLLVM\\-O0}} & SUB & 0.730 & 0.847 & 0.876 & 0.264 & 0.409 & 0.482 & 0.248 \\
		& & BCF & 0.501 & 0.570 & 0.605 & 0.254 & 0.368 & 0.423 & 0.095 \\
		& & FLA & 0.507 & 0.556 & 0.631 & 0.148 & 0.239 & 0.296 & 0.160 \\
		\hline
		\hline
		\multicolumn{3}{|c|}{Average Accuracy} & 0.638 & 0.736 & 0.784 & 0.410 & 0.560 & 0.613 & 0.374 \\
		\hline
	\end{tabular}
\end{table*}

In this section, we conduct experiments to compare normal binary programs~($E_r$) with their corresponding obfuscated code~($E_t$).
We adopt OLLVM to obfuscate binary code which is optimized with \mbox{-O3} and \mbox{-O0} separately~(OLLVM bases the compilation on Clang).
Because obfuscation would insert much redundant code, resulting in huge length differences of signatures for comparison, we disable the pruning strategy in this part of experiments, i.e.,~$\mathcal{P}=+\infty$.

The experimental results are shown in Table~\ref{tab:obf}.
Results of Asm2Vec and BinDiff are also presented as references.
OLLVM provides three techniques to fulfill the obfuscation.
\emph{Instruction substitution}~(SUB) replaces standard operators~(e.g.,~addition operators) with sequences of functionality-equivalent, but more complex instructions.
It obfuscates code on the \emph{syntax} level, affecting the comparison accuracy of Asm2Vec which treats binaries as documents, but posing fewer threats to \binCMP\ which is semantics-based.

\emph{Bogus control flow}~(BCF) adds opaque predicates to a basic block, which breaks the original basic block into two.
\emph{Control flow flattening}~(FLA) generally breaks a function up into basic blocks, then encapsulates the blocks with a selective structure~(e.g.,~the switch structure)~\cite{laszlo2009obfuscating}.
It creates a state variable for the selective structure to decide which block to execute next at runtime via conditional comparisons.
BCF and FLA both change the \emph{structure} of the original function, i.e.,~modifying the control flow.
They insert extra code which is irrelevant to the functionality of the original function, generating redundant semantic features which are indistinguishable from normal ones~(e.g.,~comparison operand values of opaque predicates).
Thus, they affect the comparison accuracy of \binCMP.
For the settings of \mbox{GCC/Clang/ICC -O3 vs. OLLVM -O0}, 
when comparing with functions obfuscated by BCF, the average Top~\emph{1} accuracy is \emph{51.3\%},
and \emph{48.1\%} for FLA, while that of \mbox{GCC/Clang/ICC -O3 vs. Clang -O0} is \emph{66.8\%}.
However, 
\binCMP\ still achieves more than \emph{1.5} times the average Top~\emph{1} accuracy of Asm2Vec, and \emph{1.7} times of BinDiff,
i.e.,~\emph{63.8\%} of \binCMP, \emph{41.0\%} of Asm2Vec, and \emph{37.4\%} of BinDiff.
On average, the Top~\emph{1} accuracy of \binCMP\ still outperforms the Top~\emph{10} accuracy of Asm2Vec.

Additionally, because the pruning strategy is disabled, 
\binCMP\ spends \emph{21.678} seconds on average to process each reference function,
while that of comparisons with pruning is \emph{4.151} seconds.
The results indicate the importance of the pruning strategy for improving the efficiency.

\subsection{Analysis across Architectures}\label{sec:eva:arch}

\begin{table}[t]
	\renewcommand{\arraystretch}{1.3}
	\centering
	\caption{Performance of cross-architecture comparison. All executables are compiled with GCC -O3, but with different instruction set architectures. @\emph{K} represents the Top~\emph{K} accuracy.}
	\label{tab:cross_arch}
	\begin{tabular}{|c|c|c|c|c|}
		\hline
		\multirow{2}{*}{\tabincell{c}{Settings}} & \multicolumn{3}{c|}{\textbf{\binCMP}} & \multirow{2}{*}{CACompare} \\
		\cline{2-4}
		& @1 & @5 & @10 & \\
		\hline
		\hline
		x86 vs. ARM & 0.628 & 0.743 & 0.798 & 0.807 \\
		\hline
		x86 vs. MIPS & 0.721 & 0.827 & 0.866 & 0.770 \\
		\hline
		ARM vs. MIPS & 0.703 & 0.816 & 0.867 & 0.810 \\
		\hline
		\hline
		Average Accuracy & 0.667 & 0.789 & 0.839 & 0.795 \\
		\hline
		Time (s) / Function & \multicolumn{3}{c|}{3.451} & 4.694 \\
		\hline
	\end{tabular}
\end{table}

In this section, we evaluate the capacity of \binCMP\ to compare the similarity of binary functions of variant ISAs, across x86, ARM and MIPS.
All the executables for comparison are compiled with \mbox{GCC -O3}.
The ARM and MIPS binaries are generated or executed in the environments emulated by \texttt{QEMU}~\cite{bellard2005qemu}.

The experimental results are presented in Table~\ref{tab:cross_arch}, which are compared to those of CACompare~\cite{hu2017binary}, the state-of-the-art cross-architecture similar binary function detector.
The average Top~\emph{5} accuracy of \binCMP\ is comparable to that of CACompare.
Besides, for each reference function, \binCMP\ is 1.2 seconds faster than CACompare on average.
Since there are \emph{3,078} reference functions, \binCMP\ then saves about \emph{1} hour for the experiments.
In fact, when all the similarity comparisons are performed with Jaccard Index and LCS, i.e.,~$\mathcal{L}=+\infty$, \binCMP\ is able to achieve the average Top~\emph{1} accuracy of \emph{86.2\%} which indicates the upper bound capacity of \binCMP.
Nevertheless, it needs to spend \emph{137.760} seconds on average in processing each reference function.
Thus, \binCMP\ has the ability to become more accurate than CACompare, but it also requires more time.
According to the scenarios and requirements, users could choose the suitable $\mathcal{L}$ for it to balance accuracy and efficiency.

CACompare samples a function with random values as inputs, and extracts the code features via emulation as well.
Illegal memory accessing is also tackled by providing random values.
However, the random values lack semantics.
They could hardly bypass the input checks of a function, and usually trigger paths which handle exceptions.
In contrast, 
\binCMP\ captures the signatures of target functions by emulating them with runtime values migrated from real executions.
As a result, in some cases, \binCMP\ is more robust than \mbox{CACompare}.
Besides, it could generate results with higher accuracy if there is no strict time limit.

\subsection{Threats to Validity}
We construct the dataset of the experiments by compiling binaries from nine open-source projects~(\S\ref{sec:eva:dataset}).
Besides, we conduct experiments to infer suitable parameter values for \binCMP~(\S\ref{sec:eva:setup:parameter}).
Although the dataset consists of various types of programs, it cannot cover all cases in the real world, neither can the corresponding parameter values of $\mathcal{L}$ and $\mathcal{P}$.

\binCMP\ adopts \mbox{\texttt{IDA Pro}} to acquire information of binary functions~(\S\ref{sec:implementation:disassembling}).
However, function boundary identification of \mbox{\texttt{IDA Pro}} is not perfect, which is actually still an issue of reverse engineering~\cite{bao2014byte,shin2015recognizing,andriesse2016depth}.
Additionally, \binCMP\ is implemented with \texttt{Valgrind} and \texttt{angr} which both adopt VEX-IR as the intermediate representation~(\S\ref{sec:implementation:instr_emult}).
However, VEX-IR is not perfect that 16\% x86 instructions could not be lifted, although only a small subset of instructions is used in executables in practice and VEX-IR could handle most cases~\cite{kim2017testing}.
The incompleteness of \mbox{VEX-IR} might affect the accuracy of semantics signature extraction, while \binCMP\ still produces promising results in above experiments.

\section{Discussion and Future Work}\label{sec:discussion}

\subsection{Application Scope and Scenarios}
In this paper, \binCMP\ is implemented to process 32-bit code.
The solution could be applied to 64-bit code as well.
To fulfill the target, there might exist the following problems:
\begin{itemize}
	\item \emph{Calling Conventions:}
	\binCMP\ identifies and assigns the arguments of a binary function according to its calling convention~(\S\ref{sec:approach:execution} and \S\ref{sec:approach:emulation:argument}).
	64-bit instruction set architectures commonly prepare arguments with specific registers, e.g.,~RDI, RSI, RDX, RCX, R8, R9 of x86-64 on Linux, and additional ones are passed via the stack.
	Thus, we need to consider those specific registers firstly, then analyze the stack if necessary.
	\item \emph{Floating-point Numbers:}
	Different instruction set architectures employ instructions of various precision to process floating-point numbers, such as x87 and SSE of x86, SSE2 of x86-64. That would affect the detection accuracy of \binCMP~(\S\ref{sec:eva:cc:compiler}).
	A possible solution is to unify the precision of floating-point values, representing the high-precision value with lower precision, e.g.,~representing double-precision values with single-precision.
	That would be left as future work.
\end{itemize}

Because \binCMP\ needs to execute the reference functions, it is more suitable for scenarios where the reference functions have available test cases, such as known vulnerability detection~(as shown in \S\ref{sec:example:movtivating}), patch analysis~\cite{zhang2018precise,duan2019automating}.
In contrast, static methods are applicable to cases which require the high coverage of the reference code, such as plagiarism detection~\cite{jhi2011value}.
Dynamic methods are appropriate for the situation where the code behaviors are emphasized or the capacity of deobfuscation is required, such as malware lineage analysis~\cite{ming2017binsim}.

\subsection{Obfuscation}
In the evaluation, \binCMP\ is shown to be effective in analyzing obfuscated binary code which is generated by \texttt{OLLVM}~(\S\ref{sec:eva:obf}).
The robust of \binCMP\ is due to the nature of dynamic analysis and the adoption of semantics-based signatures.
However, that does not mean \binCMP\ could handle all kinds of obfuscations.
Besides, the \texttt{OLLVM} code actually affects the accuracy of \binCMP\ in the experiments.
When analyzing benign code, \binCMP\ achieves better results.
For \mbox{GCC/Clang/ICC -O3 vs. Clang -O0}, the average Top~\emph{1},~\emph{5}, and~\emph{10}  accuracy is \emph{66.8\%}, \emph{80.4\%} and \emph{85.8\%}, while for \mbox{GCC/Clang/ICC -O3 vs. OLLVM -O0}, the corresponding ratio is \emph{60.9\%}, \emph{70.1\%}, and \emph{74.0\%} respectively.
In the literature, deobfuscation has been well studied~\cite{udupa2005deobfuscation,yadegari2015symbolic,yadegari2015generic,xu2018vmhunt}. Therefore, if \binCMP\ fails to detect an obfuscated function, it is a better choice to deobfuscate it firstly, then perform further analysis.

\subsection{Function Interfaces}
In this paper, we assume a pair of matched functions shares the same interface, i.e.,~the same argument number and order.
When the interface of the target function is modified, e.g.,~by obfuscation, \binCMP\ becomes ineffective.
For example, the reference function \emph{R} has the interface
\begin{center}
	\ttfamily
	\emph{R}(rarg\_0, rarg\_1, rarg\_2),
\end{center}
while the interface of its corresponding match~(target function \emph{T}) is
\begin{center}
	\ttfamily
	\emph{T}(targ\_2, targ\_1, targ\_0, targ\_3).
\end{center}
Note that not only a redundant argument \texttt{targ\_3} is added to \emph{T} via obfuscation, but also the first three arguments are disordered.
For \texttt{targ\_3}, as described in the previous section, extra analysis is necessary, such as deobfucation. That is out of the scope of this paper.
For the disordering, a possible solution is to provide \emph{T} with the permutation of \emph{R}'s argument list.
In the example, after \texttt{targ\_3} is removed, \binCMP\ generates the permutation of \emph{R}'s argument list, overall \emph{6}~($=P_3^3$) cases, then assigns them to \emph{T} and computes the similarity score separately.
The largest one among the six values, theoretically when the order is \mbox{(\texttt{rarg\_2}, \texttt{rarg\_1}, \texttt{rarg\_0})}, is considered as the final similarity score of \mbox{(\emph{R}, \emph{T})}.
It is left as future work.

\subsection{Accuracy vs. Efficiency}
It is a classical issue of program analysis.
In this paper, since the lengths of signatures extracted via~(emulated) executions are huge, we propose the hybrid method, combining LCS and SimHash for similarity comparison, to reach the compromise between efficiency and accuracy~(\S\ref{sec:approach:similarity}).
To improve the accuracy, it is possible to execute the reference function with different inputs to capture more semantics information and generate the signature.
Furthermore,
we could adopt the metrics from testing to evaluate each run of the reference function, such as delta code coverage~\cite{rawat2017vuzzer}.
Specifically, \binCMP\ only records the signature extracted from the execution covering enough new code of the function, which is not executed before.
It is left as future work.

\section{Related Work}\label{sec:relate}
Binary code similarity comparison~(or clone detection) has many important applications in fields of software engineering and security, typically including plagiarism detection~\cite{luo2014semantics,luo2017semantics}, bug detection~\cite{pewny2015cross}, malware analysis~\cite{ming2017binsim},~etc.

Syntax and structural features are widely adopted to detect binary clone code.
\mbox{S{\ae}bj{\o}rnsen~et al.}~\cite{saebjornsen2009detecting} detect binary clone code basing on opcode and operand types of instructions.
\mbox{Hemel~et al.}~\cite{hemel2011finding} treat binary code as text strings and measure similarity by data compression. The higher the compression rate is, the more similar the two pieces of binary code are.
\mbox{Khoo~et al.}~\cite{Khoo2013rendezvous} leverage n-gram to compare the control flow graph~(CFG) of binary code.
\mbox{David~et al.}~\cite{david2014tracelet} measure the similarity of binaries with the edit distances of their CFGs.
BinDiff~\cite{flake2004structural} and Kam1n0~\cite{ding2016kam1n0} extract features from the CFG and call graphs to search binary clone functions.

As discussed earlier in this paper, the main challenge of binary code similarity comparison is semantics-equivalent code transformation resulting from link-time optimization, obfuscation, etc.
Because of the transformation, representations of binary code are altered tremendously, even though the code is compiled from the same code base.
Therefore, syntax and structure-based methods become ineffective, and semantics-based methods prevail.
\mbox{Jhi~et al.}~\cite{jhi2011value} and \mbox{Zhang~et al.}~\cite{zhang2012first} leverage runtime invariants of binaries to detect software and algorithm plagiarism.
\mbox{Ming~et al.}~\cite{ming2017binsim} infer the lineage of malware by code similarity comparison with the system call traces as the semantic signature.
However, those solutions require the execution of binary programs and cannot cover all target functions.
\mbox{Egele et al. \cite{egele2014blanket}} propose blanket execution to match binary functions with full code coverage which is achieved at the cost of detection accuracy.
\mbox{Luo~et al.}~\cite{luo2014semantics,luo2017semantics} and \mbox{Zhang~et al.}~\cite{zhang2014program} detect software plagiarism by symbolic execution.
Although their methods are resilient to code transformation, symbolic execution is trapped in the performance of SMT/SAT solvers which cannot handle all cases, e.g.,~indirect calls.
\mbox{David~et al.} propose Esh~\cite{david2016statistical} which decomposes the CFG of a binary function into small blocks and measures the similarity of the small blocks basing on a statistical model.
However, the boundaries of CFG blocks would be changed by code transformation, affecting the accuracy of the method.
BinSequence~\cite{huang2017binsequence} explores the control flow graphs of binary functions and aligns the paths for similarity measurement.
Then, it would suffer from control flow transformation, e.g., control flow flattening.

More recently, with the prevalence of IoT devices, binary code similarity comparison is proposed to perform on ARM and MIPS, or even across architectures.
Multi-MH~\cite{pewny2015cross}, discovRE~\cite{eschweiler2016discovre}, Genius~\cite{feng2016scalable}, and Xmatch~\cite{feng2017extracting} are proposed to detect known vulnerabilities and bugs in multi-architecture binaries via code similarity comparison.
BinGo~\cite{chandramohan2016bingo}, CACompare~\cite{hu2017binary}, and GitZ~\cite{david2017similarity} are proposed to analyze the similarity of binary code across architectures as well.
However, discovRE and Genius still heavily depend on the CFG of a binary function.
Xmatch extracts symbolic expressions of a binary function as the features, and treats them as sets for similarity comparison, ignoring the relative order~(semantics information) of the expressions.
Multi-MH, BinGo and \mbox{CACompare} sample a binary function with random values to capture corresponding I/O values as the signature, while the random values are meaningless that they merely trigger limited behaviors of the function. Thus, it is difficult for them to cover the core semantics of a binary function.
Similar to Esh, GitZ bases the analysis on blocks of functions as well.
It lifts LLVM-IR from function blocks, and counts on the optimization strategies of Clang to normalize the representations of the IR code. Then, it measures the similarity of IR code syntactically.
Following their previous work~(Esh and GitZ), \mbox{David et al.} propose FirmUp~\cite{david2018firmup} to detect known vulnerabilities of firmware.
They consider the similarity comparison as a back-and-forth game which further ensures the accuracy of the detection.
BinArm~\cite{shirani2018binarm} is propose to detect known vulnerabilities of firmware as well.
It introduces a multi-stage strategy which firstly filters functions with the syntax and structure information, then performs graph matching of control flow graphs.

Additionally, machine learning techniques are also adopted for binary code similarity comparison.
\mbox{Xu et al.}~\cite{xu2017neural} leverage the neural network to encode CFGs of binary code into vectors.
$\alpha$Diff~\cite{liu2018diff} trains the convolutional neural network with raw bytes of binary code, generating the best parameter values for the model. Then they apply the model to comparing the similarity of binary code.
Thus, essentially, the two solutions still process with syntax and structure features.
Asm2Vec~\cite{ding2019asm2vec} considers the assembly code disassembled from the binary code as documents. It attempts to discover the semantics hidden in the co-occurrence relationships among the assembly tokens, and adopts the most representative instructions as the feature of a function.
However, Asm2Vec still originates in the text of assembly code as well.
It is not robust enough to the semantics-equivalent code transformation, as indicated in the experiments~(\S\ref{sec:eva:cc:comp}).
Inspired by the work of \mbox{Luo et al.}~\cite{luo2014semantics}, \mbox{Zuo et al.}~\cite{zuo2019neural} firstly compute the likeness of basic blocks, then align the basic blocks of a path, and finally infer the similarity of code components with multiple paths.
They regard binary code as natural language, embedding instructions basing on word2vec~\cite{mikolov2013efficient}, then adopting the neural machine
translation model to compare basic blocks in a deep learning manner.
Comparing to the previous basing on symbolic execution, the solution becomes much more efficient.

To sum up, the topic of binary code similarity comparison mainly focuses on two points:
\begin{enumerate*}[label=\roman*)]
	\item what signature to adopt, such as opcodes and operand types~(syntax), CFG~(structure) and system calls~(semantics);
	
	\item how to capture the signatures, such as statically disassembling, sampling, or dynamically running,~etc.
\end{enumerate*}
\binCMP\ leverages the combination of output values, comparison operand values, and invoked standard library functions as the signature which is able to better reveal the semantics of a binary function.
Besides, it captures the signature via both execution and emulation, which not only ensures the richness of semantics, but also covers all target functions to be analyzed.

\section{Conclusion}\label{sec:conclusion}
Binary code similarity comparison is a fundamental methodology which has many important applications in fields of software engineering and security.
In this paper, we propose \binCMP\ to compare the similarity of binary code.
\binCMP\ completely relies on semantics-based signatures which are extracted either in a static or in a dynamic manner, via~(emulated) executions.
Thus, it is able to achieve high comparison accuracy and coverage at the same time.
Besides, to balance accuracy and efficiency, in addition to the longest common subsequence algorithm, the accurate string matching method, \binCMP\ also adopts the approximate matching technique SimHash for the function signature similarity measurement.
The experimental results show that \binCMP\ not only is robust to the semantics-equivalent code transformation caused by different compilation settings, commonly-used obfuscations, and variant ISAs, but also fulfills the function comparison efficiently.
Additionally, \binCMP\ also achieves better performance than the state-of-the-art solutions as well as industrial tool of binary code similarity comparison.

\section*{Acknowledgment}
The authors would like to thank Dr. Xiangyu Zhang and the anonymous reviewers for their insightful comments which help to improve the manuscript.
This work is partially supported by 
the Key Program of National Natural Science Foundation of China~(No.~U1636217), 
the National Key Research and Development Program of China~(No.~2016YFB0801201, No.~2016QY071401), 
and the Major Project of Ministry of Industry and Information Technology of China~(No.~[2018]~282).
They would like to thank the support from the Ant Financial Services Group as well.

\InputIfFileExists{BinMatch.bbl}

\end{document}